\documentclass[%
 reprint,
superscriptaddress,
 amsmath,amssymb,
 aps,
]{revtex4-2}

\usepackage[utf8]{inputenc}
\usepackage{graphicx}
\usepackage{color}
\usepackage{physics}
\usepackage{braket}
\usepackage{comment}
\usepackage{mathrsfs}
\usepackage{enumerate}
\usepackage[version=4]{mhchem}
\usepackage[breaklinks=true]{hyperref}
\usepackage{lipsum}
\usepackage{algorithmic}
\usepackage{algorithm}

\newcommand{\mr}[1]{\mathrm{#1}}

\newcommand{\mcl}[1]{\mathcal{#1}}
\newcommand{\bbC}{\mathbb{C}}
\newcommand{\bbR}{\mathbb{R}}
\newcommand{\bbZ}{\mathbb{Z}}
\newcommand{\bbN}{\mathbb{N}}

\newcommand{\imax}{\mathrm{max}}

\newcommand{\poly}[1]{\mathrm{poly} \left( #1 \right)}
\date{\today}
\usepackage{amsthm}
\theoremstyle{definition}
\newtheorem{theorem}{Theorem}[]
\newtheorem{definition}[theorem]{Definition}
\newtheorem{proposition}[theorem]{Proposition}
\newtheorem{lemma}[theorem]{Lemma}

\newtheorem{theoremA}{Theorem}[]

\newtheorem{propositionA}[theoremA]{Proposition}

\newtheorem{theoremB}{Theorem}[]

\newtheorem{propositionB}[theoremB]{Proposition}

\newtheorem{theoremC}{Theorem}[]

\newtheorem{propositionC}[theoremC]{Proposition}

\newtheorem{theoremD}{Theorem}[]

\newtheorem{propositionD}[theoremD]{Proposition}

\newtheorem{theoremE}{Theorem}[]

\begin{document}
\title{Nearly optimal quasienergy estimation and eigenstate preparation \\ of time-periodic Hamiltonians by Sambe space formalism}

\author{Kaoru Mizuta}
\email{mizuta@qi.t.u-tokyo.ac.jp}
\affiliation{Department of Applied Physics, Graduate School of Engineering, The University of Tokyo, Hongo 7-3-1, Bunkyo, Tokyo 113-8656, Japan}
\affiliation{Photon Science Center, Graduate School of Engineering, The University of Tokyo, Hongo 7-3-1, Bunkyo, Tokyo 113-8656, Japan}
\affiliation{RIKEN Center for Quantum Computing (RQC), Hirosawa 2-1, Wako, Saitama 351-0198, Japan}

\begin{abstract}
Time-periodic (Floquet) systems are one of the most interesting nonequilibrium systems.
As the computation of energy eigenvalues and eigenstates of time-independent Hamiltonians is a central problem in both classical and quantum computation, quasienergy and Floquet eigenstates are the important targets.
However, their computation has difficulty of time dependence; the problem can be mapped to a time-independent eigenvalue problem by the Sambe space formalism, but it instead requires additional infinite dimensional space and seems to yield higher computational cost than the time-independent cases.
It is still unclear whether they can be computed with guaranteed accuracy as efficiently as the time-independent cases.
We address this issue by rigorously deriving the cutoff of the Sambe space to achieve the desired accuracy and organizing quantum algorithms for computing quasienergy and Floquet eigenstates based on the cutoff.
The quantum algorithms return quasienergy and Floquet eigenstates with guaranteed accuracy like Quantum Phase Estimation (QPE), which is the optimal algorithm for outputting energy eigenvalues and eigenstates of time-independent Hamiltonians.
While the time periodicity provides the additional dimension for the Sambe space and ramifies the eigenstates, the query complexity of the algorithms achieves the near-optimal scaling in allwable errors.
In addition, as a by-product of these algorithms, we also organize a quantum algorithm for Floquet eigenstate preparation, in which a preferred gapped Floquet eigenstate can be deterministically implemented with nearly optimal query complexity in the gap.
These results show that, despite the difficulty of time-dependence, quasienergy and Floquet eigenstates can be computed almost as efficiently as time-independent cases, shedding light on the accurate and fast simulation of nonequilibrium systems on quantum computers.

\end{abstract}

\maketitle

%=================================
% Section: Introduction
%=================================
\section{Introduction}\label{Sec:Introduction}

A time-dependent Hamiltonian $H(t)$ is called time-periodic when it satisfies $H(t)=H(t+T)$ with some period $T$.
Floquet systems, driven by time-periodic Hamiltonians, are one of the most important classes of nonequilibrium systems.
They describe typical laser-irradiated materials, and provide their various applications such as optical manipulation of phases (Floquet engineering) \cite{Oka2009photovoltaic,Oka2019review}.
They also host nonequilbrium phases absent in equilibrium systems, such as Floquet topological phases \cite{Kitagawa2010-ct,Rudner2013-yk,Harper2019-pk} and Floquet time crystals \cite{Khemani2016-dq,Else2016-mg,Khemani2019-pf}.
With various other dynamical phenomena like Floquet prethermalization \cite{Abanin2015-zg,Mori2016-tp,Kuwahara2016-yn,Abanin2017-zs,Abanin2017-li}, Floquet many-body localization \cite{Ponte_2015_MBL,Lazaridez_2015_MBL,Abanin2016-bd_MBL,Bordia2017-lq_MBL}, and Floquet quantum many-body scars \cite{Mukherjee_2020_scar,Sugiura_2021_Scar,Mizuta_2020_Scar}, computing the properties of Floquet many-body systems is of great interest in today's nonequilibrium physics.

The computation of quasienergy and a Floquet eigenstate is one of the most fundamental and significant tasks in Floquet analysis.
They give a solution of Schr\"{o}dinger equation under a time-periodic Hamiltonian, playing roles of energy eigenvalue and a energy eigenstates respectively.
Not only is this task essential for characterizing the dynamical and steady state properties, but also it becomes an extension of the fundamental problem computing energy eigenvalue and eigenstates of time-independent systems \cite{Abrams1999-ry,Aspuru-Guzik2005-nr}.
The most common approach is the so-called Sambe space formalism \cite{Sambe1973}, with which the problem can be mapped to a time-independent eigenvalue problem.
It allows us to import powerful computational techniques for time-independent systems to Floquet analysis, such as various perturbation theories in low-frequency and high-frequency regimes \cite{Sambe1973,Guerin2003-cb,Mikami2016-bw,Rodriguez-Vega2018-fm}.
However, the difficulty is conserved in a sense;while we can avoid dealing with time-dependence, the mapped time-independent problem requires infinite dimension.
In practice, a cutoff of the dimension is empirically introduced, while we suffer from the increasing computational cost both in time and space with the large cutoff or the increasing error with the small cutoff.
It has been still missing whether we can compute quasienergy and Floquet eigenstates with simultaneously supporting efficiency and accuracy. 
In particular, it has been of great interest whether their computation can be as efficient as that for time-independent systems despite the existence of time-dependency or infinite dimension.

Recently, quantum computation has brought a different perspective to Floquet analysis \cite{Fauseweh_Quantum_2023,Mizuta_Quantum_2023,Eckstein2023largescale,Mizuta_2023_multi}.
Quantum computation has provided a promising way of simulating large-scale quantum materials in the past decades, useful for quantum dynamics \cite{Lloyd1996-ko,Low2017-qsp,Low2019-qubitization}, energy eigenvalues and eigenstates \cite{Yu_Kitaev1995-oi,Cleve1998-ou,Poulin_2009_ground_st,Ge2019-cr-ground_st,Lin2020-ground_state}, and thermal equilibrium states \cite{Poulin2009-thermal,Bilgin2010-thermal,Riera2012-thermal,chowdhury2016-thermal,Gilyen2019-qsvt}.
Floquet analysis is expected to be improved by quantum algorithms in efficiency and accuracy compared to classical algorithms like them.
For instance, time-evolution of Floquet systems can be simulated as efficiently as time-independent systems with guaranteed accuracy, achieving the near optimal query complexity in time and accuracy \cite{Mizuta_Quantum_2023,Mizuta_2023_multi}.
However, such well-organized quantum algorithms that rigorously guarantee efficiency and accuracy are limited to time-evolution.
Although Ref. \cite{Fauseweh_Quantum_2023} provides a variational quantum algorithm for Floquet eigenstates based on the Sambe space space formalism, it provides heuristic solutions that are available solely on noisy intermediate-scale quantum (NISQ) devices.
The question of the computational complexity of computing quasienergy and Floquet eigenstates remains unanswered, even with the knowledge of quantum algorithms.

In this paper, we organize a nearly optimal quantum algorithms for quasienergy and Floquet eigenstates, and clarify the complexity of this task from the viewpoint of quantum computation.
Our results are presented in two steps.
First, we provide a rigorous cutoff in the Sambe space formalism to guarantee the accuracy of quasienergy and Floquet eigenstates.
This result is valid both for classical and quantum computation, and characterizes additional resources for simulating Floquet systems.
We then construct a Floquet analog of the quantum phase estimation (QPE) algorithm \cite{Yu_Kitaev1995-oi,Cleve1998-ou}, called ``Floquet QPE".
QPE is a quantum algorithm that computes energy eigenvalues and prepares the corresponding eigenstates after measurement with optimal query complexity.
With efficiently reproducing the Sambe space on quantum computers based on the rigorous cutoff, the Floquet QPE returns quasienergy with guaranteed accuracy and outputs two different types of Floquet eigenstates that are ramified by the Sambe space.
Importantly, the query complexity achieves nearly optimal scaling of the QPE, and the number of qubits differs from that of the QPE by only a small logarithmic number.
It is concluded that quasienergy and Floquet eigenstates can be computed as efficiently as energy eigenvalues and eigenstates with overcoming the difficulty of time dependence.
As an application of Floquet QPE, we provide the eigenstate preparation algorithms, in which a preferable gapped Floquet eigenstate can be deterministically prepared from an initial state with nonzero overlap.
The eigenstate preparation for time-periodic systems can also be almost as efficient as the optimal protocol for time-independent systems \cite{Rall2021-qpe,Martyn2021-grand-unif} owing to the near optimality of the Floquet QPE.
Our results will shed light on the complexity of simulating nonequilibruim quantum many-body systems, with opening up a new avenue for their fast and accurate computation by future quantum computers.

The rest of this paper is organized as follows.
In Section \ref{Sec:Preliminary}, we provide a brief review on Floquet theory and QPE in order to make this paper self-contained.
Section \ref{Sec:Summary} provides the summary of our result, and Sections \ref{Sec:Accuracy_Sambe}-\ref{Sec:Floquet_preparation} are devoted to its detail.
In Section \ref{Sec:Accuracy_Sambe}, we rigorously clarify the relation between the cutoff of the Sambe space formalism and the accuracy of quasienergy.
In Sections \ref{Sec:Floquet_QPE_physical} and \ref{Sec:Floquet_QPE_Sambe}, we explicitly organize two quantum algorithms reflecting the fact that there are two kinds of Floquet eigenstates as outputs.
Significantly, both of them achieve nearly optimal query complexity as large as time-independent cases.
The eigenstate preparation algorithm based on the Floquet QPE is provided in Section \ref{Sec:Floquet_preparation}.
We conclude this paper in Section \ref{Sec:Discussion}, where we discuss potential applications of our results.
%=================================
% Section: Preliminaries
%=================================
\section{Preliminary}\label{Sec:Preliminary}

\subsection{Notation}

Throughout the paper, we are interested in $N$-qubit quantum many-body systems, whose Hilbert space is denoted by $\mcl{H} \simeq \bbC^{2^N}$.
The norm of a bounded operator $A$ on a Hilbert space, i.e. $\norm{A}$, denotes the operator norm.
We use the Landau symbols $\order{\cdot}$, $\Theta (\cdot)$, and $\Omega (\cdot)$, where the variables in them move independently.

To express a state that deviates from an ideal state $\ket{\psi}$, we use $\ket{\psi}+\order{\delta}$ meaning a density operator $\rho$ defined on the same Hilbert space such that
\begin{equation}
    \norm{\rho-\ket{\psi}\bra{\psi}} \leq \delta,
\end{equation}
for $\delta \in [0,1)$.
Also, $O+\order{\delta}$ for an operator $O$ on the Hilbert space $\mcl{H}$ denotes a completely-positive and trace-preserving (CPTP) map whose distance from $O (\cdot) O^\dagger$ by the diamond norm is less than $\delta$.

As we will discuss later, quasienergy of time-periodic Hamiltonians has periodicity in contrast to energy eigenvalues.
To characterize it, we use the symbol $\text{($a$ mod. $b$)}$ for $b>0$, which is defined on $[-b/2,b/2)$.

\subsection{Floquet theory}\label{Subsec:Floquet_theory}

Floquet theory is a theoretical framework for time-periodic systems, in which quasienergy and Floquet eigenstates play a central role through Floquet theorem \cite{Oka2019review}.
Consider Schr\"{o}dinger equation under a time-periodic Hamiltonian on $\mcl{H}$,
\begin{equation}\label{Eq:Schrodinger_eq}
    i \dv{t} \ket{\psi(t)} = H(t) \ket{\psi(t)}, \quad H(t+T)=H(t),
\end{equation}
where $T$ is a period.
Floquet theorem states that the solution is given by
\begin{equation}\label{Eq:Floquet_theorem}
    \ket{\psi(t)} = \sum_{n=1}^{\mr{dim}(\mcl{H})} c_n e^{-i\epsilon_n t} \ket{\phi_n(t)}, \quad \ket{\phi_n(t+T)} = \ket{\phi_n(t)},
\end{equation}
with $c_n \in \bbC$, $\varepsilon_n \in \bbR$, and $\ket{\phi_n(t)} \in \mcl{H}$.
The set of states $\{ \ket{\phi_n(t)} \}_{n=1}^{\mr{dim}(\mcl{H})}$ forms a complete orthonormal basis of $\mcl{H}$ and each state is called a Floquet eigenstate.
The real value $\epsilon_n \in \bbR$ is called quasienergy.
The set of coefficients $\{ c_n \}_n$ is determined by the initial state $\ket{\psi(0)}$ as $c_n = \braket{\phi_n(0)|\psi(0)}$, and the completeness implies $\sum_n |c_n|^2 =1$.
As the solution of time-independent Schr\"{o}dinger equation is expanded by $\ket{\psi(t)} = \sum_n c_n e^{-i E_n t} \ket{\phi_n}$ with $H\ket{\phi_n}=E_n \ket{\phi_n}$ and $c_n = \braket{\phi_n|\psi(0)}$, quasienergy and Floquet eigenstate are respectively counterparts of energy eigenvalue and eigenstate in time-independent systems.

Next, we introduce the ways to compute quasienergy and Floquet eigenstates.
One way relies on the time-evolution operator
\begin{eqnarray}
    U(t;t_0) &\equiv& \mcl{T} \exp \left( -i \int_{t_0}^t \dd t^\prime H(t^\prime) \right) \label{Eq:time_evolution_op} \\
    &=& \sum_n e^{-i \epsilon_n (t-t_0)} \ket{\phi_n(t)}\bra{\phi_n(t_0)}, \label{Eq:time_evolution_eigenbasis}
\end{eqnarray}
where the second equality comes from Floquet theorem, Eq. (\ref{Eq:Floquet_theorem}).
The time-periodicity of $\ket{\phi(t)}$ states that pairs of $(\epsilon_n, \ket{\phi_n(0)})$ are obtained by diagonalizing the so-called Floquet operator,
\begin{equation}\label{Eq:Floquet_op}
    U(T;0) = \sum_n e^{-i \epsilon_n T} \ket{\phi_n(0)}\bra{\phi_n(0)}.
\end{equation}
The quasienergy $\epsilon_n$ is defined modulo $\omega$ and the instantaneous Floquet eigenstate $\ket{\phi_n(t)}$ is obtained by $\ket{\phi_n(t)} = e^{i \epsilon_n t} U(t;0) \ket{\phi_n(0)}$.
However, it is preferable to avoid computing the time-evolution operators since the time-ordered product requires fine time discretization.

The most common approach is the Sambe space formalism \cite{Sambe1973}, which relies on the Fourier transform,
\begin{eqnarray}
    H(t) &=& \sum_{m \in \bbZ} H_m e^{-im\omega t}, \\
    \ket{\phi_n(t)} &=& \sum_{l \in \bbZ} e^{-il\omega t} \ket{\phi_n^l}, \label{Eq:def_Floquet_state} 
\end{eqnarray}
with the frequency $\omega \equiv 2\pi/T$.
Substituting these relations into Eqs. (\ref{Eq:Schrodinger_eq}) and (\ref{Eq:Floquet_theorem}), we get the following eigenvalue problem,
\begin{eqnarray}\label{Eq:Sambe_eigenequation}
    H_\mr{F} \ket{\Phi_n} = \epsilon_n \ket{\Phi_n},
\end{eqnarray}
by adding a set of states $\{ \ket{l}_f \}_{l \in \bbZ}$ labeling Fourier indices $l \in \bbZ$.
Here, the time-independent Hamiltonian $H_\mr{F}$ is called a Floquet Hamiltonian given by
\begin{equation}\label{Eq:Floquet_Hamiltonian}
    H_\mr{F} = \sum_{l,m \in \bbZ} \ket{l+m}\bra{l}_f \otimes H_m - \sum_{l \in \bbZ} l \omega \ket{l}\bra{l}_f \otimes I.
\end{equation}
The eigenstate $\ket{\Phi_n}$ is described by
\begin{equation}\label{Eq:Eigenstate_Sambe}
    \ket{\Phi_n} = \sum_{l \in \bbZ} \ket{l}_f \ket{\phi_n^l},
\end{equation}
giving all the Fourier components $\{ \ket{\phi_n^l} \}_{l \in \bbZ}$ without integration.
The state $\ket{\Phi_n}$ is also called a Floquet eigenstate since it has one-to-one correspondence with $\ket{\phi_n(t)}$.
The time-independent problem by Eq. (\ref{Eq:Sambe_eigenequation}) is defined on a $L^2$ space,
\begin{equation}
    \mcl{H}^\infty = \left\{ \ket{\Psi} \in \mr{span} (\{ \ket{l}_f \}_{l \in \bbZ}) \otimes \mcl{H} \, | \, \norm{\ket{\Psi}} < \infty \right\},
\end{equation}
called the Sambe space.
Instead of getting rid of time-dependency of the problem, its difficulty is translated into the infinite-dimensionality of the Sambe space $\mcl{H}^\infty$.

We remark the equivalence of quasienergy.
Using $(\epsilon_n-l\omega, e^{il\omega t}\ket{\phi_n(t)})$ in Eq. (\ref{Eq:Floquet_theorem}) instead of $(\epsilon_n,\ket{\phi_n(t)})$ also gives the same solution for arbitrary $l \in \bbZ$.
The quasienergy $\epsilon_n$ is uniquely characterized modulo $\omega$, and it is sufficient to consider quasienergy contained in a single Brillouin zone (BZ) defined by
\begin{equation}
    \mr{BZ}_l = [(-l-1/2)\omega, (-l+1/2)\omega ),
\end{equation}
for a certain integer $l \in \bbZ$.
We often use $\mr{BZ} \equiv \mr{BZ}_0 = [-\omega/2,\omega/2)$.
In the Sambe space formalism, it appears as equivalent pairs of eigenvalues and eigenstates $(\epsilon_n-l\omega, \ket{\Phi_n^l})$ for $l \in \bbZ$, satisfying
\begin{eqnarray}
    && H_\mr{F} \ket{\Phi_n^l} = (\epsilon_n-l\omega) \ket{\Phi_n^l}, \label{Eq:Sambe_eigenequation_shifted} \\
    && \ket{\Phi_n^l} = \mr{Add}_l \ket{\Phi_n} = \sum_{l^\prime \in \bbZ} \ket{l^\prime + l}_f \ket{\phi_n^{l^\prime}}. \label{Eq:Eigenstate_Sambe_shifted}
\end{eqnarray}
The eigenstate $\ket{\Phi_n^l}$ stores Fourier components of the equivalent Floquet eigenstate $e^{il\omega t} \ket{\phi_n(t)}$.
It is sufficient to pick up $\mr{dim}(\mcl{H})$ different eigenstates, and we usually use eigenvectors of $H_\mr{F}$ with $\epsilon_n \in \mr{BZ}$, which we denote by $\{ \ket{\Phi_n} \}_{n=1}^{\mr{dim}(\mcl{H})}$.

\subsection{Quantum phase estimation}\label{Subsec:QPE}

Quantum phase estimation (QPE) is a fundamental quantum algorithm that efficiently computes some pairs of eigenvalues and eigenstates of a Hamiltonian $H$ with an initial state having large overlap with target eigenstates.
We hereby review its recent versions \cite{Rall2021-qpe,Martyn2021-grand-unif} based on quantum singular value transformation (QSVT) \cite{Gilyen2019-qsvt} instead of the primitive versions in the 1990s \cite{Yu_Kitaev1995-oi,Cleve1998-ou}.
Suppose we have an initial state $\ket{\psi}$ expanded by
\begin{equation}
    \ket{\psi} = \sum_n c_n \ket{\phi_n}, \quad c_n = \braket{\phi_n | \psi},
\end{equation}
where each state $\ket{\phi_n}$ is an eigenstate of $H$ with an eigenvalue $E_n$.
The Hamiltonian $H$ is rescaled so that every eigenvalue $E_n$ belongs to $[-1/2,1/2)$.

Here, we consider two types of QPE algorithms.
The first one transforms the initial state by
\begin{equation}\label{Eq:standard_QPE_result}
    \ket{0}_b \ket{\psi} \to \sum_n c_n \ket{(E_n)_b}_b \ket{\phi_n} + \order{\delta}.
\end{equation}
The $b$-qubit register $\ket{\cdot}_b$ stores a $b$-bit binary $(E_n)_b$, given by
\begin{equation}\label{Eq:stored_binary}
    (E_n)_b = \frac{\left\lfloor 2^b E_n \right\rfloor}{2^b} \in \left\{ -\frac12 , -\frac12 + 2^{-b}, \hdots, \frac12 - 2^{-b} \right\}.
\end{equation}
If the size of the register is set to $b \in \Theta(\log (1/\varepsilon))$, the eigenvalue is approximated by $E_n$ by $|E_n-(E_n)_b| \leq \varepsilon$.
The parameters $\varepsilon,\delta \in (0,1)$ represent errors in the eigenvalue and output state, respectively.
The measurement on the register in computational basis probabilistically returns one of the eigenvalues $E_n$ within an allowable error $\varepsilon$, and then the resulting state is projected to the eigenstate $\ket{\phi_n}$.
In addition, the transformation by Eq. (\ref{Eq:standard_QPE_result}) is compatible with uncomputation, and thus it can be employed as a subroutine of various algorithms such as quantum linear system problems \cite{Harrow2009-hhl}.
Such a protocol is known to be available if we assume rounding promise \cite{Rall2021-qpe};

\begin{definition}\label{Def:rounding_promise}
\textbf{(Rounding promise)}

A Hamiltonian $H$ has rounding promise $\nu \in (0,1)$ if every eigenvalue $E_n$ avoids width-$\nu$ fractions as
\begin{equation}
    E_n \notin \bigcup_{x=-2^{b-1}}^{2^{b-1}} \left[ \frac{x-\nu/2}{2^b}, \frac{x+\nu/2}{2^b} \right),
\end{equation}
\end{definition}

The second type of QPE works without the assumption of rounding promise, which is given by
\begin{eqnarray}
    \ket{0}_b \ket{\psi} &\to& \sum_n c_n \ket{\overline{(E_n)_b}}_b \ket{\phi_n} + \order{\delta}, \label{Eq:Standard_QPE_wo_rounding1}\\
    \ket{\overline{(E_n)_b}}_b &=& p_0^n \ket{(E_n)_{b0}}_b + p_1^n \ket{(E_n)_{b1}} \label{Eq:Standard_QPE_wo_rounding2}
\end{eqnarray}
with some weights $p_0^n, p_1^n \in \bbC$ such that $|p_0^n|^2 + |p_1^n|^2 = 1$.
The register stores a superposition of two different $b$-bit binaries $(E_n)_{b0}$ and $(E_n)_{b1}$, both of which are guaranteed to approximate $E_n$ up to the additive error $\varepsilon$.
Although this protocol is not compatible with uncomputation, it is still useful for extracting accurate energy eigenvalues.

These QPE algorithms can be executed by QSVT, whose building block is a controlled time-evolution $C[e^{2\pi iH}]$ or a controlled block-encoding $C[O_H]$ given by
\begin{eqnarray}
    C[e^{2\pi iH}] &=& \ket{0}\bra{0} \otimes I + \ket{1}\bra{1} \otimes e^{2\pi iH}, \\
    C[O_H] &=& \ket{0}\bra{0} \otimes I + \ket{1}\bra{1} \otimes O_H.
\end{eqnarray}
Here, the block-encoding $O_H$ is a unitary gate satisfying
\begin{equation}
    (\bra{0}_a \otimes I) O_H (\ket{0}_a \otimes I) = H,
\end{equation}
which is implemented with a $n_a$-qubit ancilla referenvce state $\ket{0}_a$.
Block encoding is constructed when $H$ is a linear combination of unitary, a sparse-access matrix, and so on \cite{Low2019-qubitization}.
The QPE algorithm has following preferable properties in computing an eigenenergy and  eigenstate.
\begin{enumerate}[(a)]
    \item Guaranteed accuracy; \\
    We can achieve guaranteed accuracy with arbitrarily small $\varepsilon,\delta \in (0,1)$ for energy eigenvalue by $|E_n-(E_n)_b| \leq \varepsilon$ and for output states by Eq. (\ref{Eq:standard_QPE_result}).

    \item Efficiency of algorithm; \\
    The query complexity in $C[e^{2\pi i H}]$ or $C[O_H]$ is optimal or nearly optimal in the allowable errors $\varepsilon,\delta$ and the rounding promise $\nu$.
    Namely, the query complexity $q_\mr{QPE}$ achieves the scaling
    \begin{equation}\label{Eq:query_standard_QPE}
        \qquad q_\mr{QPE} \in \begin{cases} \Theta \left( \frac{1}{\varepsilon \nu} \log (1/\delta) \right) & \text{in $C[e^{2\pi i H}]$}, \\
        \Theta \left( \left( \frac{1}{\varepsilon \nu} + \frac{\log \nu}\nu \right) \log (1/\delta) \right) & \text{in $C[O_H]$}, 
        \end{cases}
    \end{equation}
    as shown in Table \ref{Table:comparison_algorithms} (The cost for the case without the rounding promise corresponds to $\nu \in \order{1}$).
    The number of ancilla qubits is at most logarithmic in all the parameters.
    
    \item Measurement outcome; \\
    Measurement of the register projects the state onto the subspace of eigenstates with the measured energy eigenvalue.
    Moreover, the probability of an outcome $\ket{\phi_n}$ is given by the weight in the initial state as
    \begin{equation}
        p_n = |c_n|^2 = |\braket{\phi_n|\psi}|^2.
    \end{equation}
\end{enumerate}

Combining the properties (a)--(c), the total query complexity for obtaining a preferable eigenstate (e.g. ground state, low-energy excited state) is $q_\mr{QPE}$ multiplied by $\order{p_n^{-1}}$ (based on iteration until success) or by $\order{p_n^{-1/2}}$ (based on quantum amplitude amplification, QAA \cite{Brassard_2002_qaa}).
In general, identifying a preferable energy eigenvalue or eigenstate of local Hamiltonians with polynomially small accuracy is as difficult as a QMA-hard problem \cite{Kitaev2002-kz} and $p_n$ is expected to be exponentially small in the system size $N$.
The property (c) dictates that making a good guess on a target eigenstate $\ket{\phi_n}$ with the initial state $\ket{\psi}$ having large overlap $|c_n| \geq 1/\poly{N}$ (although still difficult) leads to efficient computation of $(E_n,\ket{\phi_n})$.
\begin{table*}
    \centering
    \begin{tabular}{|c|c|c|c|c|c|}
        QPE & Initial state & Output & Oracle & Query complexity & Ancilla qubits \\ \hline \hline
       \begin{tabular}{c} Time-indep., $H$ \\
      (Refs. \cite{Martyn2021-grand-unif,Rall2021-qpe}) \end{tabular} & $\sum_n c_n \ket{\phi_n}$ & $(E_n,\ket{\phi_n})$ & $C[e^{2\pi i H}]$ &
      $\frac{1}{\varepsilon \nu} \log (1/\delta)$
        & $\order{\log (1/\varepsilon)}$ 
       \\ \hline
       \begin{tabular}{c} Time-indep., $H$ \\
      (Refs. \cite{Martyn2021-grand-unif,Rall2021-qpe}) \end{tabular} & $\sum_n c_n \ket{\phi_n}$ & $(E_n,\ket{\phi_n})$ & $C[O_H]$&
      $\left(\frac{1}{\varepsilon \nu} + \frac{\log (1/\nu)}{\nu} \right) \log (1/\delta)$
        & $n_a + \order{\log (1/\varepsilon)}$ 
       \\ \hline
       \begin{tabular}{c} Floquet, $H(t)$ \\ (Section \ref{Sec:Floquet_QPE_physical}) \end{tabular} & $\sum_n c_n \ket{\phi_n(0)}$ & $(\epsilon_n, \ket{\phi_n(t)})$ &$C[O_{H_m}]$ & $\frac{\alpha T + \log (1/\min (\varepsilon, \delta) \nu)}{\min (\varepsilon,\delta) \nu}  \log (1/\delta)$ & $n_a + \order{\log (\alpha T/\min (\varepsilon, \delta)) + \log \log (1/\nu)}$ 
       \\ \hline
       \begin{tabular}{c} Floquet, $H(t)$ \\ (Section \ref{Sec:Floquet_QPE_Sambe}) \end{tabular} & $\sum_n c_n \ket{\phi_n(0)}$ & $(\epsilon_n, \ket{\Phi_n})$ & $C[ O_{H_m}]$ & $\frac{\alpha T + N + \log (1/\varepsilon \nu \delta)}{\varepsilon \nu} \log (1/\delta) $
      & $n_a + \order{\log (\alpha T/\varepsilon) + \log N + \log \log (1/\nu \delta)}$  \\ \hline
    \end{tabular}
    \caption{Computational cost of the standard QPE and the Floquet QPE with rounding promise $\nu$. The first two rows shows QPE for time-independent $H$ with rescaling $\norm{H} \leq 1/2$. The third and fourth row shows our main result which computes quasienergy and various types of Floquet eigenstates as Theorems \ref{Thm:algorithm_phys} and \ref{Thm:algorithm_Sambe}.
    The symbol $n_a$ is the qubit number required for block-encoding.
    When the rounding promise $\nu$ is set by $\nu \in \order{1}$, the computational costs become those without rounding promise.
    }
    \label{Table:comparison_algorithms}
\end{table*}

%=================================
% Section: Summary of Results
%=================================
\section{Summary of results}\label{Sec:Summary}

In this section, we briefly summarize our results.
Throughout the paper, we consider a bounded time-periodic Hamiltonian,
\begin{equation}\label{Eq:Periodic_Hamiltonian}
    H(t) = \sum_{m=-M}^M H_m e^{-im\omega t}, \quad \norm{H_m} \leq \alpha,
\end{equation}
defined on an $N$-qubit quantum many-body system.
Here, the cutoff $M$ is assumed to satisfy $M \in \order{1}$.
The parameter $\alpha$ gives the energy scale of the whole system, and scales as $\alpha \in \poly{N}$ for local Hamiltonians.
It gives an upper bound on $H(t)$ by
\begin{equation}\label{Eq:norm_Ht}
    \norm{H(t)} \leq \sum_{m=-M}^M \norm{H_m} \leq (2M+1)\alpha, \quad ^\forall t \in \bbR.
\end{equation}
This setup covers generic Floquet many-body systems composed of spins, fermions, and bosons with finite and conserved particle numbers. 
In addition, while we focus on Eq. (\ref{Eq:Periodic_Hamiltonian}) in the main text, our results can be extended for time-periodic Hamiltonians such that
\begin{equation}
    H(t) = \sum_{m \in \bbZ} H_m e^{-im \omega t}, \quad \norm{H_m} \leq e^{- \order{|m|}},
\end{equation}
which are useful for wave packets of lasers for example \cite{Mizuta_Quantum_2023}.
See Appendix \ref{A_Sec:Exp_decay} for the extension.

Our central results are nearly optimal quantum algorithms that return pairs of quasienergy and a Floquet eigenstate with guaranteed accuracy, like QPE.
The underlying strategy is quite simple; we employ the standard QPE as a subroutine with using the Floquet Hamiltonian $H_\mr{F}$ derived by the Sambe space formalism.
This is an intuitive reason why the algorithm for quasienergy and a Floquet eigenstate can almost achieve the optimal scaling of the QPE, but we have several points to be distinguished from time-independent cases.
The first one is the infinite dimensionality of the Sambe space $\mcl{H}^\infty$.
To execute computation with finite resource, we have to truncate the dimension by restricting the Fourier index $l \in \bbZ$ to
\begin{equation}
    [L] = \{ -L+1, -L+2, \hdots, L-1, L \},
\end{equation}
with some cutoff $L \in \bbN$.
We use the truncated Floquet Hamiltonian defined by
\begin{equation}
    H_\mr{F}^L = \sum_{|m| \leq M} \sum_{l \in [L]; l+m \in [L]} \ket{l+m}\bra{l}_f - \sum_{l\in [L]} l\omega \ket{l}\bra{l}_f \otimes I,
\end{equation}
and then the truncation causes inaccuracies in quasienergy and Floquet eigenstates.
The second point is the output of the algorithm.
In contrast to time-independent systems, we have options in eigenstates at the end of the algorithms, i.e., $\ket{\phi_n(t)}$, which lives in the original Hilbert space $\mcl{H}$, or $\ket{\Phi_n}$ in the Sambe space $\mcl{H}^\infty$.
Although the standard QPE is efficient as a subroutine, it is nontrivial whether we can keep this efficiency and the guaranteed accuracy when we deal with the two points above.
We resolve this problem in the following ways and prove that the optimal scaling for time-independent cases is almost achievable for time-periodic cases.  

\textbf{Accuracy of the Sambe space formalism.---} 
We begin with finding a proper cutoff $L$ for the Sambe space $\mcl{H}^\infty$.
The cutoff is determined so that the estimated quasienergy from the truncated Sambe space can approximate the exact one with an allowable error $\varepsilon$.
In Section \ref{Sec:Accuracy_Sambe}, we prove the following property of the truncated Floquet Hamiltonian.

\begin{theorem}\label{Thm:Accuracy_quasienergy}
\textbf{(Accuracy of quasienergy, informal)}

We consider a time-periodic Hamiltonian $H(t)$, given by Eq. (\ref{Eq:Periodic_Hamiltonian}), with $M \in \order{1}$.
We set the cutoff of the Sambe space by
\begin{equation}\label{Eq:Cutoff}
    L \in \Theta \left( \alpha T + \log (1/\varepsilon) \right).
\end{equation}
Then, the existence of quasienergy $\epsilon_n \in \mr{BZ}$ implies the existence of eigenvalue $\tilde{\epsilon}_n^L$ of the truncated Floquet Hamiltonian $H_\mr{F}^L$ such that
\begin{equation}
    \frac{|\tilde{\epsilon}_n^L - \epsilon_n|}{\omega} \leq \varepsilon.
\end{equation}
Conversely, the existence of an eigenvalue $\tilde{\epsilon}_n^L \in [-\omega, \omega)$ in $H_\mr{F}^L$ also implies quasienergy $\epsilon_n$ such that 
\begin{equation}
\left| \left( \frac{\epsilon_n - \tilde{\epsilon}_n^L}{\omega} \right)  \quad \text{mod. $1$ } \right| \leq \varepsilon.
\end{equation}
\end{theorem}

The above theorem says that, when we want to reproduce quasienergy and a Floquet Hamiltonian from the truncated Sambe space within an error $\varepsilon$, it is sufficient to prepare $\Theta (\alpha T + \log (1/\varepsilon))$ additional dimensions.
This result gives $\Theta (\log L)$ as the number of additional qubits for computing the Floquet eigenstate $\ket{\Phi_n}$, which is irrelevant cost for the standard QPE.

\textbf{Floquet QPE.---} In Sections \ref{Sec:Floquet_QPE_physical} and \ref{Sec:Floquet_QPE_Sambe}, we explicitly organize quantum algorithms for quasienergy and a Floquet eigenstate.
Associating the solution of Eq. (\ref{Eq:Floquet_theorem}) with that of time-independent systems, the initial state $\ket{\psi}$ is expanded by
\begin{equation}
    \ket{\psi} = \sum_n c_n \ket{\phi_n(0)}, \quad c_n = \braket{\phi_n(0)|\psi}.
\end{equation}
This means that the initial guess on Floquet eigenstates is based on the physical Hilbert space, but not on the Sambe space lacking of physical interpretation.
In contrast, time-periodicity ramifies candidates of the output.
One is a pair of $(\epsilon_n, \ket{\phi_n(t)})$, where the transformation is given by
\begin{equation}
    \ket{0}_b \ket{\psi} \to \sum_n c_n \ket{(\epsilon_n)_b}_b \ket{\phi_n(t)} + \order{\delta},
\end{equation}
in the presence of rounding promise.
The $b$-qubit register stores a $b$-bit binary $(\epsilon_n)_b$ that approximates $\epsilon_n \in \mr{BZ}$ within an error $\varepsilon$ by
\begin{equation}
    \left| \left( \frac{\epsilon_n - (\epsilon_n)_b}{\omega} \right) \quad \text{mod. $1$ }\right| \leq \varepsilon.
\end{equation}
The other Floquet eigenstate $\ket{\Phi_n}$ in the Sambe space, which has one-to-one correspondence with $\ket{\phi_n(t)}$ can also be the output.
The quantum algorithm for this output executes the transformation,
\begin{equation}
    \ket{0}_b \ket{\psi} \to \sum_n c_n \ket{(\epsilon_n)_b}_b \ket{\Phi_n} + \order{\delta},
\end{equation}
which returns pairs of $(\epsilon_n,\ket{\Phi_n})$.
This version is advantageous for extracting Fourier components $\ket{\phi_n^l}$ and computing integration in time \cite{Oka2019review}.

Based on the Sambe space formalism with proper truncation, we organize quantum algorithms for both Floquet eigenstates $\ket{\phi_n(t)}$ and $\ket{\Phi_n}$, equipped with all the favorable properties of the standard QPE, (a)--(c) (See Section \ref{Subsec:QPE}).
In Section \ref{Sec:Floquet_QPE_physical}, the quantum algorithm returns pairs of $(\epsilon_n , \ket{\phi_n(t)})$ based on the QPE for the Floquet operator $U(T;0)$, where the cutoff $L$ is determined by the Lieb-Robinson bound of the Sambe space \cite{Mizuta_Quantum_2023}.
In Section \ref{Sec:Floquet_QPE_Sambe}, the quantum algorithm returns pairs of $(\epsilon_n , \ket{\Phi_n})$ based on the QPE for the truncated Floquet Hamiltonian $H_\mr{F}^L$, where the cutoff $L$ is determined by the bound Eq. (\ref{Eq:Cutoff}) derived in Section \ref{Sec:Accuracy_Sambe}.
These algorithms run with queries to controlled unitary gates $C[O_{H_m}]$, where each block-encoding $O_{H_m}$ embeds a Fourier component Hamiltonian $H_m$ by
\begin{equation}\label{Eq:block_encoding_Hm}
    \braket{0|O_{H_m}|0}_a = \frac{H_m}{\alpha_m}, \quad \norm{H_m} \leq \alpha_m.
\end{equation}
Without loss of generality, we replace the factor $\alpha$ of Eq. (\ref{Eq:Periodic_Hamiltonian}) by
\begin{equation}
    \alpha = \max_{|m| \leq M} (\alpha_m),
\end{equation}
which does not substantially change the scaling of the computational cost.
With or without rounding promise $\nu$, which will be properly extended for Floquet systems later, we obtain query complexity and ancilla qubits required for the algorithm as Table \ref{Table:comparison_algorithms} (or see Theorem \ref{Thm:algorithm_phys} and Theorem \ref{Thm:algorithm_Sambe} formally).
Notably, as discussed respectively in Sections \ref{Sec:Floquet_QPE_physical} and \ref{Sec:Floquet_QPE_Sambe}, these quantum algorithms are as efficient as the standard QPE except for logarithmic corrections, both in terms of query complexity and ancilla qubits in $\varepsilon,\delta,\nu$.

As applications of the Floquet QPE, we also provide the eigenstate preparation algorithms for Floquet systems in Section \ref{Sec:Floquet_preparation}.
In these quantum algorithms, a preferable gapped Floquet eigenstate, either $\ket{\phi_n(t)}$ or $\ket{\Phi_n}$, can be prepared from an initial state having an overlap with $\ket{\phi_n(0)}$.
Owing to the near optimality of the Floquet QPE, these eigenstate preparation algorithms can also be executed almost as efficiently as the optimal one for time-independent systems \cite{Rall2021-qpe}, as shown in Table \ref{Table:eigenstate_preparation} later.
Through the lens of the Floquet QPE or its application, the Floquet eigenstate preparation, we conclude that the computation of quasienergy and various Floquet eigenstates is as complicated as time-independent problems, despite the existence of time-dependency or the infinite-dimensionality caused by it.
At the same time, our results will provide promising tools for exploring nonequilibrium materials immediately after realization of the standard QPE on future quantum computers.
%=================================
% Section: Accuracy of Sambe space formalism
%=================================
\section{Accuracy of Sambe space formalism}\label{Sec:Accuracy_Sambe}

\subsection{Explicit cutoff for acurate quasienergy}

\begin{figure}
    \includegraphics[height=4cm, width=9cm]{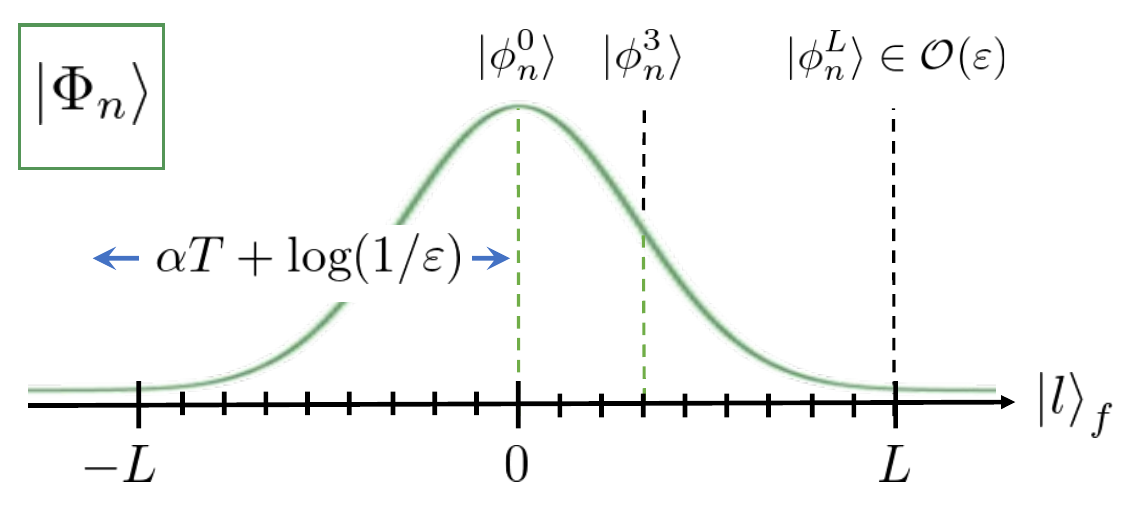}
    \caption{Decomposition of a Floquet eigenstate in the Sambe space, $\ket{\Phi_n}$. At each Fourier index $l \in \bbZ$, it has a component $\ket{\phi_n^l}$ whose magnitude rapidly decays in $l$ by Theorem \ref{Thm:Tail_Eigenstates}.} 
    \label{Fig_eigenstate}
\end{figure}

In this section, we derive the accuracy of the Sambe space formalism with truncation.
Namely, we derive the proper cutoff $L \in \Theta (\alpha T + \log (1/\varepsilon))$ so that quasienergy $\epsilon_n$ can be reproduced with an error up to $\varepsilon$, as shown by Propositions \ref{Prop:Accuracy_Floquet_Hamiltonian_eigenvalue} and \ref{Prop:Accuracy_quasienergy} in Section \ref{Sec:Accuracy_Sambe}.
The derivation relies on the generic property of every Fourier component of a Floquet eigenstate $\ket{\phi_n^l}$, i.e., the fact that the norm $\norm{\ket{\phi_n^l}}$ decays rapidly in $|l|$.
This property is summarized in the following theorem.

\begin{theorem}\label{Thm:Tail_Eigenstates}
\textbf{(Tails of Floquet eigenstates)}

Suppose that a Floquet eigenstate $\ket{\phi_n(t)}$ or equivalently $\ket{\Phi_n}$ has quasienergy $\epsilon_n \in \mr{BZ}=[-\omega/2,\omega/2)$ under the Hamiltonian, Eq. (\ref{Eq:Periodic_Hamiltonian}).
Then, every Fourier component exponentially decays as
\begin{equation}\label{Eq:Floquet_Eigenstate_Tails}
    \norm{\ket{\phi_n^l}} \leq  \exp \left( -\frac{|l|-1/2}{2M+1} + \frac{\sinh 1}{2\pi} \alpha T \right).
\end{equation}
\end{theorem}

This property can be derived by the analogy with a one-dimensional local quantum systems under a linear potential \cite{Zener1934,Wannier1960-oq}.
In this picture, the Fourier index $l$ plays a role of a coordinate, and the Fourier component $H_m$ and the linear term $-l\omega$ in $H_\mr{F}$ are respectively interpreted as the hopping by $m$ sites and the linear potential.
The exponential decay of $\norm{\ket{\phi_n^l}}$ is rigorously derived for single-particle quantum systems in Refs. \cite{Hone_1997_decay} or roughly concluded for generic systems with associating the localization under a one-dimensional linear potential in Refs. \cite{Lindner_2017_chiral,Rudner2020-handbook}.
However, we are not able to find the explicit bound valid for generic many-body systems like Eq. (\ref{Eq:Floquet_Eigenstate_Tails}), and hence, we provide its detailed derivation in Appendix \ref{A_Sec:Tails_Eigenstate} for this paper to be self-contained.

Theorem \ref{Thm:Tail_Eigenstates} states that amplitudes of every Fourier component become smaller than $\varepsilon$ for Fourier indices $l$ such that $|l| \geq \Theta (\alpha T + \log (1/\varepsilon))$, as shown in Fig. \ref{Fig_eigenstate}.
We note that this property holds for generic quasienergy $\epsilon_n - l^\prime \omega$ ($l^\prime \in \bbZ$), since shifting $\ket{\Phi_n^l}$ by Eq. (\ref{Eq:Eigenstate_Sambe_shifted}) results in the corresponding Floquet eigenstate $\ket{\Phi_n^{l^\prime}}$.
The same exponential decay holds with the shift $l \to l-l^\prime$.
Owing to the exponential decay, neglecting such large Fourier indices is expected to preserve quesienergy and Floquet eigenstates.
Indeed, the contributions of large Fourier indices $l \in \bbZ \backslash [L]$ are bounded by
\begin{eqnarray}
    \sum_{l \in \bbZ \backslash [L]} \norm{\ket{\phi^l}} &\leq& 2 \sum_{l=L}^\infty \exp \left( -\frac{l-1/2}{2M+1} + \frac{\sinh 1}{2\pi} \alpha T \right) \nonumber \\
    &=& \frac{2e^{\frac{3}{2(2M+1)}}}{e^{\frac{1}{2M+1}}-1} \exp \left( -\frac{L}{2M+1} + \frac{\sinh 1}{2\pi} \alpha T \right) \nonumber \\
    &\leq& 9 (2M+1) \exp \left( -\frac{L}{2M+1} + \frac{\sinh 1}{2\pi} \alpha T \right), \nonumber  \\
    && \label{Eq:phi_n_l_contribute_out}
\end{eqnarray}
whose scaling is $e^{-\Theta (L-\alpha T)}$ as well.
Theorem \ref{Thm:Accuracy_quasienergy} rigorously supports neglecting them.
The theorem has two parts: One is that the truncated Floquet Hamiltonian has an eigenvalue sufficiently close to every quasienergy under the cutoff $L \in \Theta (\alpha T + \log (1/\varepsilon))$, and the other is its converse.
We prove them respectively by Propositions \ref{Prop:Accuracy_Floquet_Hamiltonian_eigenvalue} and \ref{Prop:Accuracy_quasienergy} as follows.

\begin{proposition}\label{Prop:Accuracy_Floquet_Hamiltonian_eigenvalue}
\textbf{(Former part of Theorem \ref{Thm:Accuracy_quasienergy})}

When a time-periodic Hamiltonian has quasinergy $\epsilon_n \in \mr{BZ}$, the corresponding truncated Floquet Hamiltonian $H_\mr{F}^L$ has an eigenvalue $\tilde{\epsilon}_n^L$ satisfying
\begin{equation}\label{Eq:Accuracy_Floquet_eigenvalue}
    \frac{|\tilde{\epsilon}_n^L - \epsilon_n|}{\omega} \leq 8 (2M+1)^2 \alpha T \exp \left( -\frac{L}{2M+1} + \frac{\sinh 1}{2\pi} \alpha T \right).
\end{equation}
In addition, for a series of integers $l \in \bbZ$, it also has an eigenvalue $\tilde{\epsilon}_n^L$ that approximates $\epsilon_n-l\omega$ by
\begin{eqnarray}\label{Eq:Accuracy_Floquet_eigenvalue_shifted}
    && \frac{|\tilde{\epsilon}_n^L - (\epsilon_n-l\omega)|}{\omega} \nonumber \\
    && \quad \leq 8 (2M+1)^2 \alpha T \exp \left( -\frac{L-|l|}{2M+1} + \frac{\sinh 1}{2\pi} \alpha T \right). \nonumber \\
    &&
\end{eqnarray}
\end{proposition}

\textit{Proof of Proposition \ref{Prop:Accuracy_Floquet_Hamiltonian_eigenvalue}.---}
We use the fact that if we have an approximate eigenstate $\ket{\psi}$ of $H$ such that $\norm{(H-\braket{\psi|H|\psi})\ket{\psi}} \leq \varepsilon$ and $\norm{\ket{\psi}}=1$, then $H$ has an eigenvalue $E$ such that $|E-\braket{\psi|H|\psi}| \leq \varepsilon$ \cite{Bhatia2013-kn}.
In this case, we derive the bound on $\norm{(H_\mr{F}^L - \braket{\Phi_n|H_\mr{F}^L|\Phi_n}) \ket{\Phi_n}}$, and show that the value $\braket{\Phi_n|H_\mr{F}^L|\Phi_n}$ is well approximated by $\epsilon_n$.

First, the action of the truncated Floquet Hamiltonian $H_\mr{F}^L$ on the state $\ket{\Phi_n}$ is calculated as follows,
\begin{eqnarray}
    && (H_\mr{F}^L - \epsilon_n) \ket{\Phi_n} \nonumber \\
    && =  \sum_{m} \sum_{\substack{l \in [L] \\ ; l+m \in [L]}} \ket{l+m}_f H_m \ket{\phi_n^l} - \sum_{l \in [L]} (\epsilon_n+l\omega) \ket{l}_f\ket{\phi_n^l} \nonumber \\
    && = \sum_{l \in [L]} \ket{l}_f \left( \sum_{|m| \leq M} H_m \ket{\phi_n^{l-m}} - (\epsilon_n + l \omega) \ket{\phi_n^l}\right) \nonumber \\
    && \qquad + \sum_{|m|\leq M} \left( \sum_{\substack{l \in [L] \\ ; l+m \in [L]}} - \sum_{l \in -m + [L]}\right) \ket{l+m}_f H_m \ket{\phi_n^l}. \nonumber \\
    && \label{PropEq:Accuracy_Floquet_Hamiltonian_eigenvalue_1}
\end{eqnarray}
In the last inequality, $-m+[L]$ denotes the set $\{ -m-L+1, -m-L+2, \hdots, -m +L \}$.
Since the first term of Eq. (\ref{PropEq:Accuracy_Floquet_Hamiltonian_eigenvalue_1}) vanishes due to the eigenvalue equation Eq. (\ref{Eq:Sambe_eigenequation}), we arrive at the upper bound,
\begin{eqnarray}
    && \norm{(H_\mr{F}^L - \epsilon_n) \ket{\Phi_n}} \nonumber \\
    && \quad \leq \sum_{|m| \leq M} \sum_{|l| \geq L-M} \norm{H_m} \cdot \norm{\ket{\phi_n^l}} \nonumber \\
    && \quad \leq 9 (2M+1)^2 \alpha \exp \left( -\frac{L-M}{2M+1} + \frac{\sinh 1}{2\pi} \alpha T \right),  \label{PropEq:Accuracy_Floquet_Hamiltonian_eigenvalue_2}
\end{eqnarray}
where we use Eq. (\ref{Eq:phi_n_l_contribute_out}) as a result of Theorem \ref{Thm:Tail_Eigenstates}.
The same upper bound holds also for $(\epsilon_n-\braket{\Phi_n|H_\mr{F}^L|\Phi_n})$ due to
\begin{eqnarray}
    |\epsilon_n-\braket{\Phi_n|H_\mr{F}^L|\Phi_n}| &=& |\braket{\Phi_n| (H_\mr{F}^L - \epsilon_n)|\Phi_n}| \nonumber  \\
    &\leq& \norm{(H_\mr{F}^L - \epsilon_n) \ket{\Phi_n}}.
\end{eqnarray}

From these inequalities, we immediately observe that $\norm{(H_\mr{F}^L - \braket{\Phi_n|H_\mr{F}^L|\Phi_n}) \ket{\Phi_n}}$ is less than the twice of the bound of Eq. (\ref{PropEq:Accuracy_Floquet_Hamiltonian_eigenvalue_2}).
Reflecting that $\braket{\Phi_n|H_\mr{F}^L|\Phi_n}$ is approximated by $\epsilon_n$ with the same bound of Eq. (\ref{PropEq:Accuracy_Floquet_Hamiltonian_eigenvalue_2}), the truncated Floquet Hamiltonian $H_\mr{F}^L$ has an eigenvalue satisfying
\begin{equation}
    |\tilde{\epsilon}_n^L - \epsilon_n| \leq 27 (2M+1)^2 \alpha \exp \left( -\frac{L-M}{2M+1} + \frac{\sinh 1}{2\pi} \alpha T \right).
\end{equation}
Considering $\omega = 2\pi/T$ and $e^{\frac{M}{2M+1}} \leq e^{1/2}$, this eigenvalue satisfies the inequality, Eq. (\ref{Eq:Accuracy_Floquet_eigenvalue}).
In the latter part, the existence of an eigenvalue approximating $\epsilon_n-l\omega$ can be proved in a similar way using the eigenstate $\ket{\Phi_n^l}$, defined by Eq. (\ref{Eq:Eigenstate_Sambe_shifted}). $\quad \square$

Proposition \ref{Prop:Accuracy_Floquet_Hamiltonian_eigenvalue} states that eigenvalues and eigenstates of $H_\mr{F}$ accurately predict those of $H_\mr{F}^L$.
Or equivalently, owing to the equivalence of $\ket{\phi_n(t)}$ and $\ket{\Phi_n(t)}$, diagonalizing the Floquet operator $U(T;0)$ can accurately reproduce $\tilde{\epsilon}_n^L$.
Since the scaling of Eq. (\ref{Eq:Accuracy_Floquet_eigenvalue}) is expressed by $\exp (- \Theta (L-\alpha T))$, choosing $L \in \Theta (\alpha T + \log (1/\varepsilon))$ is sufficient to suppress the error up to $\varepsilon$.
Next, we prove the remaining part of Theorem \ref{Thm:Accuracy_quasienergy}, which states the oppsite, as the following proposition.

\begin{proposition}\label{Prop:Accuracy_quasienergy}
\textbf{(Latter part of Theorem \ref{Thm:Accuracy_quasienergy})}

Suppose that the truncated Floquet Hamiltonian $H_\mr{F}^L$ has an eigenvalue $\tilde{\epsilon}_n^L$.
Then, the existence of quasienergy $\epsilon_n$ satisfying
\begin{eqnarray}
    && \left| \left( \frac{\epsilon_n - \tilde{\epsilon}_n^L}{\omega} \right)  \quad \text{mod. $1$ } \right| \nonumber \\
    && \quad \leq 9 (2M+1)^2 \alpha T \exp \left( -\frac{L-|\tilde{\epsilon}_n^L|/\omega}{2M+1} + \frac{\sinh 1}{2\pi} \alpha T \right) \nonumber \\
    && \label{Eq:Accuracy_quasienergy}
\end{eqnarray}
is guaranteed.
\end{proposition}

\textit{Proof of Proposition \ref{Prop:Accuracy_quasienergy}.---}
We prove that the corresponding eigenstate $\ket{\tilde{\Phi}_n^L}$ for the Hamiltonian $H_\mr{F}^L$ is an approximate eigenstate of $H_\mr{F}$ in a manner similar to Proposition \ref{Prop:Accuracy_Floquet_Hamiltonian_eigenvalue}.
When the state $\ket{\tilde{\Phi}_n^L}$ is expanded by $\ket{\tilde{\Phi}_n^L} = \sum_{l \in [L]} \ket{l}_f \ket{\tilde{\phi}_n^{l,L}}$, the Fourier component shows an exponential decay as
\begin{equation}\label{Eq:tails_approx_eigenst}
    \norm{\ket{\tilde{\phi}_n^{l,L}}} \leq  \exp \left( -\frac{|l|-|\tilde{\epsilon}_n^L|/\omega}{2M+1} + \frac{\sinh 1}{2\pi} \alpha T \right),
\end{equation}
as well as $\ket{\phi_n^l}$ (See Proposition \ref{A_Prop:norm_approximate_eigenst} in Appendix \ref{A_Sec:Tails_Eigenstate} for its detailed derivation).
\begin{eqnarray}
    \norm{(H_\mr{F}-\tilde{\epsilon}_n^L)\ket{\tilde{\Phi_n^L}}} &=& \norm{(H_\mr{F}-H_\mr{F}^L)\ket{\tilde{\Phi_n^L}}} \nonumber \\
    &=& \norm{\sum_{|m| \leq M} \sum_{\substack{l \in [L] \\ ; l+m \notin [L]}} \ket{l+m}_f H_m \ket{\tilde{\phi}_n^{l,L}}} \nonumber \\
    &\leq& 2(2M+1)\alpha \sum_{l=L-M}^\infty \norm{\ket{\tilde{\phi}_n^{l,L}}}.
\end{eqnarray}
We can organize an inequality similar to Eq. (\ref{Eq:phi_n_l_contribute_out}) based on Eq. (\ref{Eq:tails_approx_eigenst}), giving the following inequality,
\begin{eqnarray}
    && \frac{\norm{(H_\mr{F}-\tilde{\epsilon}_n^L)\ket{\tilde{\Phi_n^L}}}}\omega \nonumber \\
    && \quad \leq 3 (2M+1)^2 \alpha T \exp \left( -\frac{L-|\tilde{\epsilon}_n^L|/\omega}{2M+1} + \frac{\sinh 1}{2\pi} \alpha T \right). \nonumber \\
    && \label{Eq:Approx_H_F_Psi_tilde}
\end{eqnarray}
Similar to Proposition \ref{Prop:Accuracy_Floquet_Hamiltonian_eigenvalue}, this inequality gives the upper bounds for $|\tilde{\epsilon}_n^L- \braket{\tilde{\Phi_n^L}|H_\mr{F}|\tilde{\Phi_n^L}}|$ and $\norm{(H_\mr{F}-\braket{\tilde{\Phi_n^L}|H_\mr{F}|\tilde{\Phi_n^L}}) \ket{\tilde{\Phi_n^L}} }$.
Then, it ensures the existence of an eigenvalue $\epsilon_n - l \omega$ in the spectrum of the Floquet Hamiltonian $H_\mr{F}$, whose difference from $\tilde{\epsilon}_n^L$ is at most $3\omega$-times as large as the right-hand side of Eq. (\ref{Eq:Approx_H_F_Psi_tilde}). 
This immediately implies the satisfaction of Eq. (\ref{Eq:Accuracy_quasienergy})$. \quad \square$

Suppose that we are interested in quasienergy $\epsilon_n$ in BZ.
Then, we compute an eigenvalue $\tilde{\epsilon}_n^L$ in or around BZ such that $|\tilde{\epsilon}_n^L|/\omega \leq 1$.
In this case, Proposition \ref{Prop:Accuracy_quasienergy} ensures that, with the choice of the cutoff $L$ by
\begin{widetext}
\begin{equation}\label{Eq:L_choice}
    L = \left\lceil (2M+1) \left( \frac{\sinh 1}{2\pi} \alpha T + \log (1/\varepsilon) + \log \left(9(2M+1)^2 \alpha T \right)  \right) \right\rceil + 1 \in \Theta \left( \alpha T + \log (1/\varepsilon) \right),
\end{equation}
\end{widetext}
there exists a quasienergy $\epsilon_n \in \mr{BZ}$ such that
\begin{equation}
    \left| \left( \frac{\epsilon_n - \tilde{\epsilon}_n^L}{\omega} \right) \text{ mod. $1$ }\right| \leq \varepsilon.
\end{equation}
Therefore, this characterizes the proper cutoff for the Sambe space formalism by $L \in \Theta (\alpha T + \log (1/\varepsilon))$.
Similarly, a series of the equivalent values $\{ \epsilon_n - l \omega \}_{l \in \bbZ}$ can be embedded in the spectrum of $H_\mr{F}^L$ with the same scaling.
For instance, when we choose the cutoff $2L$ with $L$ given by Eq. (\ref{Eq:L_choice}), the eigenvalue $\tilde{\epsilon}_n^L$ in $\mr{BZ}_l$ can approximate $\epsilon_n-l\omega$ for $l \in [L]$.
In other words, if the target quasienergy is $\Theta (\alpha T + \log (1/\varepsilon))$ away from the boundary of the truncated Sambe space, it will be reproduced by the truncated Floquet Hamiltonian within an error of $\omega \varepsilon$.

\subsection{Relation to the time-evolution in the Sambe space formalism}

Theorem \ref{Thm:Accuracy_quasienergy} states that the cutoff $L \in \Theta (\alpha T + \log (1/\varepsilon))$ is sufficient to reproduce the exact quasienergy within an error $\omega \varepsilon$ from $H_\mr{F}^L$.
Here, we discuss its relation to the similar bound on the time-evolution operators in the Sambe space formalism.

According to the Sambe space formalism \cite{Levante1995-tx}, the time-evolution operator $U(t;0)$ defined by Eq. (\ref{Eq:time_evolution_op}) is also expressed by the Floquet Hamiltonian $H_\mr{F}$ as
\begin{equation}\label{Eq:Time_evol_in_Sambe}
    U(t;0) = \sum_{l \in \bbZ} e^{-i l \omega t} \braket{l|e^{-i H_\mr{F} t}|0}_f.
\end{equation}
This formalism avoids the use of the Dyson series expansion but instead use the Sambe space $\mcl{H}^\infty$.
Like quasienergy and Floquet eigenstates here, it requires truncation of the Sambe space.
Using the Lieb-Robinson bound \cite{Lieb1972-uo,Nachtergaele2006-ok,Gong2022-bound}, Refs. \cite{Mizuta_Quantum_2023,Mizuta_2023_multi} have provided a proper cutoff $L_\mr{LR}$ which satisfies
\begin{eqnarray}
    & \norm{ U(t;0) - \sum_{l \in [L_\mr{LR}]} e^{-i l \omega t} \braket{l|e^{-i H_\mr{F}^{L_\mr{LR}} t}|0}_f} \leq \varepsilon, \label{Eq:Ut_LR_error}\\
    & L_\mr{LR} \in \Theta (\alpha t + \log (1/\varepsilon)). \label{Eq:Cutoff_LR}
\end{eqnarray}
In order to compute quasienergy by the Floquet operator $U(T;0)$, we set $t=T$ in the above.
Note that in this case the cutoff $L_\mr{LR}$ has the same scaling $\Theta (\alpha T + \log (1/\varepsilon))$ as the cutoff $L$ in Theorem \ref{Thm:Accuracy_quasienergy}.

We remark that the origins of the cutoffs $L$ and $L_\mr{LR}$ are different despite their common scaling.
The former originates from the static property of the Floquet Hamiltonian $H_\mr{F}$.
The cutoff $L$ is determined by the spread of each Fourier components $\ket{\phi_n^l}$ by Theorem \ref{Thm:Tail_Eigenstates}.
This can be interpreted as localization in a one-dimensional static system under linear potential \cite{Zener1934,Wannier1960-oq}.
In contrast, the latter cutoff $L_\mr{LR}$ reflects the dynamic property of $H_\mr{F}$.
It comes from the Lieb-Robinson bound in the Sambe space \cite{Mizuta_Quantum_2023},
\begin{equation}\label{Eq:LR_bound_in_Sambe}
    \norm{\braket{l|e^{-i H_\mr{F} t}|0}} \leq e^{- \Theta (|l|-\alpha t)},
\end{equation}
which states that the propagation from $0$ to $l$ brought by the dynamics is exponentially suppressed depending on the distance $|l|$.

The common scaling of $L$ and $L_\mr{LR}$ results in some desirable properties.
First, it ensures that computing an eigenstate $\ket{\tilde{\Phi}_n^L}$ of the truncated Floquet Hamiltonian $H_\mr{F}^L$ is also valid for estimating the Floquet eigenstate in the physical Hilbert space, $\ket{\phi_n(t)}$.
As Floquet theory gives the relations Eqs. (\ref{Eq:time_evolution_eigenbasis}) and (\ref{Eq:def_Floquet_state}), the state $\ket{\tilde{\Phi}_n^L}$ is expected to generate an approximate Floquet eigenstate by
\begin{equation}
    \ket{\tilde{\phi}_n^L(t)} \equiv e^{i \tilde{\epsilon}_n^L t} U(t;0) \sum_{l \in [L]} (\bra{l}_f \otimes I) \ket{\tilde{\Phi}_n^L}.
\end{equation}
Because of the common scaling, the choice of $L \in \Theta (\alpha T + \log (1/\varepsilon))$ actually guarantees this expectation within an error $\varepsilon$ by
\begin{equation}
    \frac{\norm{\left(U(t+T; t)- e^{-i \tilde{\epsilon}_n T}\right) \ket{\tilde{\phi}_n^L(t)}}}{\norm{\ket{\tilde{\phi}_n^L(t)}}} \leq \varepsilon.
\end{equation}

The second advantage of the common scaling is the common computational complexity for different types of Floquet eigenstates $\ket{\Phi_n}$ and $\ket{\phi_n(t)}$.
As discussed in Sections \ref{Sec:Floquet_QPE_physical} and \ref{Sec:Floquet_QPE_Sambe}, the cutoffs $L$ and $L_\mr{LR}$ will be respectively used for the quantum algorithm that returns $(\epsilon_n,\ket{\Phi_n})$ and the one for $(\epsilon_n,\ket{\phi_n(t)})$.
While their computational costs are respectively characterized by $L$ and $L_\mr{LR}$, their common scaling leads to essentially the same costs in $\varepsilon$ and $\alpha T$.
Although they have different static and dynamical origins, this corresponds to the one-to-one correspondence of the Floquet eigenstates $\ket{\Phi_n}$ and $\ket{\phi_n(t)}$.

\subsection{Cost of classical computation}

The relation between the truncation of the Sambe space and the exact quasi-energy, by Theorem \ref{Thm:Accuracy_quasienergy}, is correct regardless of whether the computation is classical or quantum.
Here, we briefly discuss what Theorem \ref{Thm:Accuracy_quasienergy} implies for classical computational cost, before moving on to quantum algorithms.

When we do not rely on the Sambe space formalism, we usually compute pairs of quasienergy and Floquet eigenstates via time discretization.
Introducing discretized time $t= 0, \Delta t, \hdots, L \Delta t$ with $\Delta t \equiv T/L$, we compute the Floquet operator by
\begin{equation}\label{Eq:Floquet_op_discretized}
    U(T;0) = \prod_{j=1}^{L-1} e^{-i H(j \Delta t) \Delta t} + \order{\frac{(\alpha T)^2}{L}}.
\end{equation}
The error term $\order{(\alpha T)^2/L}$ comes from neglecting the time-dependence in each time bin $\Delta t$, bounded by the time derivative $\norm{H^\prime (t)} \leq (2M+1)M \omega \alpha$.
To suppress the error up to $\varepsilon$ in quasienergy, the discretization number $L$ should be in $\order{(\alpha T)^2/\varepsilon}$.
Then, the computation of quasienergy by diagonalizing Eq. (\ref{Eq:Floquet_op_discretized}) requires $L \in \order{(\alpha T)^2/\varepsilon}$ times diagonalization and multiplication of size-$(\mr{dim}(\mcl{H}))$ matrices.
In general, where we assume no structures on every $H_m$, this yields $\order{((\alpha T)^2/\varepsilon) (\mr{dim}(\mcl{H}))^3}$ time of classical computation.
On the other hand, let us consider the case where we use the truncated Sambe space.
We choose the cutoff $L$ by $L \in \Theta (\alpha T + \log (1/\varepsilon))$ according to Theorem \ref{Thm:Accuracy_quasienergy}, and resort to single diagonalization of the size-$(L \times \mr{dim}(H))$ matrix $H_\mr{F}^L$.
This requires classical computational time $\order{(\alpha T + \log (1/\varepsilon))^3 (\mr{dim}(\mcl{H}))^3}$, whose scaling in the inverse error $1/\varepsilon$ is exponentially smaller than the standard method without the Sambe space.

However, we note that the Sambe space approach is not necessarily advantageous compared to the time discretization approach in classical computation.
Although it has good scaling in $1/\varepsilon$, its scaling in $\alpha T \in \poly{N}$ is rather worse.
It is also problematic that we need $\order{L^2}$ times more classical memory.
In addition, when we assume the locality of the Hamiltonian $H(t)$ as its internal structure, the time-discretization method can avoid diagonalization in every time step by Trotterization \cite{Lloyd1996-ko,childs2021-trotter}.
On the other hand, in quantum computation, the treatment of time-dependent Hamiltonians is more complicated than that for time-independent ones, exempified by the availability of QSVT \cite{Gilyen2019-qsvt,Martyn2021-grand-unif}, and the additional memory only requires $\order{\log L}$ qubits.
Thus, quantum algorithms can fully exploit the power of the Sambe space formalism, as discussed in Sections \ref{Sec:Floquet_QPE_physical} and \ref{Sec:Floquet_QPE_Sambe}.
%=================================
% Section: Quantum algorithm for Floquet eigenstates
%=================================
\section{Floquet QPE: Preliminaries}\label{Sec:Floquet_QPE}

In Sections \ref{Sec:Floquet_QPE}-\ref{Sec:Floquet_QPE_Sambe}, we organize a nearly optimal quantum algorithm called ``Floquet QPE", using the results in Section \ref{Sec:Accuracy_Sambe}.
Roughly speaking, it returns quasienergy and a Floquet eigenstate with guaranteed accuracy by
\begin{equation}
    \sum_n c_n \ket{\phi_n(0)} \to \begin{cases}
        \sum_n c_n \ket{(\epsilon_n)_b}_b \ket{\phi_n(t)} + \order{\delta}, \\
        \sum_n c_n \ket{(\epsilon_n)_b}_b \ket{\Phi_n} + \order{\delta}.
    \end{cases}
\end{equation}
Before deriving the main results on the algorithms, we here provide some preliminaries on the conditions, the block-encoding, and the rounding promise so that they can be Floquet counterparts of the standard QPE.

\subsection{Requirements of Floquet QPE}

As discussed in Section \ref{Subsec:QPE}, the standard QPE has favorable properties on accuracy, efficiency, and measurement outcome.
We construct quantum algorithms so that the following Floquet counterparts hold.
\begin{enumerate}[(a$^\prime$)]
    \item Guaranteed accuracy; \\
    Every $b$-bit outcome $(\epsilon_n)_b$ gives a good estimate for a certain quasienergy $\epsilon_n$ by
    \begin{equation}
        \left|\left(  \frac{\epsilon_n - (\epsilon_n)_b}{\omega} \right) \text{ mod. $1$ } \right|  \leq \varepsilon,
    \end{equation}
    with high probability larger than $1-\delta$.

    \item Efficiency of algorithm; \\
    The query complexity in $C[O_{H_m}]$ is at most polynomial in $\varepsilon,\nu,\delta$, and other parameters such as the system size $N$.
    The number of ancilla qubits is at-most logarithmic in the above parameters.

    \item Measurement outcome; \\
    Measurement of the register projects the state onto the subspace of Floquet eigenstates $\ket{\phi_n(t)}$ or $\ket{\Phi_n}$ with the measured quasienergy eigenvalue.
    Moreover, the probability of an outcome $\ket{\phi_n(t)}$ or $\ket{\Phi_n}$ is given by the weight in the initial state $\ket{\psi} \in \mcl{H}$ as
    \begin{equation}
        p_n = |c_n|^2 = |\braket{\phi_n(0) | \psi}|^2.
    \end{equation}
    
\end{enumerate}

The properties (a$^\prime$) and (b$^\prime$) are the minimum requirements for the accuracy and the efficiency like the standard QPE.
Since the Floquet QPE includes the standard QPE by setting $T=(2 \norm{H_0})^{-1}$ and $M=0$, its computational cost is inevitably equal to or greater than that of the time-independent cases.
Nevertheless, as will be discussed later, the efficiency can achieve near optimal scaling of time-independent cases in all the parameters.
The third property (c$^\prime$) reflects characteristics of time-periodicity.
The weight $c_n$ is determined by $\ket{\phi_n(0)}$ so that the initial guess can be made based on the physical Hilbert space but not on the virtual Sambe space.
The options in the outputs, $\ket{\phi_n(t)}$ or $\ket{\Phi_n}$, are also inherent in time-periodic Hamiltonians.
This property ensures that a preferable Floquet eigenstate can be efficiently prepared under the assumption of good initial guess on $\ket{\phi_n(0)}$ with large $|c_n|$.

\subsection{Modified Floquet Hamiltonian}\label{Subsec:Modified_H_F}

We employ block-encoding of each Fourier component Hamiltonian $O_{H_m}$, characterized by Eq. (\ref{Eq:block_encoding_Hm}), as oracles in quantum algorithms.
These oracles are used for organizing block-encoding of the truncated Floquet Hamiltonian $H_\mr{F}^L$.
However, the slightly modified Floquet Hamiltonian defined by
\begin{eqnarray}
    && H_\mr{F,pbc}^L = \nonumber \\
    && \quad \sum_{|m| \leq M} \sum_{l \in [L]} \ket{(l \oplus m)_{[L]}}\bra{l}_f \otimes H_m - \sum_{l \in [L]} l \omega \ket{l}\bra{l}_f \otimes I \nonumber \\
    && \label{Eq:H_F_pbc}
\end{eqnarray}
is more suitable for block-encoding \cite{Mizuta_Quantum_2023}.
The addition $(l \oplus m)_{[L]} \in [L]$ is defined modulo $[L]$.
The symbol ``pbc" comes from the fact that $H_\mr{F,pbc}^L$ assumes the periodic boundary condition in the Fourier index direction $\ket{l}$. 
We can organize the block-encoding of $H_\mr{F,pbc}^L$ with the following resource:
\begin{proposition}\label{Prop:block_encode_H_F_pbc}
\textbf{(Block-encoding of $H_\mr{F,pbc}^L$)}

When an ancilla system $a^\prime$ composed of the ancilla $a$ and additional $\order{\log L}$ qubits is prepared, the block-encoding $O_{H_\mr{F,pbc}^L}$ satisfying the equality,
\begin{equation}
\braket{0|O_{H_\mr{F,pbc}^L}|0}_{a^\prime} = \frac{H_\mr{F,pbc}^L}{\tilde{\alpha}},
\end{equation}
can be implemented by one query respectively for $C[O_{H_m}]$ and additional $\order{\log L}$ primitive gates for any positive number $\tilde{\alpha}$ larger than
\begin{equation}
    \alpha_\mr{F}^L \equiv (2M+1)\alpha + L \omega.
\end{equation}
\end{proposition}
See Appendix \ref{A_Sec:boundary_condition} or Ref. \cite{Mizuta_Quantum_2023} for its detailed construction.
We totally need $2M+1$ queries to one of $\{ C[O_{H_m}] \}_m$ to implement $C[O_{H_\mr{F,pbc}^L}]$.
Due to the assumption of $M \in \order{1}$, this query complexity is constant.
We also recall that the factor $\alpha_\mr{F}^L$ gives a bound on the truncated Floquet Hamiltonian by $\norm{H_\mr{F}^L} \leq \alpha_\mr{F}^L$ and $\norm{H_\mr{F,pbc}^L} \leq \alpha_\mr{F}^L$.

The difference between $H_\mr{F}^L$ and $H_\mr{F,pbc}^L$ appears only at the boundaries $l \simeq \pm L$, which have quasienergy around $\pm L \omega$, and hardly affects its center BZ.
As long as we are interested in quasienergy $\epsilon_n \in \mr{BZ}$, we can substitute $H_\mr{F,pbc}^L$ for $H_\mr{F}^L$, which has an efficient block-encoding.
Indeed, the alternative Floquet Hamiltonian $H_\mr{F,pbc}^L$ satisfies the counterpart of Theorems \ref{Thm:Accuracy_quasienergy}, which states that $\epsilon_n$ and $\ket{\Phi_n})$ are respectively an approximate eigenvalue and an approximate eigenstate of $H_\mr{F,pbc}^L$ under the same scaling of $L$ (See Appendix \ref{A_Sec:boundary_condition} for its detailed discussion).
Similar to $H_\mr{F}^L$, their errors are suppressed up to $\order{\varepsilon}$ by setting $L \in \Theta (\alpha T + \log (1/\varepsilon))$.
This ensures that running quantum algorithm based on the Floquet Hamiltonian $H_\mr{F,pbc}^L$ is essentially the same as that based on $H_\mr{F}^L$.
While we formulate the Floquet QPE algorithms based on the common Floquet Hamiltonian $H_\mr{F}^L$ in the following sections, we note that the actual algorithms run with $H_\mr{F,pbc}^L$.
We can think of the block-encoding $O_{H_\mr{F}^L}$ as being implemented by constant queries to $C[O_{H_m}]$ or its inverse.

In our algorithms, the block-encoding $C[O_{H_m}]$ or its inverse are always used for QSVT.
Since the ancilla system $a$ (or $a^\prime$) is set to the reference state $\ket{0}_a$ (or $\ket{0}_{a^\prime}$) both before and after the operations, we omit them in the following discussion.

\subsection{Rounding promise}\label{Subsec:rounding_promise_Floquet}

Next, we define rounding promise $\nu$ appropriate for time-periodic Hamiltonians.
A quasienergy in BZ satisfies $\epsilon_n/\omega \in [-1/2,1/2)$ under renormalization.
Analogous to Definition \ref{Def:rounding_promise} for time-independent Hamiltonians, we define its counterpart as follows.
\begin{definition}\label{Def:rounding_promise_Floquet}
\textbf{(Rounding promise)}

A time-periodic Hamiltonian $H(t)$ is said to have rounding promise $\nu \in (0,1)$ if every quasienergy in BZ satisfies
\begin{equation}\label{Eq:rounding_promise_quasienergy}
    \frac{\epsilon_n}{\omega} \notin \bigcup_{x=-2^{b^\prime-1}}^{2^{b^\prime-1}} \left[ \frac{x-\nu/2}{2^{b^\prime}}, \frac{x+\nu/2}{2^{b^\prime}} \right),
\end{equation}
with a certain integer $b^\prime \in \Theta (\log (1/\varepsilon))$ such that $2^{-b^\prime} \leq \varepsilon$.
\end{definition}

Note that the rounding promise of a time-periodic Hamiltonian leads to that of the truncated Floquet Hamiltonian $H_\mr{F}^L$ in the sense of Definition \ref{Def:rounding_promise}.
Let us choose the renormalization factor of $H_\mr{F}^L$ by $2^{b_\mr{F}^L} \omega$ with
\begin{equation}\label{Eq:bit_normalization}
    b_\mr{F}^L \equiv 1 + \left\lceil \log_2 \frac{2\alpha_\mr{F}^L}{\omega}\right\rceil, \quad
    2^{b_\mr{F}^L} \in \Theta \left( \alpha T + L \right),
\end{equation}
so that the norm of $H_\mr{F}^L/(2^{b_\mr{F}^L} \omega)$ can be bounded by $1/2$ for the standard QPE.
To correctly obtain an estimate of $\epsilon_n/\omega$ within the error $\varepsilon$, it is sufficient to obtain a $(b^\prime+b_\mr{F}^L)$-bit estimate of $\tilde{\epsilon}_n^L/(2^{b_\mr{F}^L} \omega)$ with $b^\prime \in \Theta (\log (1/\varepsilon))$.
Based on the fact that $\tilde{\epsilon}_n^L$ is well approximated by $\epsilon_n$ by Theorem \ref{Thm:Accuracy_quasienergy}, the rounding promise is inherited to $H_\mr{F}^L$ for the bit number $b^\prime+b_\mr{F}^L$ as follows.

\begin{proposition}\label{Prop:Inherited_rounding_promise}
\textbf{(Inherited rounding promise)}

Suppose that a time-periodic Hamiltonian $H(t)$ has rounding promise $\nu \in (0,1)$ with Eq. (\ref{Eq:rounding_promise_quasienergy}) and that we are interested in eigenvalues $\tilde{\epsilon}_n^L$ of $H_\mr{F}^L$ in $\mr{BZ}_l = [(-l-1/2)\omega, (-l+1/2)\omega)$.
When we choose the cutoff $L$ by $L \in \Omega (\alpha T + |l| + \log (1/\varepsilon \nu))$, the truncated Floquet Hamiltonian $H_\mr{F}^L$ can have rounding promise $\nu/2$ in that 
\begin{equation}\label{Eq:rounding_promise_H_F}
    \frac{\tilde{\epsilon}_n^L}{2^{b_\mr{F}^L} \omega} \notin \bigcup_{x=(-l-1/2)2^{b^\prime}}^{(-l+1/2)2^{b^\prime}} \left[ \frac{x-\nu/4}{2^{b^\prime+b_\mr{F}^L}}, \frac{x+\nu/4}{2^{b^\prime+b_\mr{F}^L}}\right)
\end{equation}
is satisfied. 
\end{proposition}

\textit{Proof of Proposition \ref{Prop:Inherited_rounding_promise}.---} According to Proposition \ref{Prop:Accuracy_Floquet_Hamiltonian_eigenvalue}, we can choose the cutoff $L \in \Theta (\alpha T + |l| + \log (1/\varepsilon \nu))$ so that every eigenvalue $\tilde{\epsilon}_n^L \in \mr{BZ}_l$ can be approximated by certain quasienergy $\epsilon_n$ as $|(\tilde{\epsilon}_n^L-\epsilon_n)/\omega ) \text{ mod. $1$ }| \leq \nu/2^{b^\prime+1}$.
Combining this relation with the rounding promise of a time-periodic Hamiltonian, Eq. (\ref{Eq:rounding_promise_quasienergy}), immediately implies Eq. (\ref{Eq:rounding_promise_H_F}). $\quad \square$

The above rounding promise is based solely on the fact that the eigenvalue $\tilde{\epsilon}_n^L$ and the eigenstates $\ket{\tilde{\Phi}_n^L}$ of $H_\mr{F}^L$ are respectively an approximate eigenvalue and eigenstate of $H_\mr{F}$ by Theorem \ref{Thm:Accuracy_quasienergy}.
Thus, the same rounding promise as Proposition \ref{Prop:Inherited_rounding_promise} holds for the modified Floquet Hamiltonian $H_\mr{F,pbc}^L$ [See Section \ref{Subsec:Modified_H_F}] as long as its eigenstate is an approximate eigenstate of $H_\mr{F}$ within an error $\order{\varepsilon \nu}$.
The inherited rounding promise will be used for the standard QPE under $H_\mr{F}^L$ or $H_\mr{F,pbc}^L$.
%=================================
% Section: Quantum algorithm for Floquet eigenstates
%=================================
\section{Floquet QPE: Eigenstates $\ket{\phi_n(t)}$ in the physical space}\label{Sec:Floquet_QPE_physical}

\subsection{Outline of the algorithm}

We construct a Floquet QPE algorithm that outputs pairs of $(\epsilon_n,\ket{\phi_n(t)})$, where the Floquet eigenstates $\ket{\phi_n(t)}$ is defined on the physical Hilbert space $\mcl{H}$.
In this algorithm, we rely on the fact that the set of the Floquet eigenstates $ \{ \ket{\phi_n(0)} \}_n$ diagonalizes the Floquet operator as Eq. (\ref{Eq:Floquet_op}).
Then, the standard QPE on $U(T;0)$ can return pairs of $(\epsilon_n,\ket{\phi_n(0)})$.
To obtain a superposition of $\ket{\phi_n(t)}$, it is sufficient to apply $U(t;0)$ based on the relation, $\ket{\phi_n(t)}=e^{i\epsilon_n t}U(t;0)\ket{\phi_n(0)}$.

We note that Ref. \cite{Fauseweh_Quantum_2023} has recently discussed a similar idea of performing the standard QPE on the Floquet operator, but they focus on variational quantum algorithms that prepare an initial state having large overlap with a preferable eigenstate $\ket{\phi_n(0)}$.
How the Floquet operator is constructed is missing, and thus it is still unclear how much resource is needed to achieve the Floquet QPE for $(\epsilon_n, \ket{\phi_n(t)})$ with satisfying conditions (a$^\prime$)--(c$^\prime$).
Here, we propose an algorithm based on the Sambe space formalism for the time-evolution operator \cite{Mizuta_Quantum_2023}.
Using the Lieb-Robinson bound, the time-evolution operator can be expressed by the truncated Floquet Hamiltonian $H_\mr{F}^{L_\mr{LR}}$ as shown in Eq. (\ref{Eq:Ut_LR_error}).
According to Ref. \cite{Mizuta_Quantum_2023}, the Hamiltonian simulation performing the transformation,
\begin{equation}\label{Eq:Hamiltonian_simulation_Floquet}
    \ket{0}_f \ket{\psi} \to \ket{0}_f U(t;0) \ket{\psi} + \order{\delta^\prime}, \quad ^\forall \ket{\psi} \in \mcl{H},
\end{equation}
can be implemented with setting $L_\mr{LR} \in \Theta (\alpha t + \log (1/\delta^\prime))$ for $t \in [0,T]$.
The ancilla $f$ has $2L_\mr{LR}$ degrees of freedom.
This algorithm requires $\Theta (\alpha t + L_\mr{LR})$ queries to $C[O_{H_m}]$ or its inverse, $n_a + \Theta (\log L_\mr{LR})$ ancilla qubits, and $\Theta (n_a + \log L_\mr{LR})$ primitive gates per query.
The implementation of $C[U(t;0)]$ yields the same cost.

\subsection{Algorithm and cost}\label{Subsec:Floquet_phys_cost}

We first consider the cases where a time-periodic Hamiltonian has rounding promise $\nu \in (0,1)$ as Eq. (\ref{Eq:rounding_promise_quasienergy}).
We construct the Floquet QPE, which transforms
\begin{equation}\label{Eq:Floquet_QPE_phys_transform}
    \ket{0}_b \ket{\psi} \to \sum_n c_n \ket{(\epsilon_n)_b}_b \ket{\phi_n(t)} + \order{\delta},
\end{equation}
under the initial state $\ket{\psi}=\sum_n c_n \ket{\phi_n(0)}$.
The $b$-bit binary $(\epsilon_n)_b$ approximates the quasienergy $\epsilon_n \in \mr{BZ}$ within an error $\omega \varepsilon$.
The algorithm combines the standard QPE and the Sambe space formalism for $U(T;0)$ based on the following steps.
\begin{enumerate}
    \item Organize a controlled operation of the Floquet operator $C[U(T;0)]+\order{\delta^\prime}$ based on Eq. (\ref{Eq:Hamiltonian_simulation_Floquet}).
    
    \item Execute the standard QPE by $C[U(T;0)]$, using the parameters $\varepsilon_\mr{QPE}, \delta,\nu$.
    The number of qubits for the register is determined by $b \in \Theta (\log (1/\varepsilon_\mr{QPE}))$.
    This step changes the initial state by
    \begin{eqnarray}
        && \qquad \ket{0}_b \ket{0}_f \ket{\psi} \nonumber \\
        && \qquad \quad \to \ket{0}_f \sum_n c_n \ket{(\epsilon_n)_b}_b \ket{\phi_n(0)} + \order{q_\mr{QPE}\delta^\prime + \delta}. \nonumber \\
        && \label{Eq:Floquet_QPE_phys_step2}
    \end{eqnarray}
    where $q_\mr{QPE} \in \order{(1/\varepsilon_\mr{QPE} \nu) \log (1/\delta)}$ denotes the query complexity of the standard QPE in Eq. (\ref{Eq:query_standard_QPE}).
    We set $\delta^\prime = \delta/q_\mr{QPE}$ so that the error term in Eq. (\ref{Eq:Floquet_QPE_phys_step2}) can scale as $\order{\delta}$.
    
    \item Time-evolution $U(t;0)+\order{\delta^\prime}$ based on Eq. (\ref{Eq:Hamiltonian_simulation_Floquet}), which results in
    \begin{equation}\label{Eq:Floquet_QPE_phys_step3}
        \qquad \quad \ket{0}_f \sum_n c_n e^{i\epsilon_n t} \ket{(\epsilon_n)_b}_b \ket{\phi_n(t)} + \order{\delta}.
    \end{equation}
    
    \item Controlled phase gate based on the estimated value $(\epsilon_n)_b$.
    This cancels the phase $e^{i \epsilon_n t}$ by changing the state Eq. (\ref{Eq:Floquet_QPE_phys_step3}) to
    \begin{eqnarray}
        && \to \sum_n c_n e^{i(\epsilon_n-(\epsilon_n)_b) t} \ket{(\epsilon_n)_b}_b \ket{\phi_n(t)} + \order{\delta} \nonumber \\
        && \quad = \sum_n c_n \ket{(\epsilon_n)_b}_b \ket{\phi_n(t)} + \order{\varepsilon_\mr{QPE}+ \delta}. \label{Eq:Floquet_QPE_phys_step4}
    \end{eqnarray}
    We set $\varepsilon_\mr{QPE}=\min (\varepsilon,\delta)$ to guarantee the errors both in the output quasisnergy $(\epsilon_n)_b$ and the output state.

\end{enumerate}

Let us evaluate the computational cost.
Based on the choice of $\delta^\prime = \delta/q_\mr{QPE}$ and $\varepsilon_\mr{QPE}=\min (\varepsilon,\delta)$, the dimension of the truncated Sambe space should be given by
\begin{equation}
    L_\mr{LR} \in \Theta \left( \alpha T + \log (1/\nu) + \log (1/\min (\varepsilon,\delta))\right).
\end{equation}
The query complexity in $C[O_{H_m}]$ or its inverse through the algorithm is given by $\Theta (q_\mr{QPE}(\alpha T + L_\mr{LR}))$.
The number of ancilla qubits other than those for the block-encoding $O_{H_m}$ is $\Theta (\log (1/\varepsilon_\mr{QPE})+\log L_\mr{LR})$, composed of the $b$-qubit register and the $f$-qubit ancilla for Fourier indices $l$.
The quantum gates other than the block-encoding are divided into those for $C[U(T;0)]$, those for the standard QPE, and those for the controlled phase gate in Step 4.
The first group is dominant, which amounts to $\Theta (q_\mr{QPE}(\alpha T + \log L_\mr{LR})(n_a+\log L_\mr{LR}))$.
Finally, we obtain the following theorem which characterizes the computation of quasienergy $\epsilon_n$ and a Floquet eigenstate $\ket{\phi_n(t)} \in \mcl{H}$ as a counterpart of QPE for time-periodic Hamiltonians.
\begin{theorem}\label{Thm:algorithm_phys}
\textbf{(Algorithm for $(\epsilon_n,\ket{\phi_n(t)})$)}

Assume the existence of rounding promise $\nu$.
The quantum algorithm of Eq. (\ref{Eq:Floquet_QPE_phys_transform}), which returns pairs of quasienergy and a Floquet eigenstate $(\epsilon_n,\ket{\phi_n(t)})$, can be executed with the following computational resources if we require the guaranteed quasisnergy error $\varepsilon$ and the guaranteed state error $\delta$;
\begin{itemize}
    \item Query complexity in $C[O_{H_m}]$ or its inverse
    \begin{equation}\label{Eq:query_Floquet_phys}
        \qquad \Theta \left( \frac{\alpha T + \log (1/\min (\varepsilon, \delta) \nu )}{\min (\varepsilon,\delta) \nu}  \log (1/\delta) \right) .
    \end{equation}
    \item Number of ancilla qubits 
    \begin{equation}\label{Eq:ancilla_Floquet_phys}
        \qquad n_a + \Theta \left( \log (\alpha T/ \min (\varepsilon,\delta) ) + \log \log (1/\nu)) \right).
    \end{equation}
    
    \item Other primitive gates per query
    \begin{equation}\label{Eq:other_gates_Floquet_phys}
        \qquad \Theta \left( n_a + \log (\alpha T/ \min (\varepsilon,\delta) ) + \log \log (1/\nu)) \right).
    \end{equation}
\end{itemize}
\end{theorem}

Note that the quantum algorithm without rounding promise $\nu$ is organized similarly the above case.
In Step 2, we run the standard QPE without rounding promise setting $\nu \in \order{1}$. Then, the $b$-qubit register in Eq. (\ref{Eq:Floquet_QPE_phys_step2}) becomes
\begin{equation}
    \ket{\overline{(\epsilon_n)_b}}_b = p_0^n \ket{(\epsilon_n)_{b0}}_b + p_1^n \ket{(\epsilon_n)_{b1}}_b,
\end{equation}
with some weights $p_0^n, p_1^n$ according to Eqs. (\ref{Eq:Standard_QPE_wo_rounding1}) and (\ref{Eq:Standard_QPE_wo_rounding2}).
Both of the $b$-bit binary numbers $(\epsilon_n)_{b0}$ and $(\epsilon_n)_{b1}$ approximate $\epsilon_n$ within an error $\omega \varepsilon$.
While the controlled phase gate in Step 4 returns the phase $e^{-i (\epsilon_n)_{b0} t}$ or $e^{-i (\epsilon_n)_{b1} t}$, both of them cancel $e^{i \epsilon_n t}$ within an error $\order{\varepsilon_\mr{QPE}}$.
As a result, performing Steps 1-4 with setting $\nu \in \order{1}$ completes the quantum algorithm in the absence of rounding promise,
\begin{equation}\label{Eq:Floquet_phys_wo_RP}
    \ket{0}_b \ket{\psi} \to \sum_n c_n \ket{\overline{(\epsilon_n)_b}}_b \ket{\phi_n(t)} + \order{\delta}.
\end{equation}
The computational cost for this task is obtained by setting $\nu \in \order{1}$ in Theorem \ref{Thm:algorithm_phys}.

The computational cost is also summarized in Table \ref{Table:comparison_algorithms}.
We emphasize that it is essentially the same as the cost of the standard QPE, excluding logarithmic corrections.
Focusing on the rounding promise $\nu$ or the quasienergy error $\varepsilon$, the query complexity achieves nearly optimal scaling in them, $\Theta (\nu^{-1} \log (1/\nu))$ and $\Theta (\varepsilon^{-1} \log (1/\varepsilon))$.
Its scaling in the state error $\delta$, given by $\Theta (\delta^{-1} \log (1/\delta))$ seems to be worse, but it does not matter.
The additional factor $\delta^{-1}$ compared to Eq. (\ref{Eq:query_standard_QPE}) comes from Step 4 only to cancel the phase of each Floquet eigenstate as Eq. (\ref{Eq:Floquet_QPE_phys_step4}).
In practice, we do not care about the phase in many cases such as the case where we measure the register and prepare the corresponding eigenstate (such an algorithm is called QPE with garbage phases \cite{Rall2021-qpe}).
We can omit Step 4 in that case with setting $\varepsilon_\mr{QPE}=\varepsilon$, and then the query complexity reproduces the scaling $\Theta (\log (1/\delta))$, which is optimal in $\delta$.
Even without this omission, the similar cost is achieved under a relatively loose constraint $\varepsilon \leq \delta$.
Finally, we mention about the coefficient $\alpha T \in \poly{N}$ in the query complexity Eq. (\ref{Eq:query_Floquet_phys}).
This factor, which is proportional to the norm of $H(t)$ by Eq. (\ref{Eq:norm_Ht}), is an artifact of normalization.
While a time-independent Hamiltonian in the standard QPE is normalized by $\norm{H} \leq 1/2$, a time-periodic Hamiltonian is not.
If a given Hamiltonian $H$ is not normalized as well for fair comparison, the standard QPE has the same factor proportional to $\norm{H}$ in the query complexity, by replacing $\varepsilon \to \varepsilon / \norm{H}$.
To summarize, we can say that the query complexity of our algorithm resembles that of the standard QPE in all the parameters with logarithmic corrections.
Including that the differences in the number of ancilla qubits and other quantum gates are respectively logarithmic, the quantum algorithm outputting pairs of quasienergy $\epsilon_n$ and a Floquet eigenstate $\ket{\phi_n(t)} \in \mcl{H}$ can be as efficient as the optimal one for time-independent cases.
%=================================
% Section: Quantum algorithm for Floquet eigenstates
%=================================
\section{Floquet QPE: Eigenstates in the Sambe space}\label{Sec:Floquet_QPE_Sambe}
\subsection{Outline of the algorithm}\label{Subsec:Floquet_QPE_Sambe_outline}

\begin{figure*}
    \includegraphics[height=5.69cm, width=18cm]{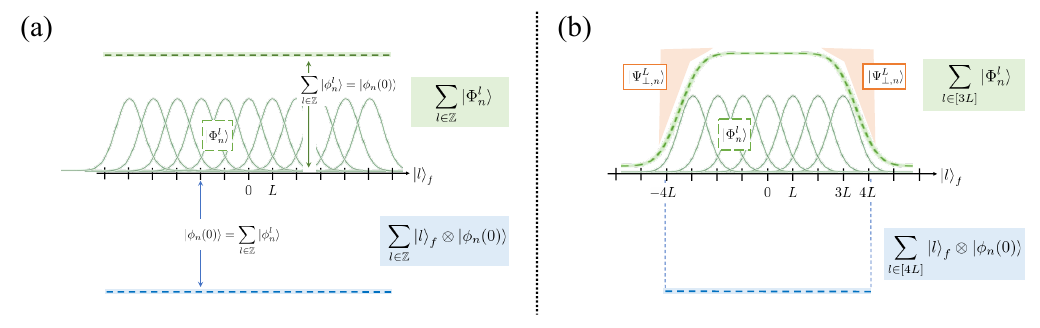}
    \caption{Decomposition of the initial state for the Floquet QPE to compute $(\epsilon_n,\ket{\Phi_n})$. (a) The initial state when the infinite-dimensional Sambe space is ideally prepared. The lower and the upper sides respectively represent the left and the right hand sides of Eq. (\ref{Eq_QPE:Initial_decomp_infinite}). (b) The initial state actually implemented in the algorithm with the truncated Sambe space. Due to the restricted width of each Floquet eigenstate $\ket{\Phi_n^l}$ as Fig. \ref{Fig_eigenstate}, the truncation causes an error only at the edges $l \sim \pm 4L$. As a consequence, the decomposition for the Sambe space holds for $|l| \lesssim 3L$ while we have an unpreferable term $\ket{\Psi_{\perp,n}^L}$ at the boundaries as shown by Proposition \ref{Prop:FloquetQPE_Initial_st}.} 
    \label{Fig_Initial_st}
\end{figure*}

In this section, we construct the Floquet QPE algorithm, which returns pairs of $(\epsilon_n, \ket{\Phi_n})$ for Floquet eigenstate $\ket{\Phi_n}$ on the Sambe space $\mcl{H}^\infty$.
We aim at the transformation, 
\begin{equation}
\sum_n c_n \ket{\phi_n(0)} \to \sum_n c_n \ket{(\epsilon_n)_b}_b \ket{\Phi_n} + \order{\delta}
\end{equation}
in this algorithm.
The strategy is to exploit the standard QPE under the Floquet Hamiltonian $H_\mr{F}^L$ (or exactly $H_\mr{F,pbc}^L$) and extract its eigenvalue $\tilde{\epsilon}_n^L$, which rigorously approximates quasienergy $\epsilon_n$ by Theorem \ref{Thm:Accuracy_quasienergy} proved in Section \ref{Sec:Accuracy_Sambe}.
However, compared to outputting $\ket{\phi_n(t)}$ as described in Section \ref{Sec:Floquet_QPE_physical}, it is difficult to output a Floquet eigenstate $\ket{\Phi_n}$ living in the space different from the initial state $\ket{\psi}$ while preserving the weights $\{ c_n \}_n$ during the process.
We solve this by considering a uniform superposition of a Fourier index $\ket{l}$.
We present a rough sketch of the algorithm with the assumption that the cutoff $L$ is large enough to neglect errors.
\begin{enumerate}
    \item Prepare an initial state in the Sambe space\\
    We prepare an initial state $\ket{\psi}=\sum_n c_n \ket{\phi_n(0)}$, and separately prepare an ancilla state with uniform distribution in Fourier indices.
    The initial state in the Sambe space is a product state,
    \begin{equation}\label{Eq_QPE:Initial_st_infinite}
        \ket{\Psi_0} \propto \left(\sum_{l} \ket{l}_f \right) \otimes \ket{\psi},
    \end{equation}
    where we omit the renormalization factor.
    This allows us to convert $\ket{\phi_n(0)}$ to $\ket{\Phi_n}$ in different spaces with keeping the weights $\{c_n\}_n$, which can be confirmed by the following simple (but yet not rigorous) calculation:
    \begin{eqnarray}
        \qquad \sum_l \ket{l}_f \ket{\psi} &=& \sum_{l} \ket{l}_f \sum_{n; \epsilon_n \in \mr{BZ}} c_n \sum_{l^\prime} \ket{\phi_n^{l^\prime}} \nonumber \\
        &=& \sum_{n; \epsilon_n \in \mr{BZ}} c_n  \sum_{l,l^\prime} \mr{Add}_{l-l^\prime} \ket{l^\prime}_f \ket{\phi_n^{l^\prime}} \nonumber \\
        &=& \sum_{n; \epsilon_n \in \mr{BZ}} c_n \sum_{l} \ket{\Phi_n^l}. \label{Eq_QPE:Initial_decomp_infinite}
    \end{eqnarray}
    This transformation is visualized by Fig. \ref{Fig_Initial_st} (a).
    
    \item QPE by the Floquet Hamiltonian \\
    The standard QPE by the time-independent Floquet Hamiltonian $H_\mr{F}$ transforms the state of Eq. (\ref{Eq_QPE:Initial_decomp_infinite}) into
    \begin{equation}\label{Eq:Floquet_QPE_Sambe_step2}
        \sum_{n; \epsilon_n \in \mr{BZ}} c_n \sum_l \ket{(\epsilon_n-l\omega)_b}_b \ket{\Phi_n^l}.
    \end{equation}
    The $b$-qubit register stores an estimate of each eigenvalue $\epsilon_n-l\omega$ in a $b$-bit binary.
    
    \item Quantum arithmetic to convert $\ket{\Phi_n^l}$ to $\ket{\Phi_n}$ \\
    Quantum division by $\omega$ allows the transformation, 
    \begin{equation}\label{Eq:quantum_division}
        \ket{(\epsilon_n-l\omega)_b}_b \ket{0}_f \to \ket{(\epsilon_n)_b}_b \ket{l}_f.
    \end{equation}
    Then, applying $\sum_l \ket{l}\bra{l}_f \otimes \mr{Add}_l^\dagger$ by quantum substitution removes $\mr{Add}_l$ from Eq. (\ref{Eq:Eigenstate_Sambe_shifted}).
    This results in the preferable output state,
    \begin{equation}\label{Eq:Floquet_QPE_Sambe_step3}
        \left( \sum_{n; \epsilon_n \in \mr{BZ}} c_n \ket{(\epsilon_n)_b}_b \ket{\Phi_n} \right) \otimes \sum_l \ket{l}_f.
    \end{equation}
\end{enumerate}

The above protocol requires the standard QPE under time-independent Hamiltonians and some elementary quantum arithmetic.
Since the cost of the latter is at most poly-logarithmic in all the parameters, the cost of the above algorithm is dominated by the standard QPE.
Therefore, conditions (a$^\prime$) and (b$^\prime$) are expected to be satisfied.
Requirement (c$^\prime$) is also promising because of the maintained coherence.
The probability of measuring the outcome $(\epsilon_n, \ket{\Phi_n})$ is $p_n = |c_n|^2 = |\braket{\phi_n(0)|\psi}|^2$ from Eq. (\ref{Eq:Floquet_QPE_Sambe_step3}).
Thus, we can prepare Floquet eigenstates in the Sambe space  $\ket{\Phi_n}$ based on a good initial guess on the physical eigenstates $\ket{\phi_n(0)}$.

However, we have to keep in mind that the above rough sketch neglects the infinite-dimensionality of the Sambe space.
We formulate the exact quantum algorithm that works on the truncated Sambe space, satisfying requirements (a)--(c).
The central ingredients for resolving the infinite-dimensionality are Theorem \ref{Thm:Accuracy_quasienergy} and \ref{Thm:Tail_Eigenstates}, which can guarantee efficiency and accuracy of the truncation.
The following sections are organized as follows.
In Section \ref{Subsec:FloquetQPE_Initial_st}, we construct a reasonable initial state in the truncated Sambe space, and show its counterpart of the decomposition, Eq. (\ref{Eq_QPE:Initial_decomp_infinite}).
Sections \ref{Subsec:FloquetQPE_with_RP} and \ref{Subsec:Floquet_QPE_wo_RP} are dedicated to the algorithms with and without rounding promise, respectively.
As an artifact of the truncation, we need QAA to remove non-negligible errors caused by it, but it does not change the scaling of the cost.
As a result, the scaling of the computational complexity is also essentially the same as that of the standard QPE even when we take the truncation into account.
The above rough sketch of the algorithm is intuitively correct, which is summarized by Theorem \ref{Thm:algorithm_Sambe}.

\subsection{Initial state in the Sambe space}\label{Subsec:FloquetQPE_Initial_st}

In this section, we organize a reasonable initial state in the truncated Sambe space from a given physical initial state $\ket{\psi} \in \mcl{H}$, which plays a role of Eq. (\ref{Eq_QPE:Initial_st_infinite}).
The serious problem caused by the truncation appears in the decomposition, Eq. (\ref{Eq_QPE:Initial_decomp_infinite}).
It relies heavily on the infinite sum over $l \in \bbZ$, and this makes the initial state Eq. (\ref{Eq_QPE:Initial_st_infinite}) unnormalized and divergent.
We solve this by showing the counterpart of the decomposition Eq. (\ref{Eq_QPE:Initial_decomp_infinite}).

Let us consider a truncated Sambe space $\mcl{H}^{2pL} = \bbC^{[2pL]} \otimes \mcl{H}$, where an integer $p \in \bbN$ and a large cutoff $L \in \bbN$ will be determined later.
The infinite dimensional Sambe space is used here for the sake of calculation, but we note that the physically accessible states are in the truncated Sambe space.
For a given initial state $\ket{\psi} = \sum_{n; \epsilon_n \in \mr{BZ}} c_n \ket{\phi_n(0)}$, we define the initial state in the truncated Sambe space by
\begin{equation}\label{Eq_QPE:initial_st_truncated}
    \ket{\Psi_0^L} = \left( \frac{1}{\sqrt{8L}} \sum_{l \in [4L]} \ket{l}_f \right) \otimes \ket{\psi} \equiv U_\mr{uni}^{4L} \ket{0}_f \ket{\psi}
\end{equation}
For this definition, $p \geq 4$ should hold.
The unitary circuit $U_\mr{uni}^{4L}$ for state preparation can be easily constructed by $\order{\log L}$ primitive gates to generate uniform distribution on the ancilla system.
We formulate the decomposition corresponding to Eq. (\ref{Eq_QPE:Initial_decomp_infinite}).
A significant difference from the infinite-dimensional case appears due to the boundaries of the summation over $l$, i.e., $l \simeq \pm 4L$.
The exponentially-decaying tails of Floquet eigenstates (Theorem \ref{Thm:Tail_Eigenstates}) imply that the computation based on the infinite-dimensional Sambe space, i.e. Eq. (\ref{Eq_QPE:Initial_decomp_infinite}), is correct in the middle of $[4L]$, but no longer valid near the boundaries $l \simeq \pm 4L$.
Namely, the decomposition of the initial states contains unwanted terms at the boundaries, as shown in Fig. \ref{Fig_Initial_st} (b).
The following proposition rigorously provides this decomposition for the truncated Sambe space.
While we focus on a single Floquet eigenstate for simplicity, we note that generic cases are easily reproduced by the linearity.

\begin{proposition}\label{Prop:FloquetQPE_Initial_st}                                   
\textbf{(Decomposing the initial state)}

Consider the initial state $\ket{\Psi_0^L}$ for a single Floquet eigenstate, given by
\begin{equation}
    \ket{\Psi_{0,n}^L} \equiv \left( \frac{1}{\sqrt{8L}} \sum_{l \in [4L]} \ket{l}_f \right) \otimes \ket{\phi_n(0)}.
\end{equation}
Then, it is decomposed by
\begin{equation}\label{Eq_QPE:Initial_decomp_finite}
    \ket{\Psi_{0,n}^L} =  \frac{1}{\sqrt{8L}} \sum_{l \in [3L]}  \ket{\Phi_n^l}+ \frac12 \ket{\Psi_{\perp,n}^L} + \ket{\Psi_{\mr{neg},n}^L},
\end{equation}
where the states $\ket{\Psi_{\perp,n}^L}$ and $\ket{\Psi_{\mr{neg},n}^L}$ on the Sambe space satisfy the following properties.
\begin{itemize}
    \item The high-frequency term $\ket{\Psi_{\perp,n}^L}$: \\
    Its norm is renormalized as $\braket{\Psi_{\perp,n}^L|\Psi_{\perp,n}^L}=1$.
    When we define a projection $P(\mr{BZ}_{L+1}^{6L})$ by
    \begin{equation}
        \qquad \qquad P(\mr{BZ}_{L+1}^{6L}) = \sum_{n; \epsilon_n \in \mr{BZ}} \sum_{l \in [6L] \backslash [L+1]} \ket{\Phi_n^l}\bra{\Phi_n^l}, 
    \end{equation}
    the state $\ket{\Psi_{\perp,n}^L}$ satisfies
    \begin{equation}\label{Eq_QPE:high_term_Proj}
        \qquad \qquad P(\mr{BZ}_{L+1}^{6L}) \ket{\Psi_{\perp,n}^L} = \ket{\Psi_{\perp,n}^L}.
    \end{equation}
    It means that this state has a large Fourier index $l \in [6L]\backslash [L+1]$, indicating the large-scale energy under $H_\mr{F}$.
    
    \item The negligible term $\ket{\Psi_{\mr{neg},n}^L}$: \\
    Its norm is bounded by
    \begin{equation}\label{Eq_QPE:Initial_decomp_negligible}
    \norm{\ket{\Psi_{\mr{neg},n}^L}} \leq e^{-\Theta(L-\alpha T)},
    \end{equation}
    and can be negligible under the large cutoff $L$.
\end{itemize}
\end{proposition}

The first term in the right-hand side of Eq. (\ref{Eq_QPE:Initial_decomp_finite}) represents a uniform superposition of Floquet eigenstates in the Sambe space, which appears as a counterpart of Eq. (\ref{Eq_QPE:Initial_decomp_infinite}).
On the other hand, the second and the third terms are drawbacks of truncating the Sambe space.

We will prove Proposition \ref{Prop:FloquetQPE_Initial_st} by decomposing the residual term defined by
\begin{equation}\label{Eq_QPE:Initial_decomp_residual}
    \ket{\Psi_{1,n}^L} \equiv \ket{\Psi_{0,n}^L} -  \frac{1}{\sqrt{8L}} \sum_{l \in [3L]} \ket{\Phi_n^l}
\end{equation}
in a few steps.
We start with the following lemma, which gives the norm of this state.

\begin{lemma}\label{Lemma:Norm_residual}
\textbf{(Norm of Residual term)}

The norm of the residual term $\ket{\Psi_{1,n}^L}$, defined by Eq. (\ref{Eq_QPE:Initial_decomp_residual}), is bounded by
\begin{eqnarray}\label{Eq_QPE:Norm_residual}
    \left| \norm{\ket{\Psi_{1,n}^L}} - \frac12 \right| &\leq& 27 (2M+1) e^{- \frac{L}{2M+1} + \frac{\sinh 1}{2\pi} \alpha T}\nonumber \\
    &\in& e^{- \Theta (L-\alpha T)}.
\end{eqnarray}
\end{lemma}

\textit{Proof of Lemma \ref{Lemma:Norm_residual}.---}
Note that $\ket{\Phi_n^l}$ and $\ket{\Phi_n^{l^\prime}}$ are orthogonal for $l \neq l^\prime$ since they are different eigenstates of the Floquet Hamiltonian $H_\mr{F}$.
The squared norm of interest is evaluated by
\begin{eqnarray}
    \braket{\Psi_{1,n}^L|\Psi_{1,n}^L} &=& \frac74 - \frac1{\sqrt{2L}} \sum_{l \in [3L]} \mr{Re}\braket{\Psi_{0,n}^L |\Phi_n^l} \nonumber \\
    &=&  \frac74 - \frac1{4L} \sum_{l \in [3L]} \sum_{l^\prime \in [4L]} \mr{Re} \braket{\phi_n(0)|\phi_n^{l^\prime - l}} \nonumber \\
    &=& \frac14 + \sum_{l \in [3L]} \frac{1 - \sum_{l^\prime \in [4L]} \mr{Re}  \braket{\phi_n(0)|\phi_n^{l^\prime-l}}}{4L}. \nonumber \\
    &&
\end{eqnarray}
We recall that each Floquet eigenstate $\ket{\phi_n(0)}$ is renormalized and that it is given by the summation $\ket{\phi_n(0)} = \sum_{l\in \bbZ} \ket{\phi_n^l}$ in the Fourier series as Eq. (\ref{Eq:def_Floquet_state}).
For every $l \in [3L]$, the second term is bounded by
\begin{eqnarray}
    && \left| 1 - \sum_{l^\prime \in [4L]} \mr{Re}  \braket{\phi_n(0)|\phi_n^{l^\prime-l}} \right| \nonumber \\
    && \quad = \left| \mr{Re} \bra{\phi_n(0)} \left( \ket{\phi_n(0)} - \sum_{l^\prime \in [4L]} \ket{\phi_n^{l^\prime-l}} \right) \right| \nonumber \\
    && \quad \leq \sum_{l^\prime \in \bbZ \backslash [L]} \norm{\ket{\phi_n^{l^\prime}}}. 
\end{eqnarray}
The left hand side of Eq. (\ref{Eq_QPE:Norm_residual}) is evaluated by
\begin{eqnarray}
    \left| \norm{\ket{\Psi_{1,n}^L}} - \frac12 \right| &\leq& 2 \left| \braket{\Psi_{1,n}^L|\Psi_{1,n}^L} - \frac14 \right| \nonumber \\
    &\leq& 3 \sum_{l^\prime \in \bbZ \backslash [L]} \norm{\ket{\phi_n^{l^\prime}}}.
\end{eqnarray}
Using Eq. (\ref{Eq:phi_n_l_contribute_out}) as the result of Theorem \ref{Thm:Tail_Eigenstates}, we arrive at the inequality, Eq. (\ref{Eq_QPE:Norm_residual}). $\quad \square$

This lemma dictates that the weight of the unwanted term in the initial state $\ket{\Psi_{0,n}^L}$ is approximately $1/4$.
Namely, the drawback of the finite-dimensionality is not negligible, which is the reason of introducing QAA later.
Anyway, we are ready to prove Proposition \ref{Prop:FloquetQPE_Initial_st} as follows.

\textit{Proof of Proposition \ref{Prop:FloquetQPE_Initial_st}.---}
We consider the decomposition of the residual term $\ket{\Psi_{1,n}^L}$ by the projection $P(\mr{BZ}_{L+1}^{6L})$ as follows:
\begin{equation}\label{PropEq:FloquetQPE_Initial_st_1}
\ket{\Psi_{1,n}^L} = P(\mr{BZ}_{L+1}^{6L}) \ket{\Psi_{1,n}^L} + (I-P(\mr{BZ}_{L+1}^{6L}))\ket{\Psi_{1,n}^L}.
\end{equation}
Then, we define the states in the Sambe space, $\ket{\Psi_{\perp,n}^L}$ and $\ket{\Psi_{\mr{neg},n}^L}$, by
\begin{eqnarray}
    \ket{\Psi_{\perp,n}^L} &=& \frac{P(\mr{BZ}_{L+1}^{6L})\ket{\Psi_{1,n}^L}}{\norm{P(\mr{BZ}_{L+1}^{6L})\ket{\Psi_{1,n}^L}}}, \label{PropEq:FloquetQPE_Initial_st_2} \\
    \ket{\Psi_{\mr{neg},n}^L} &=& (I-P(\mr{BZ}_{L+1}^{6L}))\ket{\Psi_{1,n}^L} \nonumber \\
    &&  + \left( \norm{P(\mr{BZ}_{L+1}^{6L})\ket{\Psi_{1,n}^L}}-1/2 \right) \ket{\Psi_{\perp,n}^L}. \nonumber \\
    && \label{PropEq:FloquetQPE_Initial_st_3} 
\end{eqnarray}
From the definitions, they give the decomposition of the initial state $\ket{\Psi_{0,n}^L}$ by Eq. (\ref{Eq_QPE:Initial_decomp_finite}).
About the state $\ket{\Psi_{\perp,n}^L}$, the normalization and the invariance under the projection by Eq. (\ref{Eq_QPE:high_term_Proj}) is trivial by definition as long as we can show that $P(\mr{BZ}_{L+1}^{6L})\ket{\Psi_{1,n}^L}$ is non-vanishing.
We start by focusing on the state $\ket{\Psi_{\mr{neg},n}^L}$ and prove that its norm decays as Eq. (\ref{Eq_QPE:Initial_decomp_negligible}).

We evaluate each term in Eq. (\ref{PropEq:FloquetQPE_Initial_st_3}), which is composed of the state $\ket{\Psi_{\mr{neg},n}^L}$.
Using a projection $P_l = \ket{l}\bra{l}_f \otimes I$, the norm of the first term is expressed by
\begin{eqnarray}
    && \norm{(I-P(\mr{BZ}_{L+1}^{6L})) \ket{\Psi_{1,n}^L}} \nonumber \\
    && \quad \leq \sum_{l \in \bbZ} \norm{(I-P(\mr{BZ}_{L+1}^{6L})) P_l} \cdot \norm{P_l \ket{\Psi_{1,n}^L}} \nonumber \\
    && \quad \leq \left(\sum_{l \in [2L]} + \sum_{\bbZ \backslash [5L]} \right) \norm{P_l \ket{\Psi_{1,n}^L}} \nonumber \\
    && \quad \qquad + \sum_{l \in [5L]\backslash[2L]} \norm{(I-P(\mr{BZ}_{L+1}^{6L})) P_l}. \label{Eq_QPE:neg_term_outside}
\end{eqnarray}
In the above, we use $\norm{\ket{\Psi_{1,n}^L}} \leq 1$ by Lemma \ref{Lemma:Norm_residual}.
The sum over $l \in [2L]$ is bounded by
\begin{eqnarray}
    \sum_{l \in [2L]} \norm{P_l \ket{\Psi_{1,n}^L}} &\leq& \frac{1}{\sqrt{8L}} \sum_{l \in [2L]} \norm{\ket{\phi_n(0)} - \sum_{l^\prime \in [3L]} \ket{\phi_n^{l-l^\prime}}} \nonumber \\
    &\leq& \sqrt{2L} \sum_{l^\prime \in \bbZ \backslash [L]} \norm{\ket{\phi_n^{l^\prime}}}. 
\end{eqnarray}
Similarly, considering that the state $\ket{\Psi_{0,n}^L}$ has Fourier indices only in $l \in [4L]$, the summation over $l \in \bbZ \backslash [5L]$ has a bound,
\begin{eqnarray}
    \sum_{l \in \bbZ \backslash [5L]} \norm{P_l \ket{\Psi_{1,n}^L}} &\leq& \frac{1}{\sqrt{8L}} \sum_{l \in \bbZ \backslash [5L]} \sum_{l^\prime \in [3L]} \norm{\ket{\phi_n^{l-l^\prime}}} \nonumber \\
    &\leq& \sqrt{\frac{9L}{2}} \sum_{l \in \bbZ \backslash [L]}\norm{\ket{\phi_n^l}}. 
\end{eqnarray}
The sum over $l \in [5L] \backslash [2L]$ in Eq. (\ref{Eq_QPE:neg_term_outside}) is bounded by
\begin{eqnarray}
    && \sum_{l \in [5L]\backslash[2L]} \norm{(I-P(\mr{BZ}_{L+1}^{6L})) P_l} \nonumber \\
    && \quad \leq \sum_{l \in [5L]\backslash [2L]} \norm{\sum_n \left( \sum_{l^\prime \in [L+1]} + \sum_{l^\prime \in \bbZ \backslash [6L]} \right)\ket{\Phi_n^{l^\prime}} \bra{\Phi_n^{l^\prime}} P_l} \nonumber \\
    && \quad \leq \sum_{l \in [5L]\backslash [2L]} \max_{n, l^\prime} \left( \norm{\ket{\phi_n^{l^\prime-l}}} \, | \, l^\prime \in [L+1] \cup (\bbZ \backslash [6L]) \right) \nonumber \\
    && \quad \leq 2 \sum_{l \in \bbZ \backslash [L]} \norm{\ket{\phi_n^l}}.
\end{eqnarray}
Thus, the first term in $\ket{\Psi_\mr{neg}^L}$ can be bounded by
\begin{equation}\label{Eq_QPE:neg_term_outside_bound}
    \norm{(I-P(\mr{BZ}_{L+1}^{6L}))\ket{\Psi_{1,n}^L}} \leq \left( \frac52 \sqrt{2L} + 2 \right) \sum_{l \in \bbZ \backslash [L]} \norm{\ket{\phi_n^l}}. 
\end{equation}
On the other hand, the second term of $\ket{\Phi_\mr{neg}^L}$ is immediately evaluated by the triangle inequality,
\begin{eqnarray}
    && \norm{ \left( \norm{P(\mr{BZ}_{L+1}^{6L}) \ket{\Psi_{1,n}^L}} - \frac12 \right) \ket{\Psi_{\perp,n}^L}} \nonumber \\
    && \leq \norm{(I-P(\mr{BZ}_{L+1}^{6L}))\ket{\Psi_{1,n}^L}} + \left| \norm{\ket{\Psi_{1,n}^L}} - \frac12\right|.
\end{eqnarray}
These two terms are bounded by Eqs. (\ref{Eq_QPE:neg_term_outside_bound}) and (\ref{Eq_QPE:Norm_residual}) [i.e. Lemma \ref{Lemma:Norm_residual}], respectively.
Using Eq. (\ref{Eq:phi_n_l_contribute_out}) as a result of Theorem \ref{Thm:Tail_Eigenstates}, the state $\ket{\Psi_{\mr{neg},n}^L}$ has an upper bound,
\begin{equation}
    \norm{\ket{\Psi_{\mr{neg},n}^L}} \leq 45 (2M+1)^2 \left( \sqrt{2L}+1 \right) e^{- \frac{L}{2M+1} + \frac{\sinh 1}{2\pi} \alpha T}.
\end{equation}
The scaling of the right-hand side is given by $e^{- \Theta (L-\alpha T)}$, which implies Eq. (\ref{Eq_QPE:Initial_decomp_negligible}). 

Finally, we confirm that $P(\mr{BZ}_{L+1}^{6L})\ket{\Psi_{1,n}^L}$ is non-vanishing.
This follows directly from the triangle inequality,
\begin{eqnarray}
    \norm{P(\mr{BZ}_{L+1}^{6L})\ket{\Psi_{1,n}^L}} &\geq& \norm{\ket{\Psi_{1,n}^L}} - \norm{(I-P(\mr{BZ}_{L+1}^{6L}))\ket{\Psi_{1,n}^L}} \nonumber \\
    &\geq& \frac12 - e^{-\Theta (L-\alpha T)}.
\end{eqnarray}
Therefore, the state $\ket{\Psi_{\perp,n}^L}$ defined by Eq. (\ref{PropEq:FloquetQPE_Initial_st_2}) satisfies the normalization $\braket{\Psi_{\perp,n}^L | \Psi_{\perp,n}^L } = 1$ and the invariance under the projection by Eq. (\ref{Eq_QPE:high_term_Proj}). $\quad \square$

If we start with a generic initial state $\ket{\psi} = \sum_n c_n \ket{\phi_n(0)}$, the linearity immediately provides the decomposition of $\ket{\Psi_0^L}$ [Eq. (\ref{Eq_QPE:initial_st_truncated})] given by
\begin{equation}\label{Eq_QPE:Initial_decomp_finite_generic}
    \ket{\Psi_0^L} = \sum_n c_n \left( \frac{1}{\sqrt{8L}} \sum_{l \in [3L]} \ket{\Phi_n^l} \right) + \frac12 \ket{\Psi_{\perp}^L} + \ket{\Psi_\mr{neg}^L},
\end{equation}
where we use the abbreviations $\ket{\Psi_{\perp}^L} = \sum_n c_n \ket{\Psi_{\perp,n}^L}$ and $\ket{\Psi_\mr{neg}^L} = \sum_n c_n \ket{\Psi_{\mr{neg},n}^L}$.
This decomposition is summarized in Fig. \ref{Fig_Initial_st} (b).
The first term corresponds to Eq. (\ref{Eq_QPE:Initial_st_infinite}), which gives the ideal decomposition of the Floquet eigenstates while maintaining the coherence by $c_n$.
Due to the finite dimensionality, the summation over $l \in \bbZ$ is replaced by the one over $l \in [3L]$, which implies that the boundary terms around $l \simeq \pm 4L$ are invalid, leading to the undesirable second term $\ket{\Psi_\perp^L}$.
Due to the property of $\ket{\Psi_{\perp,n}^L}$ by Eq. (\ref{Eq_QPE:high_term_Proj}), this state can be expanded by
\begin{equation}\label{Eq_QPE:high_term_decomp}
    \ket{\Psi_{\perp,n}^L} = \sum_{n; \epsilon_n \in \mr{BZ}} \sum_{l \in [6L]\backslash [L+1]} d_{nl} \ket{\Phi_n^l}
\end{equation}
with $\sum_n |d_{nl}|^2 = 1$.
The negligible state $\ket{\Psi_\mr{neg}^L}$ can be evaluated by using Cauchy-Schwartz inequality, which gives an additional factor $\sum_{n; \epsilon_n \in \mr{BZ}} |c_n| \leq \sqrt{\mr{dim}(\mcl{H})} = 2^{N/2}$.
We get the upper bound of its norm,
\begin{eqnarray}
    \norm{\ket{\Psi_\mr{neg}^L}} &\leq& \sum_n |c_n| \cdot \norm{\ket{\Psi_{\mr{neg},n}^L}} \nonumber \\
    &\leq& e^{- \Theta (L - \alpha T -N)}. \label{Eq_QPE:neg_term_generic_bound}
\end{eqnarray}
The state $\norm{\ket{\Psi_\mr{neg}^L}}$ can be arbitrarily small by redefining the cutoff $L \in \Omega (\alpha T + N + \log (1/\varepsilon))$, as shown later.
Since the energy term $\alpha T$ typically grows polynomially with $N$, the increase in the cutoff is not problematic.

\subsection{Floquet QPE under rounding promise}\label{Subsec:FloquetQPE_with_RP}

Next, we formulate the QPE protocol employing the truncated Floquet Hamiltonian, corresponding to Step 2 in Section \ref{Subsec:Floquet_QPE_Sambe_outline}.
The difficulty here is the appearance of the undesirable state $\ket{\Psi_\perp^L}$ in Eq. (\ref{Eq_QPE:Initial_decomp_finite_generic}), which results from the truncation.
We combine QPE and QAA in order to simultaneously extract accurate quasienergy and remove the unfavorable component.

First, we look at the quantum algorithm for the transformation,
\begin{equation}\label{Eq_QPE:Floquet_Sambe_transf_RP}
    \ket{0}_b \ket{\psi} \to \sum_n c_n \ket{(\epsilon_n)_b}_b \ket{\Phi_n} + \order{\delta},
\end{equation}
under the rounding promise $\nu$.
Following Proposition \ref{Prop:Inherited_rounding_promise}, the truncated Floquet Hamiltonian $H_\mr{F}^{pL}$ can have rounding promise when the quasienergy $\epsilon_n$ satisfies Eq. (\ref{Eq:rounding_promise_quasienergy}) in certain $b^\prime \in \Theta (\log(1/\varepsilon))$ bits.
We choose the cutoff by $L \in \Omega (\alpha T + \log (1/\varepsilon \nu))$ and set the integer $p$ by $p \geq 7$.
Then, the rounding promise $\nu/2$ in $b^\prime + b_\mr{F}^{pL}$ bits is satisfied for every eigenvalue $\tilde{\epsilon}_n^{pL} \in \mr{BZ}_l$ of $^\forall l \in [6L]$, where the bit number $b_\mr{F}^{pL} \in \Theta (\log (\alpha T + L))$ is given by Eq. (\ref{Eq:bit_normalization}).
To obtain an estimate of $\tilde{\epsilon}_n^{pL}/\omega$ within an error $\order{\varepsilon}$, each renormalized eigenvalue $\tilde{\epsilon}_n^L/(2^{b_\mr{F}^{pL}} \omega)$ must be computed within an error $2^{-b_\mr{F}^{pL}} \varepsilon$.
We run the standard QPE for the renormalized Floquet Hamiltonian $H_\mr{F}^{pL}/(2^{b_\mr{F}^{pL}} \omega)$, setting their parameters $\nu,\varepsilon,\delta$ by $\nu/2, 2^{-b_\mr{F}^{pL}} \varepsilon, \delta$ respectively.
Denoting the corresponding unitary circuit by $U_\mr{QPE}^L$ and considering the decomposition of the initial state $\ket{\Psi_0^L}$ by Eqs. (\ref{Eq_QPE:Initial_decomp_finite_generic}), (\ref{Eq_QPE:high_term_decomp}), and (\ref{Eq_QPE:neg_term_generic_bound}), the transformation is given by
\begin{widetext}
\begin{eqnarray}
    U_\mr{QPE}^L \ket{0}_b \ket{\Psi_0^L} &=& \sum_{n; \epsilon_n \in \mr{BZ}} \left( \frac{c_n}{\sqrt{8L}} \sum_{l \in [3L]} \ket{(\epsilon_n - l \omega)_b}_b \ket{\Phi_n^l} +  \sum_{l \in [6L] \backslash [L+1] }\frac{d_{nl}}{2} \ket{(\epsilon_n - l \omega)_b}_b \ket{\Phi_n^l}\right)
    \nonumber \\
    && \qquad \qquad \qquad \qquad \qquad \qquad \qquad \qquad \qquad \qquad   + \order{\delta_\mr{approx}} + \order{ \delta}  + e^{- \Theta (L-\alpha T - N)}. \label{Eq_QPE:QPE_H_F}
\end{eqnarray}
\end{widetext}
The register has $b=b^\prime + b_\mr{F}^{pL}$ qubits.
The error term $\delta_\mr{approx}$ arises from the fact that every Floquet eigenstate $\ket{\Phi_n}$ is an approximate eigenstate of $H_\mr{F}^{pL}$.
The first term of Eq. (\ref{Eq_QPE:QPE_H_F}) is obtained by running the QPE as if $\ket{\Phi_n}$ were an exact eigenstate of $H_\mr{F}^{pL}$.
Its drawback appears as the accumulation of $e^{-\Theta (L-\alpha T)}$ errors by Theorem \ref{Thm:Accuracy_quasienergy}, which grows with the increasing query complexity of the standard QPE, $q_\mr{QPE}$.
In fact, the error term is bounded by
\begin{equation}\label{Eq_QPE:delta_approx}
    \delta_\mr{approx} \leq q_\mr{QPE}^2 e^{- \Theta (L-\alpha T -N)}.
\end{equation}
See Appendix \ref{A_Sec:Approximate_QSVT} for its detailed derivation.
The third error term in Eq. (\ref{Eq_QPE:QPE_H_F}) comes from $\ket{\Psi_{\varepsilon}^L}$ with using Eq. (\ref{Eq_QPE:neg_term_generic_bound}).
To ensure that these three errors are bounded by $\order{\delta}$, we require that the cutoff $L$ satisfies
\begin{equation}\label{Eq_QPE:cutoff_for_state_error}
    L \in \Omega \left( \alpha T + N + \log q_\mr{QPE} + \log (1/\delta) \right).
\end{equation}

\begin{figure*}
    \includegraphics[height=7.5cm, width=18cm]{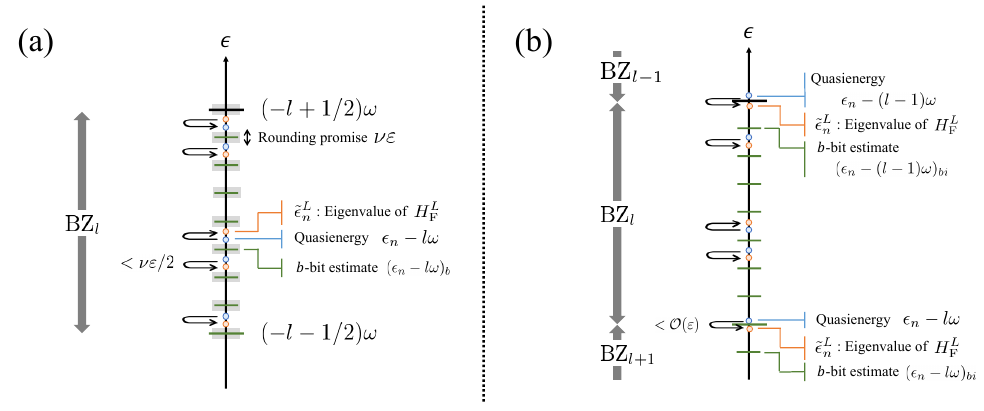}
    \caption{Relations among quasienergy, eigenvalue of the truncated Floquet Hamiltonian, and their $b$-bit estimate in the $l$-th Brillouin zone $\mr{BZ}_l$. (a) The spectrum in the presence of rounding promise. Rounding promise prohibits each quasienergy $\epsilon_n - l\omega$ by an blue dot to be located in gray regions around the ticks. The same thing holds for any eigenvalue of $H_\mr{F}^L$ designated by an orange dot owing to Proposition \ref{Prop:Inherited_rounding_promise}. Since each $b$-bit estimate $(\epsilon_n - l \omega)_b$ described by an green tick gives a floor function of them, the QPE based on $H_\mr{F}^L$ returns an $b$-bit estimate in $\mr{BZ}_l$ for every quasienergy $\epsilon_n-l\omega \in \mr{BZ}_l$. (b) The spectrum in the absence of rounding promise. At the boundaries of $\mr{BZ}_l$, some pairs of quasienergy and $b$-bit estimates can belong to the different Brillouin zones, i.e., $\mr{BZ}_{l-1}$ or $\mr{BZ}_{l+1}$. This leads to a failure of the QAA like Eq. (\ref{Eq_QPE:QSVT_QAA_wo_RP}), and therefore we consider the quasienergy estimation excluding the $\varepsilon \omega$-neighborhood of the boundaries of $\mr{BZ}$ as Eqs. (\ref{Eq_QPE:Initial_st_wo_RP}) and (\ref{Eq_QPE:BZ_wo_boundaries}).} 
    \label{Fig_spectrum}
\end{figure*}

\textbf{Measurement on the register.---} We show that Eq. (\ref{Eq_QPE:QPE_H_F}) probabilistically provides the success of the Floquet QPE for $(\epsilon_n,\ket{\Phi_n})$.
Let us consider what happens when we measure its $b$-qubit register.
Each measured $b$-bit value $(\epsilon_n - l\omega)_b$ for $l \in [6L]$ is an estimate of $\tilde{\epsilon}_n^{pL}$, which accurately approximates the exact $\epsilon_n-l\omega$.
Although the approximate value $(\epsilon_n - l\omega)_b$ may not be included in $\mr{BZ}_l$, we have $(\epsilon_n - l\omega)_b \in \mr{BZ}_l$ for every $l \in [6L]$ owing to the rounding promise.
Indeed, each renormalized quasienergy $(\epsilon_n-l\omega)/\omega$ belongs to $[-l-1/2 + \order{\varepsilon\nu}, -l+1/2-\order{\varepsilon \nu})$, and so does the corresponding eigenvalue $\tilde{\epsilon}_n^{pL}/\omega$, obtaining the error up to $\order{\varepsilon \nu}$ by Theorem \ref{Thm:Accuracy_quasienergy}.
Since the standard QPE returns the floor function of $\tilde{\epsilon}_n^{pL}/\omega$ by Eq. (\ref{Eq:stored_binary}), the measured value $(\epsilon_n - l\omega)_b$ is always contained in $\mr{BZ}_l$ as shown in Fig. \ref{Fig_spectrum} (a).
Namely, we can extract the index $l$ for the BZ without error from the stored values of the register.

Let us focus on the case where the measured binary $(\epsilon_n-l\omega)_b$ belongs to the $2L$ repetitions of the BZ,
\begin{equation}
    \mr{BZ}_{[L]} = \bigcup_{l \in [L]} \mr{BZ}_l = [(-L+1/2)\omega,(L+1/2)\omega).
\end{equation}
In other words, we post-select the measured state so that it can be projected by
\begin{equation}\label{Eq:Proj_L_range}
    \Pi_{[L]} = \sum_{x \in \mr{BZ}_{[L]}} \ket{x}\bra{x}_b \otimes I_f \otimes I.
\end{equation}
Applying the projection $\Pi_{[L]}$ to the output state Eq. (\ref{Eq_QPE:QPE_H_F}), only the summation over $l \in [L]$ survives due to $(\epsilon_n-l\omega)_b \in \mr{BZ}_l$.
This implies that this projection can delete the unwanted component $\ket{\Psi_\perp^L}$ (or $d_{nl}$) coming from the truncation.
The success probability of the projection is given by
\begin{eqnarray}
    p_\mr{suc} &=& \norm{\Pi_{[L]} U_\mr{QPE}^L \ket{\Psi_0^L}}^2 \nonumber \\
    &=& \norm{\sum_{n; \epsilon_n \in \mr{BZ}} \frac{c_n}{\sqrt{8L}} \sum_{l \in [L]} \ket{(\epsilon_n-l\omega)_b}_b \ket{\Phi_n^l} + \order{\delta}}^2  \nonumber \\
    &\geq& \frac14 - \order{\delta}.
\end{eqnarray}
After the projection, we obtain the state,
\begin{eqnarray}
    && \frac{\Pi_{[L]} U_\mr{QPE}^L \ket{0}_b\ket{\Psi_0^L}}{\sqrt{p_\mr{suc}}} \nonumber \\
    && \quad = \sum_{n; \epsilon_n \in \mr{BZ}} \frac{c_n}{\sqrt{2L}} \sum_{l \in [L]} \ket{(\epsilon_n-l\omega)_b}_b \ket{\Phi_n^l} + \order{\delta}. \nonumber \\
    && \label{Eq:QPE_Proj_L_range}
\end{eqnarray}
This state corresponds to the rough result of the standard QPE for the infinite-dimensional Sambe space given by Eq. (\ref{Eq:Floquet_QPE_Sambe_step2}).
Therefore, we can obtain the target state $\sum_n c_n \ket{(\epsilon_n)_b}_b \ket{\Phi_n}$ by the following quantum arithmetic by Eqs. (\ref{Eq:quantum_division}) and (\ref{Eq:Floquet_QPE_Sambe_step3}), if we succeed in the post-selection with probability $p_\mr{suc} \geq 1/4 - \order{\delta}$.

\textbf{Floquet QPE with QAA.---}
Finally, we formulate the protocol to obtain $\sum_n c_n \ket{(\epsilon_n)_b}_b \ket{\Phi_n}$ with certainty by exploiting QAA based on QSVT \cite{Berry2017-qaa,Martyn2021-grand-unif}.
It allows us to enhance a preferable state when we have an oracle that can distinguish it from the other orthogonal states.
Similar to the Grover's search algorithm \cite{Grover1997-prl,Hoyer2000-grover,Long2001-grover}, it has a quadratic speedup in that a state prepared with probability $p_\mr{suc}$ can be implemented by $\order{1/\sqrt{p_\mr{suc}}}$ query complexities.
Here, we formulate this based on QSVT to amplify the success probability of post-selection $1/4-\order{\delta}$.

We use two projection operators; the first one is $\Pi_0$ defined by
\begin{equation}
    \Pi_0 = \ket{0}\bra{0}_b \otimes \ket{0}\bra{0}_f \otimes I, 
\end{equation}
and the second one is $\Pi_{[L]}$ defined by Eq. (\ref{Eq:Proj_L_range}).
Then, the unitary operation $U_\mr{QPE}^L U_\mr{uni}^{4L}$ with these projections provides an approximate block-encoding,
\begin{widetext}
\begin{equation}\label{Eq_QPE:Floquet_Sambe_block_encode_ext}
    \Pi_{[L]} \left\{ U_\mr{QPE}^L U_\mr{uni}^{4L} \right\}\Pi_0  = \sum_n \frac12 \left( \frac{1}{\sqrt{2L}} \sum_{l \in [L]} \ket{(\epsilon_n - l \omega)_b}_b \ket{\Phi_n^l} \right) \bra{0}_b \bra{0}_f\bra{\phi_n(0)} + \order{\delta}, 
\end{equation}
\end{widetext}
which is derived from Eqs. (\ref{Eq_QPE:QPE_H_F}) and (\ref{Eq:QPE_Proj_L_range}).
All of the singular values of the embedded matrix are $1/2$.
To run QAA, we define a sequence of QSVT based on this block-encoding by
\begin{widetext}
\begin{equation}\label{Eq_QPE:Floquet_Sambe_QSVT}
    \mr{QSVT}[\vec{\theta}_q] = R_{\Pi_{[L]}}(\theta_1) \left\{ U_\mr{QPE}^L U_\mr{uni}^{4L}  \right\} \prod_{i=1}^{(q-1)/2} \left(  R_{\Pi_0}(\theta_{2i}) \left\{ U_\mr{QPE}^L U_\mr{uni}^{4L} \right\}^\dagger R_{\Pi_{[L]}}(\theta_{2i+1}) \left\{ U_\mr{QPE}^L U_\mr{uni}^{4L} \right\}  \right), 
\end{equation}
\end{widetext}
for an odd integer $q$, a tunable parameter set $\vec{\theta}_q \in \bbR^q$, and a unitary on auxiliary qubits,
\begin{equation}\label{Eq:Phased_iterate}
    R_\Pi (\theta) = e^{i \theta} \Pi + e^{-i\theta} (I-\Pi), \quad \text{$\Pi$; projection}.
\end{equation}
For both projections $\Pi=\Pi_0,\Pi_{[L]}$, this rotation $R_\Pi(\theta)$ can be efficiently implemented by $\order{\log L}$ primitive gates only with $\order{1}$ additional ancilla qubits \footnote{$R_{\Pi_0} (\theta)$ is implemented by $\order{\log L}$ Toffoli gates, $\order{\log L}$ additional qubits, and one controlled-phase gate. The implementation of $R_{\Pi_1} (\theta)$ exploits the quantum comparator, which transforms $\ket{x}\ket{0} \to \ket{x}\ket{\mr{bool}(x>m)}$ for a $b$-bit variable $x$ and a constant $m$.}.

Based on the quantum signal processing (QSP) \cite{Low2017-qsp}, tuning the parameter set $\vec{\theta}_q \in \bbR^q$ allows us to realize
\begin{eqnarray}
    && \Pi_{[L]} \mr{QSVT}[\vec{\theta}_q] \ket{0}_b \ket{0}_f \ket{\psi} = \nonumber \\
    && \quad \sum_n c_n \frac{f_q(1/2)}{\sqrt{2L}}\sum_{l \in [L]}\ket{(\epsilon_n - l\omega)_b}_b \ket{\Phi_n^l} + \order{q\delta}, \nonumber \\
    && \label{Eq_QPE:Floquet_Sambe_QSVT_QAA}
\end{eqnarray}
for a certain class of odd degree-$q$ polynomials $f_q(x)$. 
The amplitude of the preferable term, i.e. the superposition of $\ket{(\epsilon_n - l\omega)_b}_b \ket{\Phi_n^l}$ for $l \in [L]$, can be amplified by choosing a polynomial $f_q(x)$ such that $|f_q (1/2)|=1$.
Such a choice is well known by the Grover's search algorithm, which yields the negative Chebyshev polynomial $f_3(x)=-4x^3+3x$ with $\theta_1 = 0$, $\theta_2 = -\pi/2$, and $\theta_3 = -\pi/2$.
As a result of the QAA based on QSVT, the initial state $\ket{0}_b \ket{0}_f \ket{\psi}$ is transformed into
\begin{equation}
    \sum_n c_n \frac{1}{\sqrt{2L}}\sum_{l \in [L]}\ket{(\epsilon_n - l\omega)_b}_b \ket{\Phi_n^l} + \order{\delta}.
\end{equation}
While the QAA triples the query complexity and the error in Eq. (\ref{Eq_QPE:QPE_H_F}), it preserves their scaling in all the parameters.
Therefore, the state of Eq. (\ref{Eq:Floquet_QPE_Sambe_step2}) obtained by the ideal operation in the Sambe space [See Step 2 in Section \ref{Sec:Floquet_QPE}] can also be generated by the truncated Sambe space with essentially the same efficiency.

\textbf{Algorithm and Cost.---} We summarize the algorithm for executing the Floquet QPE under rounding promise and its cost.
Based on the rough sketch in Sec. \ref{Subsec:Floquet_QPE_Sambe_outline}, the algorithm working with the truncated Sambe space is summarized as follows.
\begin{enumerate}
    \item Prepare initial state on truncated Sambe space \\ 
    With a given initial state $\ket{\psi} = \sum_n c_n \ket{\phi_n(0)} \in \mcl{H}$, we prepare a $\lceil \log_2 (2pL)\rceil $-qubit ancilla state $\ket{0}_f$ setting the integer $p$ by $p \geq 7$ and the cutoff $L$ by
    \begin{equation}\label{Eq_QPE:Floquet_Sambe_cutoff}
        L \in \Theta \left( \alpha T + N + \log (1/\nu \varepsilon  \delta)\right).
    \end{equation}
    Then, $U_\mr{uni}^{4L}$ [Eq. (\ref{Eq_QPE:initial_st_truncated})] is applied to the ancilla $f$.
    
    \item Run the standard QPE \\
     Each eigenvalue of the truncated Floquet Hamiltonian $H_\mr{F}^{pL}/(2^{b_\mr{F}^{pL}} \omega)$ (or more precisely, $H_\mr{F,pbc}^{pL}/(2^{b_\mr{F}^{pL}} \omega)$ as described in Sec. \ref{Subsec:Modified_H_F}) is extracted in $b=b^\prime+b_\mr{F}^{pL}$ bits.
    The parameters $\nu, \varepsilon, \delta$ in the QPE are determined by
    \begin{eqnarray}
        && \nu \leftarrow \nu/2, \quad \delta \leftarrow \delta, \label{Eq_QPE:Floquet_Sambe_nu_delta}\\
        && \varepsilon \leftarrow \frac{\varepsilon}{2^{b_\mr{F}^{pL}}} \in \Theta \left( \frac{\varepsilon}{\alpha T + \log (1/\nu \varepsilon \delta)}\right). \label{Eq_QPE:Floquet_Sambe_epsilon}
    \end{eqnarray}
   
    \item Execute the QAA \\
    By $3$-times repetition of Steps 1 and 2 or their inverses [Eq. (\ref{Eq_QPE:Floquet_Sambe_QSVT})], the state proportional to $\sum_n c_n \sum_{l \in [L]} \ket{(\epsilon_n-l\omega)_b}_b \ket{\Phi_n^l} + \order{\delta}$ is obtained.
    
    \item Perform quantum arithmetic \\
    Quantum division and substitution are used to convert $\ket{\Phi_n^l} \to \ket{\Phi_n}$, resulting in the output $\sum_n c_n \ket{(\epsilon_n)_b}_b \ket{\Phi_n} + \order{\delta}$.
\end{enumerate}

The choice of the parameters is explained.
The integer $p$ is introduced for the Sambe space $\mcl{H}^{pL}$ to cover the initial state $\ket{\Psi_0^L}$.
We require $p \geq 7$ so that the rounding promise of $H_\mr{F}^{pL}$ can hold within the range $\mr{BZ}_l$ for every $l \in [6L]$, as explained below Eq. (\ref{Eq_QPE:Floquet_Sambe_transf_RP}).
The proper choice $p=8$ requires additional $3$ qubits.
The cutoff $L$ is chosen to satisfy the following three requirements.
First, it must be ensured that each eigenvalue of $H_\mr{F}^{pL}$ approximates quasienergy $\epsilon_n$ within $\order{\varepsilon}$ based on Theorem \ref{Thm:Accuracy_quasienergy}, which requires $L \in \Omega (\alpha T + \log (1/\varepsilon))$.
Second, the rounding promise $\nu$ should be inherited to $H_\mr{F}^{pL}$ based on Proposition \ref{Prop:Inherited_rounding_promise}, requiring $L \in \Omega (\alpha T + \log (1/\nu \varepsilon))$.
The last requirement is that the error terms via QPE must be suppressed up to $\order{\delta}$, which requires Eq. (\ref{Eq_QPE:cutoff_for_state_error}).
While the query complexity $q_\mr{QPE}$ includes the cutoff $L$ itself by
\begin{eqnarray}
    q_\mr{QPE} &\in& \order{\left( \frac{1}{\nu 2^{-b_\mr{F}^{pL}} \varepsilon} + \frac{\log (1/\nu)}{\nu}\right)\log (1/\delta)} \nonumber \\
    &=& \order{\left( \frac{\alpha T + L}{\nu \varepsilon} + \frac{\log (1/\nu)}{\nu}\right)\log (1/\delta)}, \label{Eq_QPE:query_std_QPE_H_F}
\end{eqnarray}
based on Eqs. (\ref{Eq:query_standard_QPE}) and (\ref{Eq_QPE:cutoff_for_state_error}), the $L$-dependence of $\log q_\mr{QPE}$ does not affect the scaling of $L$.
Therefore, the choice of the cutoff $L$ by Eq. (\ref{Eq_QPE:Floquet_Sambe_cutoff}) is sufficient.
Substituting the cutoff $L$ into Eq. (\ref{Eq:bit_normalization}) immediately provides the parameters in the QPE by Eqs. (\ref{Eq_QPE:Floquet_Sambe_nu_delta}) and (\ref{Eq_QPE:Floquet_Sambe_epsilon}).
Based on these parameters, we obtain the cost of the algorithm as follows.

\begin{theorem}\label{Thm:algorithm_Sambe}
\textbf{(Algorithm for $(\epsilon_n,\ket{\Phi_n})$)}

Assume the existence of rounding promise $\nu$.
The quantum algorithm of Eq. (\ref{Eq_QPE:Floquet_Sambe_transf_RP}), which returns pairs of quasienergy and a Floquet eigenstate $(\epsilon_n,\ket{\Phi_n})$, can be executed with the following computational resource when we demand the guaranteed quasienergy error $\varepsilon$ and the guaranteed state error $\delta$;
\begin{itemize}
    \item Query complexity in $C[O_{H_m}]$ or its inverse
    \begin{equation}\label{Eq_QPE:query_Floquet_Sambe}
        \qquad \order{\frac{\alpha T + N + \log (1/\nu \varepsilon \delta)}{\varepsilon \nu} \log (1/\delta)}. 
    \end{equation}

    \item Number of ancilla qubits
    \begin{equation}\label{Eq_QPE:ancilla_Floquet_Sambe}
        \qquad n_a + \order{\log (\alpha T/\varepsilon) + \log N + \log \log (1/\nu \delta)}.
    \end{equation}
    \item Number of other primitive gates per query 
    \begin{equation}\label{Eq_QPE:primitive_Floquet_Sambe}
        \qquad \order{n_a + \log (\alpha T/\varepsilon) + \log N + \log \log (1/\nu \delta)}.
    \end{equation}
\end{itemize}
\end{theorem}

\textit{Proof of Theorem \ref{Thm:algorithm_Sambe}.---} Queries to $C[O_{H_m}]$ or its inverse are employed for the standard QPE via the truncated Floquet Hamiltonian $H_\mr{F,pbc}^{pL}$.
The query complexity in the controlled block-encoding of $H_\mr{F,pbc}^{pL}$ is given by Eq. (\ref{Eq_QPE:query_std_QPE_H_F}). 
Since this controlled block-encoding can be realized by $2M+1 \in \order{1}$ queries to $C[O_{H_m}]$ or its inverse as Proposition \ref{Prop:block_encode_H_F_pbc}, we obtain the scaling Eq. (\ref{Eq_QPE:query_Floquet_Sambe}).
In the number of ancilla qubits, $n_a$ denotes the one required for the block-encoding.
We need $b=b^\prime + b_\mr{F}^{pL}$ qubits for the register and $\lceil \log_2 (2pL) \rceil$ qubits for the Fourier index $l$.
The number of the other ancilla qubits (e.g. those for quantum arithmetic) is negligible.
In total, the required number of ancilla qubits is represented by Eq. (\ref{Eq_QPE:ancilla_Floquet_Sambe}).
Quantum gates other than $C[O_{H_m}]$ or its inverse are composed of $U_\mr{ini}^{4L}$ in Step 1, ancilla unitary gates for the QPE (or the QSVT) in Step 2, ancilla unitary gates for the QAA in Step 3, and those for quantum arithmetic in Step 4.
Among them, only the first and second ones are repeated $\order{q_\mr{QPE}}$ times, while the others are repeated $\order{1}$ times.
The number of primitive gates per query to $C[O_{H_m}]$ or its inverse is determined by the first two, which amounts to Eq. (\ref{Eq_QPE:primitive_Floquet_Sambe}). $\quad \square$

Let us compare the costs with those of the standard QPE for time-independent systems, and the Floquet QPE for $(\epsilon_n, \ket{\phi_n(t)})$ in Section \ref{Sec:Floquet_QPE_physical}.
We first see the relation to the Floquet QPE for $(\epsilon_n, \ket{\phi_n(t)})$, which is summarized in Table \ref{Table:comparison_algorithms}.
In terms of query complexity, they share common scaling $\nu^{-1} \log (1/\nu)$ in the rounding promise $\nu$, and $\varepsilon^{-1} \log (1/\varepsilon)$ in the quasienergy error $\varepsilon$.
Although the scaling in $\delta$ appears to be different, the denominator $\delta$ for the case of $(\epsilon_n,\ket{\phi_n(t)})$ is not important as discussed in Section \ref{Subsec:Floquet_phys_cost}.
The actual dependence on $\delta$ is essentially the same.
In the numerator, the difference as large as the system size $N$ appears while the factor of $\alpha T$ is common.
From the definition of $\alpha$ by Eq. (\ref{Eq:Periodic_Hamiltonian}), the factor $\alpha T$ for generic quantum many-body systems is polynomial in $N$ or at least linear in $N$.
The difference of $N$ is negligible or absorbed into the constant coefficient.
We can see similar correspondence also in the number of ancilla qubits and the number of other primitive gates.
Therefore, we can conclude that the cost of computing pairs of $(\epsilon_n,\ket{\phi_n(t)})$ and that for $(\epsilon_n,\ket{\Phi_n})$ is essentially the same.
This seems to fit intuitively since every Floquet eigenstate $\ket{\phi_n(t)} \in \mcl{H}$ has a one-to-one correspondence with a Floquet eigenstate $\ket{\Phi_n} \in \mcl{H}^\infty$.
However, the scaling for $(\epsilon_n,\ket{\phi_n(t)})$ comes from the Lieb-Robinson bound, Eq. (\ref{Eq:LR_bound_in_Sambe}), while the scaling for $(\epsilon_n,\ket{\Phi_n})$ originates from the decay of Floquet eigenstates from Theorem \ref{Thm:Tail_Eigenstates}.
They are respectively dynamical and static aspects of the Sambe space, and hence their equivalence in the computational cost is brought about by the coincident scaling of their different origins.

Finally, we discuss the relation to the cost of the standard QPE.
As in Section \ref{Subsec:Floquet_phys_cost}, the computation of pairs of $(\epsilon_n, \ket{\Phi_n})$ can be performed as efficiently as the standard QPE except for logarithmic corrections.
The factor of $\alpha T + N$ in Eq. (\ref{Eq_QPE:query_Floquet_Sambe}) comes from the lack of normalization in the Hamiltonian, and this also appears in unnormalized time-independent Hamiltonians.
This is one of the main results of this paper.

\subsection{QPE without rounding promise}\label{Subsec:Floquet_QPE_wo_RP}

The algorithm in the absence of rounding promise is organized similarly Section \ref{Subsec:FloquetQPE_with_RP}.
All the operations are essentially the same as those with rounding promise, but a difference appears in the output.
In this case, for the success of QAA discussed later, we additionally assume that quasienergy and a Floquet eigenstate slightly away from the boundaries of BZ are of interest.
Namely, the initial state $\ket{\psi}$ is assumed to be expanded by
\begin{eqnarray}
    \ket{\psi} &=& \sum_{n; \epsilon_n \in \mr{BZ}^{-\varepsilon}} c_n \ket{\phi_n(0)}, \label{Eq_QPE:Initial_st_wo_RP}\\
    \mr{BZ}^{-\varepsilon} &=& [-(1/2-\varepsilon)\omega, (1/2-\varepsilon) \omega). \label{Eq_QPE:BZ_wo_boundaries}
\end{eqnarray}
Note that the boundaries of the BZ can be freely moved by introducing a global phase.
Therefore, we can obtain any desired quasienergy and Floquet eigenstate even with this additional assumption, and it does not lose the generality of the discussion.

\textbf{Algorithm.---}
The initial state preparation is performed in the same way as Step 1 in Section \ref{Subsec:FloquetQPE_with_RP}.
The QPE under the truncated Floquet Hamiltonian follows Step 2 except that it employs the algorithm without rounding promise which returns Eqs. (\ref{Eq:Standard_QPE_wo_rounding1}) and (\ref{Eq:Standard_QPE_wo_rounding2}).
The $b$-qubit register then stores superpositions of estimated values of $\epsilon_n-l\omega$,
\begin{equation}
    \ket{\overline{(\epsilon_n-l\omega)_b}}_b = p_0^{nl} \ket{(\epsilon_n-l\omega)_{b0}}_b + p_1^{nl} \ket{(\epsilon_n-l\omega)_{b1}}_b
\end{equation}
with some weights $p_0^{nl}$ and $p_1^{nl}$ such that $|p_0^{nl}|^2+|p_1^ {nl}|^2 = 1$.
We set the cutoff $L$ and the accuracy of the QPE such that the two $b$-bit values $(\epsilon_n-l\omega)_{b0}$ and $(\epsilon_n-l\omega)_{b1}$ approximate the exact quasienergy $\epsilon_n - l \omega$ with an error at most $\varepsilon \omega$.

Next, we run QAA following Step 3 in Section \ref{Subsec:FloquetQPE_with_RP}.
For $n$ such that $\epsilon_n \in \mr{BZ}^{-\varepsilon}$, the $b$-bit values estimated by QPE satisfy
\begin{equation}
    (\epsilon_n-l\omega)_{b0}, (\epsilon_n-l\omega)_{b1} \in [(-l-1/2)\omega, (-l+1/2)\omega),
\end{equation}
for each $l \in [L]$.
As a result, the projection of the estimated values onto $\mr{BZ}_{[L]}$ [Eq. (\ref{Eq:Proj_L_range})] deletes the terms having $\ket{\Phi_n^l}$ for $l \notin [L]$.
Namely, the projection $\Pi_{[L]}$ generates the truncated counterpart of Eq. (\ref{Eq_QPE:Initial_decomp_infinite}),
\begin{eqnarray}
    && \Pi_{[L]} U_\mr{QPE}^L U_\mr{ini}^{4L}\ket{0}_b \ket{0}_f \ket{\psi} \nonumber \\
    && \quad = \frac12 \sum_n \frac{c_n}{\sqrt{2L}} \sum_{l \in [L]} \ket{\overline{(\epsilon_n-l\omega)_b}}_b \ket{\Phi_n^l} + \order{\delta}, \label{Eq_QPE:block_encode_ext_wo_RP}
\end{eqnarray}
in a similar way to Eq. (\ref{Eq_QPE:Floquet_Sambe_block_encode_ext}) for the case with rounding promise.
The QSVT by the negative Chebyshev polynomial $f_3(x)$ remains valid.
The state after QAA as in Eq. (\ref{Eq_QPE:Floquet_Sambe_QSVT_QAA}) becomes
\begin{eqnarray}
    && \mr{QSVT}[\vec{\theta}_q] \ket{0}_b \ket{0}_f \ket{\psi} = \nonumber \\
    && \quad \sum_{n; \epsilon_n \in \mr{BZ}^{-\varepsilon}} c_n \frac{1}{\sqrt{2L}}\sum_{l \in [L]}\ket{\overline{(\epsilon_n - l\omega)_b}}_b \ket{\Phi_n^l} + \order{\delta}. \nonumber \\
    && 
\end{eqnarray}
for the initial state $\ket{\psi}$ given by Eq. (\ref{Eq_QPE:Initial_st_wo_RP}).

The quantum arithmetic in Step 4 of Section \ref{Subsec:FloquetQPE_with_RP} is performed in the same way.
Since both of the estimated values $(\epsilon_n - l \omega)_{b0}$ and $(\epsilon_n - l \omega)_{b1}$ belong to $\mr{BZ}_l$, the quantum division by $\omega$ returns the same quotient $l$.
The transformation is written as
\begin{equation}
    \ket{\overline{(\epsilon_n - l\omega)_b}}_b \to (p_0^{nl} \ket{(\epsilon_n)_{b0}}_b + p_1^{nl} \ket{(\epsilon_n)_{b1}}_b ) \ket{l},
\end{equation}
where $(\epsilon_n)_{b0}$ and $(\epsilon_n)_{b1}$ are $b$-bit estimates that approximate the exact $\epsilon_n$ within an error $\omega \varepsilon$.
The quantum substitution by the quotient results in the following output for every $n$;
\begin{eqnarray}
    && \frac{1}{\sqrt{2L}}\sum_{l \in [L]}\ket{\overline{(\epsilon_n - l\omega)_b}}_b \ket{\Phi_n^l} \nonumber \\
    && \quad \to \frac{1}{\sqrt{2L}}\sum_{l \in [L]} (p_0^{nl} \ket{(\epsilon_n)_{b0}}_b + p_1^{nl} \ket{(\epsilon_n)_{b1}}_b ) \ket{l} \ket{\Phi_n}. \nonumber \\
    &&
\end{eqnarray}
The state for the quotient $l$ can be summarized as the garbage states defined by
\begin{equation}
    p_i^n = \sqrt{\frac{\sum_{l \in [L]} |p_i^{nl}|^2}{2L}}, \quad \ket{g_i^n} = \frac{\sum_{l \in [L]} p_i^{nl} \ket{l}}{\sqrt{\sum_{l \in [L]} |p_i^{nl}|^2}},
\end{equation}
for each of $i=0,1$.
The output after the quantum substitution is given by
\begin{eqnarray}
    \sum_n c_n \left( \sum_{i=0,1} p_i^n \ket{g_i^n} \ket{(\epsilon_n)_{bi}}_b \right)\ket{\Phi_n} + \order{\delta}.
\end{eqnarray}
When we measure the $b$-qubit register, we obtain $(\epsilon_n)_{b0}$ or $(\epsilon_n)_{b1}$ with total probability $p_n = |c_n|^2$.
The projected state is the Floquet eigenstate $\ket{\Phi_n}$ (exactly the eigenstate of the truncated Floquet Hamiltonian) with the garbage state $\ket{g_0^n}$ or $\ket{g_1^n}$.

We also mention about the case where the initial state $\ket{\psi}$ contains a Floquet eigenstate with quasienergy out of interest.
Namely, we assume that $c_n \neq 0$ for some $n$ such that $\epsilon_n \in \mr{BZ} \backslash \mr{BZ}^{-\varepsilon}$, instead of Eq. (\ref{Eq_QPE:Initial_st_wo_RP}).
For such $n$, the $b$-bit estimate of $\epsilon_n - l \omega \in \mr{BZ}_l$ may be outside of $\mr{BZ}_l$ as shown in Fig. \ref{Fig_spectrum} (b).
At the same time, those for some quasienergy in $\mr{BZ}_{l \pm 1}$ may also belong to $\mr{BZ}^l$.
Such deviations leads to the violation of the fixed weight $1/2$ under the projection $\Pi_{[L]}$ like Eq. (\ref{Eq_QPE:block_encode_ext_wo_RP}).
The weight $s_n$ is located between $\sqrt{(2L-1)/(8L)}$ and $\sqrt{(2L+1)/8L}$ depending on the number of the $b$-bit estimates $(\epsilon_n-l\omega)_{bi}$ belonging to $\mr{BZ}_{[L]}$ for $l \in [L]$.
The QAA arranged to satisfy $f_3(1/2)=1$ becomes incomplete then, leading to
\begin{eqnarray}
    && \mr{QSVT}[\vec{\theta}_q] \ket{0}_b \ket{0}_f \sum_{n; \epsilon_n \notin \mr{BZ}^{-\varepsilon}} c_n \ket{\phi_n (0)} = \nonumber \\
    && \quad \sum_{n; \epsilon_n \notin \mr{BZ}^{-\varepsilon}} c_n (f_3(s_n) \ket{\Xi_n} + \sqrt{1-f_3(s_n)^2} \ket{\Xi_n^\perp})  + \order{\delta}. \nonumber \\
    &&  \label{Eq_QPE:QSVT_QAA_wo_RP}
\end{eqnarray}
with some states $\ket{\Xi_n}, \ket{\Xi_n^\perp}$ satisfying $\Pi_{[L]} \ket{\Xi_n} = \ket{\Xi_n}$ and $\Pi_{[L]} \ket{\Xi_n^\perp} = 0$.
However, the $b$-qubit register in the above state stores $b$-bit estimates of $\epsilon_n - l \omega$ for $\epsilon_n \notin \mr{BZ}^{-\varepsilon}$ or eigenvalues out of $\mr{BZ}_{[L]}$ coming from the high-energy term $\ket{\Psi_{n,\perp}^L}$ in $\ket{\Xi_n^\perp}$ (See Proposition \ref{Prop:FloquetQPE_Initial_st}).
Modifying the quantum substitution in the last step by 
\begin{equation}
\sum_{l \in [L]} \ket{l}\bra{l} \otimes \mr{Add}_l^\dagger + \sum_{l \notin [L]} \ket{l}\bra{l} \otimes I,
\end{equation}
the state of Eq. (\ref{Eq_QPE:QSVT_QAA_wo_RP}) does not affect $b$-bit values in $\mr{BZ}^{-\varepsilon}$ stored in the register.
When the measurement outcome in $\mr{BZ}^{-\varepsilon}$ is post-selected, preferable quasienergy $\epsilon_n \in \mr{BZ}^{-2 \varepsilon}$ can be extracted with probability $p_n = |c_n|^2$.
The factor $2$ in $\mr{BZ}^{-2 \varepsilon}$ comes from discarding the results in $\mr{BZ} \backslash \mr{BZ}^{-\varepsilon}$.
Thus, even in the presence of undesirable components in the initial state, the algorithm works well for target quasienergy and Floquet eigenstates.

\textbf{Cost.---}
We derive the cost in the absence of rounding promise.
The standard QPE without rounding promise is used for the truncated Floquet Hamiltonian, which yields the query complexity,
\begin{equation}
    q_\mr{QPE} \in \Theta \left( \frac{\alpha T + L}{\varepsilon} \log (1/\delta) \right),
\end{equation}
where the parameters $\nu,\varepsilon,\delta$ in Eq. (\ref{Eq:query_standard_QPE}) are replaced by $\order{1}, 2^{-b_\mr{F}^{pL}}\varepsilon, \delta$, respectively.
The cutoff $L$ is required to satisfy $\Omega (\alpha T + \log (1/\varepsilon))$ and Eq. (\ref{Eq_QPE:cutoff_for_state_error}), respectively, so that the quasienergy error by the truncation is less than $\omega \varepsilon$ and the state error is less than $\delta$.
The cutoff $L$ that satisfies these conditions can be
\begin{equation}
    L \in \Theta \left( \alpha T + N + \log (1/\varepsilon \delta) \right),
\end{equation}
and we get the following cost.
\begin{itemize}
    \item Query complexity in $C[O_{H_m}]$ or its inverse
    \begin{equation}
        \order{\frac{\alpha T + N + \log (1/\varepsilon \delta)}{\varepsilon} \log (1/\delta)}.
    \end{equation}

    \item Number of ancilla qubits
    \begin{equation}
        \qquad n_a + \order{\log (\alpha T/\varepsilon) + \log N + \log \log (1/\delta)}.
    \end{equation}
    \item Number of other primitive gates per query
    \begin{equation}
          \qquad \order{n_a + \log (\alpha T / \varepsilon) + \log N + \log \log (1/\delta)}.
    \end{equation}
\end{itemize}
The above results are summarized by Table \ref{Table:comparison_algorithms} or Theorem \ref{Thm:algorithm_Sambe}, setting $\nu \in \order{1}$ in the case with rounding promise.
Importantly, the computational cost is essentially the same as the standard QPE for time-independent Hamiltonians, even in the case without rounding promise.

%===================================================
% Section
%===================================================
\section{Floquet eigenstate preparation}\label{Sec:Floquet_preparation}

Consider a time-independent Hamiltonian $H = \sum_n E_n \ket{\phi_n}\bra{\phi_n}$.
Eigenstate preparation is a quantum algorithm to realize a single preferable eigenstate $\ket{\phi_n}$ from a given initial state $\ket{\psi}=\sum_n c_n \ket{\phi_n}$ under the assumption that we have a promised gap $\Delta$ around $E_n$, a bound $\gamma$ on the overlap $c_n$ such that $|c_n| \geq \gamma$, and the value of $E_n$.
The standard QPE algorithm can efficiently execute eigenstate preparation to obtain an accurate eigenstate $\ket{\phi_n}+\order{\delta}$ ($\delta \in (0,1)$) \cite{Poulin_2009_ground_st}, and the optimal query complexity in the controlled block-encoding and the state preparation unitary for $\ket{\psi}$ has recently been achieved by the QSVT \cite{Ge2019-cr-ground_st,Lin2020-ground_state} (See Table \ref{Table:eigenstate_preparation} \footnote{We note that the factor $\log (1/\delta)$ is added to the query complexity in Ref. \cite{Lin2020-ground_state} for fair comparison. While Ref. \cite{Lin2020-ground_state} executes the QAA so that they can succeed in the state preparation with some constant property, we require the success probability to be greater than $1-\order{\delta}$. The cost of achieving it by QAA appears as an additional factor of $\log (1/\delta)$.}).
If the initial state $\ket{\psi}$ has a large overlap with the target eigenstate $\ket{\phi_n}$ such that $|c_n| \geq 1/\poly{N}$, eigenstate preparation can efficiently implement the eigenstate $\ket{\phi_n}$ on quantum computers (Note that this is generally difficult due to its QMA-hardness \cite{Kitaev2002-kz}).

Similarly, the Floquet QPE algorithm for time-periodic Hamiltonians allows us to perform nearly optimal eigenstate preparation for a target Floquet eigenstate $\ket{\phi_n(t)}$ or $\ket{\Phi_n}$.
Here we assume that we know the exact value of certain preferable quasienergy $\epsilon_n$ and that $H(t)$ has a quasienergy gap $\Delta$ around $\epsilon_n$ by
\begin{equation}
    \left| \left( \frac{\epsilon_{n^\prime}-\epsilon_n}{\omega} \right) \text{ mod. 1 }\right| \leq \Delta , \quad ^\forall n^\prime \neq n.
\end{equation}
Then, the Floquet eigenstate preparation with allowable state error $\delta \in (0,1)$ indicates a quantum algorithm performing the transformation,
\begin{equation}
    \sum_{n^\prime} c_{n^\prime} \ket{\phi_{n^\prime} (0)} \to \ket{\phi_n(t)}+\order{\delta} \text{ or } \ket{\Phi_n}+\order{\delta},
\end{equation}
where $\ket{\psi} = \sum_{n^\prime} c_{n^\prime} \ket{\phi_{n^\prime} (0)}$ is an initial state having overlap $c_n \geq \gamma$ with a single preferable Floquet eigenstate $\ket{\phi_n(0)}$ with a gap $\Delta$.
Without loss of of generality, we assume that the overlap $c_n$ is real and positive.
We require the success probability to be larger than $1- \order{\delta}$. 
Its computational cost is measured by the query complexity in the controlled block-encoding $C[O_{H_m}]$, the state preparation unitary $U_\psi$ such that $U_\psi \ket{0} = \ket{\psi}$, and their inverses.

\subsection{Eigenstate preparation of $\ket{\phi_n(t)}$}\label{Subsec:Prep_phys}

We organize the Floquet eigenstate preparation algorithm whose output is a preferable Floquet eigenstate in the physical space, $\ket{\phi_n(t)}+\order{\delta}$.
The key ingredient for this algorithm is the Floquet QPE for $(\epsilon_n, \ket{\phi_n(t)})$.

First, we execute Floquet QPE without rounding promise in Section \ref{Sec:Floquet_QPE_physical} from Step 1 to Step 3.
Setting the quasienergy error $\varepsilon \leftarrow \Delta/2$ and the state error $\delta \leftarrow \gamma \delta$, we obtain a unitary operation $U_\mr{FQPE}^\mr{phys}$ such that
\begin{eqnarray}
    && U_\mr{FQPE}^\mr{phys} \ket{0}_b \ket{\psi} = \nonumber \\
    && \qquad \sum_{n^\prime} c_{n^\prime} \ket{\overline{(\epsilon_{n^\prime})_b}}_b \ket{\phi_{n^\prime}(0)} + \order{\gamma \delta},
\end{eqnarray}
where we omit the ancilla $f$ since it remains $\ket{0}_f$.
The query complexity for $U_\mr{FQPE}^\mr{phys}$ in $C[O_{H_m}]$ or its inverse is given by
\begin{equation}\label{Eq_PREP:query_FQPE_phys}
    \order{\frac{\alpha T + \log (1/\Delta)}{\Delta} \log (1/\gamma\delta)}.
\end{equation}
When the estimation error is guaranteed to be less than $\Delta \omega/2$ under the promised gap $\Delta$, then an estimated quasienergy $\overline{(\epsilon_{n^\prime})_b}$ belongs to $(\epsilon_n-\Delta\omega/2, \epsilon_n+\Delta\omega/2)$ if and only if $n^\prime = n$.
In other words, when we apply the projection,
\begin{equation}\label{Eq_Prep:Proj_filtering}
    \Pi_{n,\Delta} = \sum_{x; |x-\epsilon_n| < \Delta \omega/2} \ket{x}\bra{x}_b \otimes I,
\end{equation}
to the output of the algorithm
\begin{equation}
    \sum_n c_n \ket{\overline{(\epsilon_n)_b}}_b \ket{\phi_n(0)} + \order{\gamma \delta},
\end{equation}
only the preferable component having $\ket{\phi_n(0)}$ survives.
As a result, we get the following relation,
\begin{eqnarray}
    && \Pi_{n,\Delta} U_\mr{FQPE}^\mr{phys} \Pi_\psi = \nonumber \\
    && \quad c_n \ket{\overline{(\epsilon_n)_b}}_b \ket{\phi_n(0)} \bra{0}_b \bra{\psi} + \order{\gamma\delta}, \label{Eq_PREP:block_encode_phys}
\end{eqnarray}
where the projection $\Pi_\psi$ is defined by $\Pi_\psi = \ket{0}\bra{0}_b \otimes \ket{\psi}\bra{\psi}$.

Since Eq. (\ref{Eq_PREP:block_encode_phys}) forms a block-encoding with a singular value $c_n$ like Eq. (\ref{Eq_QPE:Floquet_Sambe_block_encode_ext}), we can perform QAA based on QSVT.
Similar to Eq. (\ref{Eq_QPE:Floquet_Sambe_QSVT_QAA}), we implement a odd polynomial function $f_q (x)$ such that $f_q (x) = 1 - \order{\delta}$ for any $x \in [\gamma,1]$ with using the block-encoding $U_\mr{FQPE}^\mr{phys}$ and the parametrized unitaries $R_{\Pi_{n,\Delta}}(\theta)$ and $R_{\Pi_\psi}(\theta)$.
Then, the transformation of $\ket{0}_b \ket{\psi} \to \ket{\overline{(\epsilon_n)_b}}_b \ket{\phi_n(0)}+ \order{\delta}$, with yielding $\order{q}=\order{\gamma^{-1}\log (1/\delta)}$ queries to the unitaries $U_\mr{FQPE}^\mr{phys}$, $R_{\Pi_{n,\Delta}}(\theta)$, and $R_{\Pi_\psi}(\theta)$ respectively.
Finally, by applying the time-evolution operator $U(t;0)+\order{\delta}$ based on the Sambe space formalism as Eq. (\ref{Eq:Hamiltonian_simulation_Floquet}) and neglecting the global phase, the Floquet eigenstate preparation of $\ket{\psi} \to \ket{\phi_n(t)}+\order{\delta}$ is completed.

The cost of the Floquet eigenstate preparation is evaluated as follows.
The controlled block-encoding $C[O_{H_m}]$ or its inverse are used in $U_\mr{FQPE}^\mr{phys}$ by Eq. (\ref{Eq_PREP:query_FQPE_phys}).
Since the QAA requires $\order{q}=\order{\gamma^{-1}\log (1/\delta)}$ queries to $U_\mr{FQPE}^\mr{phys}$, the query complexity in them is $\order{q}$ times as large as Eq. (\ref{Eq_PREP:query_FQPE_phys}).
While the time-evolution $U(t;0)+\order{\delta}$ with $t \in [0,T)$ also requires $\order{\alpha T + \log (1/\delta)}$ queries to $C[O_{H_m}]$, $C[O_{H_m}]^\dagger$, this scaling is negligible compared to the above process of the Floquet QPE and the QAA.
The parametrized unitary $R_{\Pi_{n,\Delta}}(\theta)$ defined by Eq. (\ref{Eq:Phased_iterate}) can be implemented by basic quantum arithmetic like $R_{\Pi_{[L]}}$ in Section \ref{Subsec:FloquetQPE_with_RP}, requiring only $\order{b} \subset \order{\log (\alpha T) + \log \log (1/\Delta)}$ primitive gates.
Since the unitary $R_{\Pi_\psi}$ is represented by the state preparation unitary by
\begin{equation}
    R_{\Pi_\psi}(\theta) = (I_b \otimes U_\psi) e^{i \phi (2 \ket{0}_b\bra{0}_b \otimes \ket{0}\bra{0} - I_b \otimes I)} (I_b \otimes U_\psi^\dagger),
\end{equation}
it requires $\order{\log N + b}$ primitive gates and one query respectively for $U_\psi$ and $U_\psi^\dagger$.
Thus, the query complexity in $U_\psi$ or its inverse is $\order{q}=\order{\gamma^{-1}\log (1/\delta)}$.
We summarize these results in the second row of Table \ref{Table:eigenstate_preparation} with a comparison of the optimal eigenstate preparation for time-independent Hamiltonians \cite{Lin2020-ground_state}.

For time-independent Hamiltonians, the optimal query complexity in $C[O_H]$ is proved to be $\Omega (\Delta^{-1})$ in terms of the gap $\Delta$ and $\Omega (\gamma^{-1})$ in terms of the overlap $\gamma$ by associating the problem with an unstructured search \cite{Lin2020-ground_state}.
The optimal scaling for the state preparation $U_\psi$ is known to be $\Omega (\gamma^{-1})$ then.
This is also true for time-periodic Hamiltonians since they include time-independent ones.
The Floquet eigenstate preparation algorithm yields the scaling $\order{\Delta^{-1} \log \Delta^{-1}}$ in the gap $\Delta$ and the scaling $\order{\gamma^{-1} \log \gamma^{-1}}$ in the overlap $\gamma$, and thus it achieves nearly optimal scaling for time-independent systems.
The difference of the polynomial factor $\alpha T \in \poly{N}$ comes from renormalization.
If a time-independent Hamiltonian is not normalized, we have to renormalize the gap $\Delta$, and the query complexity for time-independent cases has a similar factor.
This difference is not significant, as is the comparison with QPE.
The query complexity in the state preparation unitary coincides with that of the time-independent cases, and is therefore optimal in $\gamma$.

\begin{table*}
    \centering
    \begin{tabular}{|c|c|c|c|c|}
        Target & Initial state $\ket{\psi}$ & Output & \begin{tabular}{c} Queries to Hamiltonian \\ ($C[O_H]$ or $C[O_{H_m}]$) \end{tabular} & \begin{tabular}{c} Queries to state preparation \\ (Unitary $U_\psi$) \end{tabular} \\ \hline \hline
       \begin{tabular}{c} Time-indep., $H$ \\
      (Ref. \cite{Lin2020-ground_state}) \end{tabular} & $\sum_{n^\prime} c_{n^\prime} \ket{\phi_{n^\prime}}$, $|c_n| \geq \gamma$ & $\ket{\phi_n} + \order{\delta}$ & $\frac{1}{\gamma \Delta}  \log \left(\frac{1}{ \delta}\right) \log \left(\frac{1}{\gamma \delta}\right)$ &
      $\frac{1}{\gamma}  \log \left(\frac{1}{ \delta}\right)$
       \\ \hline
       \begin{tabular}{c} Floquet, $H(t)$ \\ (Section \ref{Subsec:Prep_phys}) \end{tabular} & $\sum_{n^\prime} c_{n^\prime} \ket{\phi_{n^\prime}(0)}$, $|c_n| \geq \gamma$ & $\ket{\phi_n(t)}+\order{\delta}$ & $\frac{\alpha T + \log (1/\Delta)}{\gamma \Delta} \log \left( \frac{1}{\delta} \right) \log \left( \frac{1}{\gamma \delta} \right)$ & $\frac{1}{\gamma} \log \left(\frac{1}{ \delta}\right)$ 
       \\ \hline
       \begin{tabular}{c} Floquet, $H(t)$ \\ (Section \ref{Subsec:Prep_Sambe}) \end{tabular} & $\sum_{n^\prime} c_{n^\prime} \ket{\phi_{n^\prime}(0)}$, $|c_n| \geq \gamma$ & $\ket{\Phi_n}+\order{\delta}$ & $\frac{\alpha T + N+ \log (1/\Delta \gamma \delta)}{\gamma \Delta} \log \left( \frac{1}{\delta} \right) \log \left( \frac{1}{\gamma \delta} \right)$ & $\frac{1}{\gamma} \log \left(\frac{1}{ \delta}\right)$
        \\ \hline
    \end{tabular}
    \caption{Cost of eigenstate preparation for time-independent and time-periodic Hamiltonians under a promised gap $\Delta$.  A preferable Floquet eigenstate, either $\ket{\phi_n(t)} \in \mcl{H}$ or $\ket{\Phi_n} \in \mcl{H}^\infty$ can be prepared in nearly optimal query complexity in the parameters. The difference from time-independent cases is at most logarithmic corrections in all the parameters $\gamma,\Delta,\delta$. }
    \label{Table:eigenstate_preparation}
\end{table*}

\subsection{Eigenstate preparation of $\ket{\Phi_n}$}\label{Subsec:Prep_Sambe}

The preparation of a preferable Floquet eigenstate $\ket{\Phi_n}$ living in the Sambe space is also formulated by the Floquet QPE.
Let us assume without loss of generality that the preferable quasienergy $\epsilon_n$ is located in $[(-1/2+\Delta)\omega, (1/2-\Delta)\omega)$ (If not, we move the origin of the BZ by multiplying a global phase).
Then, we can use the Floquet QPE without rounding promise in Section \ref{Subsec:Floquet_QPE_wo_RP}, where the parameters are chosen by $\varepsilon \leftarrow \Delta/2$ and $\delta \leftarrow \gamma \delta$.
We get a unitary operation $U_\mr{FQPE}^\mr{Sambe}$ such that
\begin{eqnarray}
    && U_\mr{FQPE}^\mr{Sambe} \ket{0}_g \ket{0}_b \ket{0}_f \ket{\psi} = \nonumber \\
    && \qquad \sum_{n^\prime} c_{n^\prime} \left( \sum_{i=0,1} p_i^{n^\prime} \ket{g_i^{n^\prime}}_g \ket{(\epsilon_{n^\prime})_{bi}}_b \right) \ket{\Phi_{n^\prime}} + \order{\gamma \delta}. \nonumber \\
    &&
\end{eqnarray}
The promised gap $\Delta$ and the accuracy of the Floquet QPE ensure that the projection $\Pi_{n,\Delta} \otimes I_g \otimes I_f$ by Eq. (\ref{Eq_Prep:Proj_filtering}) keeps only the preferable component with  $\ket{\Phi_n}$ in the above state, and it allows us to form the block-encoding $U_\mr{FQPE}^\mr{Sambe}$ like Eq. (\ref{Eq_PREP:block_encode_phys}).
Again, we can run QAA based on QSVT, which does the transformation,
\begin{equation}
    \ket{\psi} \to \left( \sum_{i=0,1} p_i^{n^\prime} \ket{g_i^{n}}_g \ket{(\epsilon_{n})_{bi}}_b \right) \otimes \ket{\Phi_{n}} + \order{\delta}.
\end{equation}
Discarding the ancilla systems other than $f$ completes the accurate preparation of the preferable Floquet eigenstate $\ket{\Phi_n}$ (or more presicely, the eigenstate of the truncated Floquet Hamiltonian).

The cost is evaluated in a similar way as in Section \ref{Subsec:Prep_phys}.
The query complexity in $C[O_{H_m}]$ or its inverse amounts to the product of that of the Floquet QPE,
\begin{equation}
    \order{\frac{\alpha T + N + \log (1/\Delta \gamma \delta)}{\Delta} \log (1/\gamma \delta)}, 
\end{equation}
and that of QAA, $\order{q}=\order{\gamma^{-1}\log (1/\delta)}$.
The query complexity in the state preparation unitary $U_\psi$ or its inverse, which are used for the counterpart of the parametrized unitary $R_{\Pi_\psi}(\theta)$, is $\order{q}=\order{\gamma^{-1}\log (1/\delta)}$.
These results are summarized in the last row of Table \ref{Table:eigenstate_preparation}.
Like the preparation of $\ket{\phi_n(t)}$ in Section \ref{Subsec:Prep_phys}, the preparation of $\ket{\Phi_n}$ can also be performed in nearly optimal query complexity.
The scaling in the gap $\Delta$ is $\order{\Delta^{-1} \log \Delta^{-1}}$ and that in the overlap $\gamma$ is $\order{\gamma^{-1} \log^2 \gamma^{-1}}$, both of which differ from the optimal algorithm for time-independent cases by at most logarithmic corrections.
%===================================================
% Section
%===================================================
\section{Conclusion and Discussion}\label{Sec:Discussion}

Throughout the paper, we have focused on the problem of computing quasienergy and Floquet eigenstates under time-periodic Hamiltonians, and have organized efficient quantum algorithms for it.
Our algorithms can output pairs of accurate quasienergy and a Floquet eigenstate in a coherent way, referred to as ``Floquet quantum phase estimation (Floquet QPE)", and also deterministically prepare a preferable gapped Floquet eigenstate based on the Floquet QPE.
Time-periodic systems have the difficulty of time-dependency or equivalently infinite-dimensionality of the Sambe space compared to time-independent cases.
Nevertheless, these quantum algorithms achieve nearly optimal query complexity whose difference from the optimal algorithms for time-independent cases is at most logarithmic in all the parameters.
Note that this efficiency comes from the interplay of nonequilibrium many-body physics and quantum algorithms; 
The guaranteed accuracy is derived from the Sambe space formalism in Floquet theory, and the Lieb-Robinson bound or the localization of Floquet eigenstates.
The computational resources increased by the Sambe space can be small in quantum algorithms according to the QPE and the QAA, which allows us to solve the time-dependent problems almost as efficiently as the time-independent problems.

The computation of energy eigenvalues and eigenstates has been a central problem in fundamental quantum many-body physics, and offers various applications in condensed matter physics and quantum chemistry.
Our quantum algorithms first achieve nearly optimal query complexity for its natural extension to time-periodic systems, which will provide a deep insight into the complexity of nonequilibrium systems like the sampling complexity of time-periodic systems \cite{Tangpanitanon2023-complexity-Floquet}.
We also expect that it will provide a powerful tool for exploring nonequilibrium phases of matter:
For instance, in Floquet time crystalline phases \cite{Khemani2016-dq,Else2016-mg,Khemani2019-pf}, every low-entangled state (e.g. product state) becomes a superposition of Floquet eigenstates which are vulnerable to noise and measurement (e.g. cat states). 
Since these Floquet eigenstates have equally large weight and a quasienergy gap $\omega/n$ ($n=2,3,\hdots$), our quantum algorithms will be useful for the identification of time crystals by confirming their properties.
Similarly, various nonequilibrium phenomena such as Floquet many-body localization \cite{Ponte_2015_MBL,Lazaridez_2015_MBL,Abanin2016-bd_MBL,Bordia2017-lq_MBL} and Floquet quantum many-body scars \cite{Sugiura_2021_Scar,Mizuta_2020_Scar} are characterized by Floquet eigenstates, and hence they are also in the scope.
On the other hand, generic nonintegrable time-periodic Hamiltonians are believed to satisfy Floquet eigenstate thermalization hypothesis \cite{Alessio2014-FloquetETH,Lazaridez2014-FloquetETH}, where every Floquet eigenstate is locally indistinguishable from a trivial infinite temperature state.
Although this is an undesired phenomenon in condensed matter physics, which is known as heating, our quantum algorithms for nearly optimal Floquet eigenstate preparation will serve as a source of randomness appearing in time-periodic Hamiltonians \cite{Tangpanitanon2023-complexity-Floquet}. 

\section*{Acknowledgment}

We thank T. N. Ikeda for fruitful discussion.
We also thank S. Kitamura for giving us some useful information about Floquet theory.
K. M. is supported by JST PRESTO Grant No. JPMJPR235A.

\bibliography{bibliography.bib}

%===================================================
% Appendix
%===================================================
\appendix
\begin{center}
\bf{\large Appendix}
\end{center}

\section{Exponential tails of Floquet eigenstates}\label{A_Sec:Tails_Eigenstate}

Theorem \ref{Thm:Tail_Eigenstates} in the main text states the exponential decay of every Fourier component $\ket{\phi_n^l}$ in $l \in \bbZ$ of Floquet eigenstates.
This is related to the localization on one-dimensional lattice under linear potential, where the Fourier index $l \in \bbZ$ plays a role of the coordinate.
While several references such as Refs. \cite{Hone_1997_decay,Lindner_2017_chiral,Rudner2020-handbook} mention about the exponential decay in Floquet eigenstates, we cannot find an explicit bound that fits our setup.
For this paper to be self-contained, we provide a rigorous proof of Theorem \ref{Thm:Tail_Eigenstates}. 
The theorem is restated as follows.

\renewcommand{\thetheorem}{\arabic{theorem}}
\setcounter{theorem}{2}
\begin{theorem}
\textbf{(Tails of Floquet eigenstates)}

Suppose that a Floquet eigenstate $\ket{\phi_n(t)}$ or equivalently $\ket{\Phi_n}$ has quasienergy $\epsilon_n \in \mr{BZ}=[-\omega/2,\omega/2)$ under the Hamiltonian, Eq. (\ref{Eq:Periodic_Hamiltonian}).
Then, every Fourier component exponentially decays as
\begin{equation}
    \norm{\ket{\phi_n^l}} \leq  \exp \left( -\frac{|l|-1/2}{2M+1} + \frac{\sinh 1}{2\pi} \alpha T \right).
\end{equation}
\end{theorem}

\renewcommand{\thetheoremA}{\ref*{A_Sec:Tails_Eigenstate}\arabic{theoremA}}

The proof consists of three main steps.
In the first step, we show the exponential decay in each eigenstate of the truncated Floquet Hamiltonian $H_\mr{F}^L$ instead of the exact one (Proposition \ref{A_Prop:decay_truncated_eigenstates}).
The second step is to show that the eigenspace of the truncated Floquet Hamiltonian well approximates the subspace spanned by the exact Floquet eigenstates (Proposition \ref{A_Prop:Relation_eigenspaces}).
Finally, we combine them and prove that the Floquet eigenstates must also show exponential decay as Theorem \ref{Thm:Tail_Eigenstates}.
We hereby provide the first step and its derivation.

\begin{propositionA}\label{A_Prop:decay_truncated_eigenstates}
\textbf{(Truncated Floquet eigenstates)}

Let $\ket{\tilde{\Phi}_n^L} \in \mcl{H}^L$ be an eigenstate of the truncated Floquet Hamiltonian $H_\mr{F}^L$ as $H_\mr{F}^L \ket{\tilde{\Phi}_n^L} = \tilde{\epsilon}_n^L \ket{\tilde{\Phi}_n^L}$.
We define projections to a certain Fourier index $l \in \bbZ$ and an eigenspectra $E \subset \bbR$ respectively by
\begin{eqnarray}
    P_l &=& \ket{l}\bra{l}_f \otimes I, \label{Eq:Proj_l} \\
    \tilde{P}^L(E) &=& \sum_{n: \tilde{\epsilon}_n^L \in E} \ket{\tilde{\Phi}_n^L} \bra{\tilde{\Phi}_n^L}.  \label{Eq:Proj_truncated}
\end{eqnarray}
Then, for arbitrary $l \in [L]$,
\begin{equation}\label{Eq:decay_truncated_eigenstates}
    \norm{P_l \tilde{P}^L(E)} \leq \exp \left( -\frac{|l|-\epsilon_\imax /\omega }{2M+1} + \frac{\sinh 1}{2\pi} \alpha T \right)
\end{equation}
is satisfied with $\epsilon_\imax = \max_{\epsilon \in E}(|\epsilon|)$.
\end{propositionA}

\textit{Proof of Proposition \ref{A_Prop:decay_truncated_eigenstates}.---}
We start with the case $l \geq 0$.
For arbitrary $\lambda \in \bbR$, we obtain the following inequality:
\begin{eqnarray}
    \norm{P_l \tilde{P}^L(E)} &=& \norm{P_l e^{\lambda H_\mr{F}^L} e^{-\lambda H_\mr{F}^L} \tilde{P}^L(E)} \nonumber \\
    &\leq&  \norm{P_l e^{\lambda H_\mr{F}^L}}
    \cdot \norm{\sum_{\tilde{\varepsilon}_n \in E} e^{-\lambda \tilde{\varepsilon}_n}\ket{\tilde{\Phi}_n^L} \bra{\tilde{\Phi}_n^L}} \nonumber \\
    &\leq& e^{\max_{\varepsilon \in E} (|\lambda \varepsilon|)} \sqrt{\norm{\braket{l|e^{2\lambda H_\mr{F}^L} |l}_f}}. \label{PropEq:decay_truncated_1}
\end{eqnarray}
We should evaluate a bound on $\norm{\braket{l|e^{2\lambda H_\mr{F}^L} |l}_f}$.
By splitting the Floquet Hamiltonian into $H_\mr{F}^L = H_\mr{Add}^L - H_\mr{LP}^L$ with
\begin{eqnarray}
    H_\mr{Add}^L &=& \sum_{|m| \leq M} \sum_{l \in [L]; l+m \in [L]} \ket{l+m}\bra{l}_f \otimes H_m, \\
    H_\mr{LP}^L &=& \sum_{l \in [L]} l \omega \ket{l}\bra{l}_f \otimes I,
\end{eqnarray}
we use an interaction picture based on $H_\mr{LP}^L$.
The imaginary-time evolution $e^{2\lambda H_\mr{F}^L}$ is rewritten by
\begin{eqnarray}
    e^{2\lambda H_\mr{F}^L} &=& e^{- 2\lambda H_\mr{LP}^L} \mcl{T} \exp \left( \int_0^{2\lambda} H_{\mr{Add},I}^L(\tau) \dd \tau \right).
\end{eqnarray}
Here, $H_{\mr{Add},I}^L(\tau)$ denotes the interaction Hamiltonian described by
\begin{eqnarray}
    H_{\mr{Add},I}^L(\tau) &=& e^{\tau H_\mr{LP}^L} H_\mr{Add}^L e^{-\tau H_\mr{LP}^L} \nonumber \\
    &=& \sum_{|m| \leq M} \sum_{l \in [L]; l+m \in [L]} \ket{l+m}\bra{l}_f \otimes e^{m\omega\tau} H_m. \nonumber \\
    && 
\end{eqnarray}
and its norm is bounded by 
\begin{eqnarray}
    \norm{H_{\mr{Add},I}^L(\tau)} &\leq& \sum_{|m| \leq M} e^{m \omega \tau} \alpha \nonumber \\
    &=& \frac{\sinh (2M+1)\omega \tau/2}{\sinh (\omega \tau/2)}\alpha \nonumber \\
    &\leq& (2M+1) \alpha \sinh 1, \label{A_Eq:H_add_int_bound}
\end{eqnarray}
for $\tau \in [0,2/\{(2M+1)\omega \} ]$.
Finally, by setting $\lambda = 1/ \{ (2M+1) \omega \}$, we get the bound, 
\begin{equation}
    \norm{\braket{l|e^{2\lambda H_\mr{F}^L} |l}_f} \leq \exp \left( \frac{-2l\omega + 2(2M+1) \alpha \sinh 1}{(2M+1) \omega}\right),
\end{equation}
and combining this inequality with Eq. (\ref{PropEq:decay_truncated_1}) immediately implies the upper bound, Eq. (\ref{Eq:decay_truncated_eigenstates}).
We get the same statement for a negative integer $l$ by inserting $e^{-\lambda H_\mr{F}^L} e^{\lambda H_\mr{F}^L}$ instead of $e^{\lambda H_\mr{F}^L} e^{-\lambda H_\mr{F}^L}$ in Eq. (\ref{PropEq:decay_truncated_1}). $\quad \square$

For a single eigenstate $\ket{\tilde{\Phi}_n^L}$, this proposition gives the inequality for $\ket{\tilde{\phi}_n^{l,L}}=(\bra{l}_f \otimes I)\ket{\tilde{\Phi}_n^L}$,
\begin{equation}\label{A_Eq:psi_tilde_decay}
    \norm{\ket{\tilde{\phi}_n^{l,L}}} \leq  \exp \left( -\frac{|l|-\tilde{\epsilon}_n^L /\omega }{2M+1} + \frac{\sinh 1}{2\pi} \alpha T \right),
\end{equation}
which gives rise to the exponential decay.
Although we would like to prove a similar statement for the Floquet Hamiltonian $H_\mr{F}$, we note that the truncation is essential in the above proof.
In Eq. (\ref{PropEq:decay_truncated_1}), we insert the exponential functions $e^{\pm \lambda H_\mr{F}^L}$, and they are well defined only when the exponent is bounded.
To extend this result to the Floquet eigenstates, we next show that the eigenspace of the truncated Floquet Hamiltonian approximates the exact one as follows.
We also prove Theorem \ref{Thm:Tail_Eigenstates} following to this.

\begin{propositionA}\label{A_Prop:Relation_eigenspaces}
\textbf{(Relation of eigenspaces)}

For every Floquet eigenstate $\ket{\Phi_n}$ having quasinergy $\epsilon_n \in \mr{BZ}$,
\begin{equation}
    \lim_{L \to \infty} (1-\tilde{P}^L(\mr{BZ}^\varepsilon))  \ket{\Phi_n} = 0,
\end{equation}
is satisfied for arbitrary $\varepsilon > 0$, where $\mr{BZ}^\varepsilon$ is defined by
\begin{equation}
    \mr{BZ}^\varepsilon = [-(1/2+\varepsilon)\omega, (1/2+\varepsilon)\omega).
\end{equation}

\end{propositionA}

\textit{Proof of Proposition \ref{A_Prop:Relation_eigenspaces}.---} From the definition of the projection, Eq. (\ref{Eq:Proj_truncated}), the norm is represented by
\begin{equation}\label{PropEq:Relation_eigenspaces_1}
     \norm{(1-\tilde{P}^L(\mr{BZ}^\varepsilon))\ket{\Phi_n}} = \sqrt{\sum_{n^\prime: \epsilon_{n^\prime} \notin \mr{BZ}^\varepsilon} |\braket{\tilde{\Phi}_{n^\prime}^L | \Phi_n}|^2 }.
\end{equation}
We evaluate an upper bound on each inner product $|\braket{\tilde{\Phi}_{n^\prime}^L | \Phi_n}|$.
Plugging the truncated Floquet Hamiltonian into it, we obtain 
\begin{eqnarray}
    |\braket{\tilde{\Phi}_{n^\prime}^L | \Phi_n}| &\leq& \frac{ | \braket{\tilde{\Phi}_{n^\prime}^L | H_\mr{F}^L | \Phi_n} - \epsilon_n \braket{\tilde{\Phi}_{n^\prime}^L | \Phi_n}|}{|\tilde{\epsilon}_{n^\prime}^L - \epsilon_n|} \nonumber \\
    &\leq& \frac{\norm{(H_\mr{F}^L - \epsilon_n) \ket{\Phi_n}}}{\varepsilon \omega}, \label{PropEq:Relation_eigenspaces_2}
\end{eqnarray}
where $\tilde{\epsilon}_{n^\prime}^L \in \mr{BZ}^\varepsilon$ denotes an eigenvalue corresponding to $\ket{\tilde{\Phi}_{n^\prime}^L}$.
The denominator can be evaluated by
\begin{eqnarray}
    && (H_\mr{F}^L - \epsilon_n) \ket{\Phi_n} \nonumber \\
    && =  \sum_{m} \sum_{\substack{l \in [L] \\ ; l+m \in [L]}} \ket{l+m}_f H_m \ket{\phi_n^l} - \sum_{l \in [L]} (\epsilon_n+l\omega) \ket{l}_f\ket{\phi_n^l} \nonumber \\
    && = \sum_{l \in [L]} \ket{l}_f \left( \sum_m H_m \ket{\phi_n^{l-m}} - (\epsilon_n + l \omega) \ket{\phi_n^l}\right) \nonumber \\
    && \qquad + \sum_{|m|\leq M} \left( \sum_{\substack{l \in [L] \\ ; l+m \in [L]}} - \sum_{l \in -m + [L]}\right) \ket{l+m}_f H_m \ket{\phi_n^l}. \nonumber \\
    && \label{PropEq:Relation_eigenspaces_3}
\end{eqnarray}
The first term is exactly zero due to the eigenvalue equation, Eq. (\ref{Eq:Sambe_eigenequation}), for Floquet eigenstates.
In the second term, we use integration by part based on the analyticity of $\ket{\phi_n(t)}$, which implies
\begin{eqnarray}
    \norm{\ket{\phi_n^l}} &\leq& \frac{1}{(|l|\omega)^2} \max_{t \in [0,T]} \left( \norm{\dv[2]{t} \ket{\phi_n(t)}}\right) \nonumber \\
    &\leq&  \frac{((2M+1)\alpha+\omega/2)^2 + M (2M+1) \omega \alpha}{(|l|\omega)^2}. \nonumber \\
    &&
\end{eqnarray}
In the second inequality, we use $\ket{\phi_n(t)}=e^{i\epsilon_n t} U(t;0)\ket{\phi_n(0)}$ and the relations, $\norm{H(t)} \leq (2M+1) \alpha$, $\norm{H^\prime(t)} \leq M (2M+1) \omega \alpha$, and $\epsilon_n \in \mr{BZ}$.
As a result, we arrive at the inequality,
\begin{eqnarray}
    \norm{(H_\mr{F}^L - \epsilon_n) \ket{\Phi_n}} &\leq& \sum_{|m| \leq M} \sum_{|l| \geq L-M} \norm{H_m \ket{\phi_n^l}} \nonumber \\
    &\leq& \mr{Const.} \times \sum_{l=L-M}^\infty l^{-2} \nonumber \\
    &\leq& \frac{\mr{Const.}}{L-M}.
\end{eqnarray}
In the above, the constant does not depend on the cutoff $L$, but includes $M$, $\alpha$, and $\omega$.
Combining this with Eqs. (\ref{PropEq:Relation_eigenspaces_1}) and (\ref{PropEq:Relation_eigenspaces_2}), we finally obtain the following relation,
\begin{eqnarray}
    \norm{(1-\tilde{P}^L(\mr{BZ}^\varepsilon))\ket{\Phi_n}} &\leq& \frac{\mr{Const.}}\varepsilon \sqrt{\frac{2L \cdot \mr{dim}(\mcl{H})}{(L-M)^2}} \nonumber \\
    &\substack{\to \\ L \to \infty}& 0,
\end{eqnarray}
where the coefficient in the first line comes from the dimension of the truncated Sambe space.
This completes the proof of Proposition \ref{A_Prop:Relation_eigenspaces}. $\quad \square$

\textit{Proof of Theorem \ref{Thm:Tail_Eigenstates}.---}
For an eigenstate in the Sambe space $\ket{\Phi_n}$, corresponding to a Floquet eigenstate $\ket{\phi_n(t)}$, we choose a arbitrarily large cutoff $L$ such that $l \in [L]$.
Then, the norm of each Fourier component is evaluated by
\begin{eqnarray}
    \norm{\ket{\phi_n^l}} &=& \norm{P_l \ket{\Phi_n}} \nonumber \\
    &\leq& \norm{\tilde{P}^L(\mr{BZ}^\varepsilon) P_l} + \norm{(1-\tilde{P}^L(\mr{BZ}^\varepsilon)) P_l \ket{\Phi_n}}. \nonumber \\
    &&
\end{eqnarray}
The first term is bounded by Eq. (\ref{Eq:decay_truncated_eigenstates}) regardless of $L$ from Proposition \ref{A_Prop:decay_truncated_eigenstates}.
The second term goes to zero by $L \to \infty$ from Proposition \ref{A_Prop:Relation_eigenspaces}.
As a result, the inequality
\begin{equation}
    \norm{\ket{\phi_n^l}} \leq  \exp \left( -\frac{|l|-1/2 - \varepsilon}{2M+1} + \frac{\sinh 1}{2\pi} \alpha T \right)
\end{equation}
holds for arbitrary $\varepsilon >0$, and this completes the proof of Theorem \ref{Thm:Tail_Eigenstates}. $\quad \square$

%=============================
% Appendix: Detailed proof
%=============================
\section{Approximate Floquet eigenstate from the truncated Sambe space}\label{A_Sec:approx_truncated_sambe}

\renewcommand{\thetheoremB}{\ref*{A_Sec:approx_truncated_sambe}\arabic{theoremB}}

In Section \ref{Sec:Accuracy_Sambe}, we show that the state $\ket{\tilde{\Phi}_n^L} \in \mcl{H}^L$, an eigenstate of the truncated Floquet Hamiltonian $H_\mr{F}^L$, becomes an approximate eigenstate of the Floquet Hamiltonian $H_\mr{F}$ by Proposition \ref{Prop:Accuracy_quasienergy}.
Namely, the truncated Sambe space provides an accurate estimate for each Floquet eigenstate on the Sambe space, $\ket{\Phi_n} \in \mcl{H}^\infty$.
Here, we prove that it also gives an accurate estimate on each Floquet eigenstate on the physical space, $\ket{\phi_n(t)}$.

Let $\tilde{\epsilon}_n^L$ be an eigenvalue of $H_\mr{F}^L$ for the eigenstate $\ket{\tilde{\Phi}_n^L} \in \mcl{H}^L$.
From the exact relation between $\ket{\phi_n(t)}$ and $\ket{\Phi_n}$ shown in Eqs. (\ref{Eq:time_evolution_eigenbasis}) and (\ref{Eq:def_Floquet_state}), the Floquet eigenstate in the physical space is expected to be reproduced by
\begin{equation}\label{A_Eq:Est_Eigenst_from_truncated}
    \ket{\tilde{\phi}_n^L(t)} \equiv e^{i \tilde{\epsilon}_n^L t} U(t;0) \sum_{l \in [L]} (\bra{l}_f \otimes I) \ket{\tilde{\Phi}_n^L}
\end{equation}
Each Floquet eigenstate $\ket{\phi_n(t)}$ is characterized as an eigenstate of the time-evolution operator $U(t+T;t)$ with the eigenvalue $e^{-i \epsilon_n T}$.
The state $\ket{\tilde{\phi}_n^L(t)}$ provides an accurate estimate on the exact Floquet eigenstate $\ket{\phi_n(t)}$ in that it becomes an approximate eigenstate of $U(t+T;t)$, as mentioned in Section \ref{Subsec:Floquet_theory}.
We prove this fact by the following theorem, based on the coincidence of the cutoff $L$ for Floquet eigenstates (Theorem \ref{Thm:Accuracy_quasienergy}) and the Lieb-Robinson bound in the Sambe space [Eq. (\ref{Eq:Cutoff_LR})].

\begin{propositionB}\label{A_Prop:Est_Eigenst_from_truncated}
\textbf{}

We define a state $\ket{\tilde{\phi}_n^L(t)} \in \mcl{H}$ by Eq. (\ref{A_Eq:Est_Eigenst_from_truncated}) with the eigenvalue $\tilde{\epsilon}_n^L$ and the eigenstate $\ket{\tilde{\Phi}_n^L}$ of the truncated Floquet Hamiltonian $H_\mr{F}^L$.
Then, it can be an approximate eigenstate of the time-evolution $U(t+T;t)$ with an error $\varepsilon \in (0,1)$ by
\begin{equation}\label{A_Eq:Error_approx_eigenst}
    \frac{\norm{\left(U(t+T; t)- e^{-i \tilde{\epsilon}_n^L T}\right) \ket{\tilde{\phi}_n^L(t)}}}{\norm{\ket{\tilde{\phi}_n^L(t)}}} \leq \varepsilon
\end{equation}
under the choice of the cutoff $L \in \Theta (\alpha T + |\tilde{\epsilon}_n^L|/\omega + \log (1/\varepsilon))$.

\end{propositionB}

We first focus on the case $t=0$, where the state is given by
\begin{equation}\label{PropEq:Accuracy_quasienergy_1}
    \ket{\tilde{\phi}_n^L(0)} = \sum_{l \in [L]} \ket{\tilde{\phi}_n^{l,L}}, \quad \ket{\tilde{\phi}_n^{l,L}} \equiv (\bra{l}_f \otimes I)\ket{\tilde{\Phi}_n^L},
\end{equation}
and show that this gives an approximate eigenstate of the Floquet operator $U(T;0)$.
Applying the Floquet operator in the Sambe space formalism by Eq. (\ref{Eq:Time_evol_in_Sambe}) to $\ket{\tilde{\phi}_n^L(0)}$, the numerator of Eq. (\ref{A_Eq:Error_approx_eigenst}) can be transformed into
\begin{eqnarray}
   && U(T;0) \ket{\tilde{\phi}_n^L(0)} - e^{-i \tilde{\epsilon}_n^L T} \ket{\tilde{\phi}_n^L(0)} \nonumber \\
   && \quad =  \sum_{l \in \bbZ} \sum_{l^\prime \in [L]} \bra{l} \left( e^{-i H_\mr{F} T}- e^{-i H_\mr{F}^L T} \right) \ket{l^\prime}_f \ket{\tilde{\phi}_n^{l^\prime,L}}. \nonumber \\
   && \label{PropEq:Accuracy_quasienergy_3}
\end{eqnarray}
Before going to the proof of Proposition \ref{A_Prop:Est_Eigenst_from_truncated}, we provide two propositions: One is about the numerator associated with the Lieb-Robinson bound (Proposition \ref{A_Prop:Lieb_Robinson}), and the other is about the denominator, i.e., the norm of $\ket{\tilde{\phi_n^L}(0)}$ (Proposition \ref{A_Prop:norm_approximate_eigenst}).
The first proposition says that each amplitude $\bra{l} \left( e^{-i H_\mr{F} T}- e^{-i H_\mr{F}^L T} \right) \ket{l^\prime}_f$ in Eq. (\ref{A_Eq:Error_approx_eigenst}) is exponentially suppressed in the distance between $l$ and $l^\prime$.

\begin{propositionB}\label{A_Prop:Lieb_Robinson}
\textbf{}

Consider a time-periodic Hamiltonian $H(t)$, satisfying Eq. (\ref{Eq:Periodic_Hamiltonian}).
For a Fourier index $l^\prime \in [L]$, the inequality
\begin{equation}\label{A_Eq:Lieb_Robinson_difference}
    \norm{\braket{l|(e^{-iH_\mr{F} t} - e^{-i H_\mr{F}^L t })|l^\prime}_f} \leq 2 e^{(2M+1) e\alpha t - d(l,l^\prime)/M }
\end{equation}
is satisfied, where $d(l,l^\prime)$ is defined by
\begin{equation}
    d(l,l^\prime) = \begin{cases}
        2L-|l|-|l^\prime| & (l \in [L]) \\
        |l|-|l^\prime| & (l \notin [L]).
    \end{cases}
\end{equation}
\end{propositionB}

\textit{Proof of Proposition \ref{A_Prop:Lieb_Robinson}.---}
We use an interaction picture based on $H_\mr{LP} = \sum_{l \in \bbZ} l \omega \ket{l}\bra{l}_f \otimes I$.
Using the Dyson series expansion under the interaction Hamiltonian $H_{\mr{Add},I}(t) = \sum_{|m| \leq M} \sum_{l \in \bbZ} \ket{l+m}\bra{l}_f \otimes e^{im\omega t} H_m$,
\begin{equation}
   U_{\mr{Add},I}(t) = \sum_{n=0}^\infty \int_0^t \dd t_1 \hdots \int_0^{t_{n-1}} \dd t_n \prod_{i=1}^n H_{\mr{Add},I}(t_i)  
\end{equation}
and that for the truncated Floquet Hamiltonian, we get the relation,
\begin{equation}
    (\text{l.h.s of Eq. (\ref{A_Eq:Lieb_Robinson_difference})}) = \norm{\braket{l|U_{\mr{Add},I}(t)-U_{\mr{Add},I}^L(t)|l^\prime}_f}.
\end{equation}
By plugging the completeness $\sum_{l_i \in \bbZ} \ket{l_i}\bra{l_i}_f \otimes I$ into the Dyson series for $i=1,2,\hdots,n-1$, each term of the right-hand side represents the transition amplitude via the path $l^\prime \equiv l_n \to l_{n-1} \to \hdots \to l_1 \to l_0 \equiv l$ as
\begin{eqnarray}
    && \braket{l|U_{\mr{Add},I}(t)|l^\prime}_f \nonumber \\
    && \quad = \sum_{n=0}^\infty \int_0^t \dd t_1 \hdots \int_0^{t_{n-1}} \dd t_n \prod_{i=1}^n \bra{l_{i-1}} H_{\mr{Add},I}(t_i) \ket{l_i}_f. \nonumber \\
    &&
\end{eqnarray}
Each transition $l_i \to l_{i-1}$ is allowed only when $|l_i - l_{i-1}| \leq M$ is satisfied due to the assumption.
The difference from the truncated one appears when the path goes across the region $\bbZ \backslash [L]$, i.e., low order terms with $n < d(l,l^\prime)$ in the Dyson series vanish.
Considering the bound on the interaction Hamiltonian $\norm{H_{\mr{Add},I}(t)} \leq \max_{t \in [0,T]} \norm{H(t)} \leq (2M+1)\alpha$ and that for the truncated one \cite{Mizuta_Quantum_2023}, we arrive at the inequality,
\begin{eqnarray}
    (\text{l.h.s of Eq. (\ref{A_Eq:Lieb_Robinson_difference})}) &\leq& 2 \sum_{n= \lceil d(l,l^\prime)/M \rceil}^\infty \frac{t^n}{n!} ((2M+1)\alpha)^n \nonumber \\
    &\leq& 2 \exp \left( (2M+1) e \alpha t - \frac{d(l,l^\prime)}{M}\right). \nonumber \\
    &&
\end{eqnarray}
In the last line, we use the relation $\sum_{n=n_0}^\infty (x/n)^n \leq e^{ex-n_0}$ for arbitrary $x\geq 0$. $\quad \square$

We substitute $t=T$ in the above proposition to evaluate the numerator of Eq. (\ref{A_Eq:Error_approx_eigenst}) later.
Note that this bound comes from the Lieb-Robinson bound in the Sambe space \cite{Mizuta_Quantum_2023}, i.e., the decay of the propagation in Fourier indices.
Indeed, when the distance $d(l,l^\prime)$ is greater than $L_\mr{LR} \in \Theta (\alpha t+ \log (1/\varepsilon))$, the transition amplitude becomes smaller than $\varepsilon$.
Next, we prove the second proposition, which provides the norm of the state $\ket{\tilde{\phi}_n^L(t)}$ in the denominator of Eq. (\ref{A_Eq:Error_approx_eigenst}).

\begin{propositionB}\label{A_Prop:norm_approximate_eigenst}
\textbf{}

The state $\ket{\tilde{\phi}_n^L(0)} = \sum_{l \in [L]} \ket{\tilde{\phi}_n^{l,L}}$ is approximately normalized in the sense that
\begin{equation}\label{A_Eq:norm_normalized_error}
    1 - \varepsilon_\mr{norm}^L \leq \norm{\ket{\tilde{\phi}_n^{L}(0)}} \leq 1 + \varepsilon_\mr{norm}^L
\end{equation}
is satisfied.
Here, the value $\varepsilon_\mr{norm}^L$ is defined by
\begin{eqnarray}
    \varepsilon_\mr{norm}^L &=& 6 (2M+1)^2 \alpha T \log (2eL) \nonumber \\
    && \times \exp \left( -\frac{L-|\tilde{\epsilon}_n^L|/\omega}{2M+1} + \frac{\sinh 1}{2\pi} \alpha T \right),  \label{A_Eq:varepsilon_norm}
\end{eqnarray}
whose scaling is $e^{-\Theta (L-|\tilde{\epsilon}_n^L|/\omega-\alpha T)}$.
\end{propositionB}
\textit{Proof of Proposition \ref{A_Prop:norm_approximate_eigenst}.---} 
As we define the equivalent Floquet eigenstate under translation by $\ket{\Phi_n^l}= \mr{Add}_l \ket{\Phi_n}$ [See Eq. (\ref{Eq:Eigenstate_Sambe_shifted})], we organize its approximate counterpart by
\begin{equation}
    \ket{\tilde{\Phi}_n^{l,L}} = \mr{Add}_l \ket{\tilde{\Phi}_n^L} = \sum_{l^\prime \in [L]} \ket{l^\prime + l}_f \ket{\tilde{\phi}_n^{l^\prime,L}}.
\end{equation}
Then, the norm of the state $\ket{\tilde{\phi}_n^L(0)}$ is evaluated by
\begin{eqnarray}
    \braket{\tilde{\phi}_n^L(0) | \tilde{\phi}_n^L(0)} &=& \sum_{k,l \in \bbZ} \braket{\tilde{\Phi}_n^L|(\ket{k}\bra{l}_f \otimes I)|\tilde{\Phi}_n^L} \nonumber \\
    &=& \sum_{k,l \in \bbZ} \braket{\tilde{\Phi}_n^L| P_k \mr{Add}_{k-l}|\tilde{\Phi}_n^L} \nonumber \\
    &=& 1 + \sum_{l \in \bbZ \backslash \{0\} } \braket{\tilde{\Phi}_n^L |
    \tilde{\Phi}_n^{l,L}}. \label{A_Eq:Norm_squared_eval}
\end{eqnarray}
We derive the approximate orthogonality of $\ket{\tilde{\Phi}_n^L}$ and $\ket{\tilde{\Phi}_n^{l,L}}$ for $l \neq 0$ based on the fact that they are approximate eigenvectors of $H_\mr{F}$ with different eigenvalues.
As discussed in Proposition \ref{Prop:Accuracy_quasienergy}, the error $\norm{(H_F - \tilde{\epsilon}_n^L)\ket{\tilde{\Phi}_n^L}}$ is suppressed up to $e^{- \Theta (L-|\tilde{\epsilon}_n^L|/\omega - \alpha T)}$ by Eq. (\ref{Eq:Approx_H_F_Psi_tilde}).
In addition, the translation symmetry of the Floquet Hamiltonian, $\mr{Add}_l^\dagger H_\mr{F} \mr{Add}_l = H_\mr{F} - l\omega$, implies the relation,
\begin{equation}
    \norm{(H_F - \tilde{\epsilon}_n^L+l\omega)\ket{\tilde{\Phi}_n^{L,l}}} = \norm{(H_\mr{F}-\tilde{\epsilon}_n^L) \ket{\tilde{\Phi}_n^L}},
\end{equation}
which gives the same upper bound as Eq. (\ref{Eq:Approx_H_F_Psi_tilde}).
The inner products appearing in Eq. (\ref{A_Eq:Norm_squared_eval}) can be evaluated by
\begin{eqnarray}
    |\braket{\tilde{\Phi}_n^L | \tilde{\Phi}_n^{l,L}}| &=& \frac{\left| \braket{\tilde{\Phi}_n^L | (\tilde{\epsilon}_n^L - l\omega - \tilde{\epsilon}_n^L)|\tilde{\Phi}_n^{l,L}} \right|}{|l| \omega} \nonumber \\
    &\leq& \frac{\norm{(H_\mr{F}-\tilde{\epsilon}_n^L) \ket{\tilde{\Phi}_n^L}}}{|l| \omega},
\end{eqnarray}
for $l \in [2L] \backslash \{0\}$.
We note $\braket{\tilde{\Phi}_n^L | \tilde{\Phi}_n^{l,L}}=0$ for $l \in \bbZ \backslash [2L]$ by definition.
Using the inequality $\sum_{l=1}^{2L} l^{-1} \leq 1 + \log (2L)$, we arrive at the inequality,
\begin{eqnarray}
    \left| 1- \norm{\ket{\tilde{\phi}_n(0)}}^2 \right| &\leq& 2  \log (2eL) \times \frac{\norm{(H_\mr{F}-\tilde{\epsilon}_n^L) \ket{\tilde{\Phi}_n^L}}}{\omega} \nonumber \\
    &=& 6 (2M+1)^2 \alpha T \log (2eL) \nonumber \\
    && \times \exp \left( -\frac{L-|\tilde{\epsilon}_n^L|/\omega}{2M+1} + \frac{\sinh 1}{2\pi} \alpha T \right). \nonumber \\
    &&
\end{eqnarray}
This immediately implies the relation of Eqs. (\ref{A_Eq:norm_normalized_error}) and (\ref{A_Eq:varepsilon_norm}). $\quad \square$

Now, we are ready for proving Proposition \ref{A_Prop:Est_Eigenst_from_truncated}.
We go back to showing that the state $\ket{\tilde{\phi}_n^L(0)}$ organized by the truncated Sambe space appropriately provides an approximate eigenstate of the Floquet operator $U(T;0)$.
The proof for generic $t \in \bbR$ follows it.

\textit{Proof of Proposition \ref{A_Prop:Est_Eigenst_from_truncated}.---}
Beginning with Eq. (\ref{PropEq:Accuracy_quasienergy_3}), we evaluate the bound on its norm given by
\begin{eqnarray}
   && \norm{\left( U(T;0) -  e^{-i \tilde{\epsilon}_n^L T} 
 \right) \ket{\tilde{\phi}_n^L(0)}} \nonumber \\
   && \quad \leq  \sum_{l \in \bbZ} \sum_{l^\prime \in [L]} \norm{ \bra{l} \left( e^{-i H_\mr{F} T}- e^{-i H_\mr{F}^L T} \right) \ket{l^\prime}_f} \cdot \norm{ \ket{\tilde{\phi}_n^{l^\prime,L}}} . \nonumber \\
   &&
\end{eqnarray}
Let us first focus on the summation over $l \in [L]$ in the above.
The bound from Proposition \ref{A_Prop:Lieb_Robinson} and the one for the state $\ket{\tilde{\phi}_n^{l,L}}$ by Eq. (\ref{A_Eq:psi_tilde_decay}) indicates the inequality
\begin{eqnarray}
    && \sum_{l \in [L]} \sum_{l^\prime \in [L]} \norm{ \bra{l} \left( e^{-i H_\mr{F} T}- e^{-i H_\mr{F}^L T} \right) \ket{l^\prime}_f} \cdot \norm{ \ket{\tilde{\phi}_n^{l^\prime,L}}} \nonumber \\
    && \quad \leq 8 \sum_{l,l^\prime=0}^L \exp \left( (2M+1) e\alpha T - \frac{2L-l-l^\prime}{M}\right) \nonumber \\
    && \qquad \qquad \qquad \times \exp \left( - \frac{l^\prime - |\tilde{\epsilon}_n^L|/\omega}{2M+1} + \frac{\sinh1}{2\pi} \alpha T \right) \nonumber \\
    && \quad \leq 16 (2M+1)^2 \exp \left( - \frac{L-|\tilde{\epsilon}_n^L|/\omega}{2M+1} + 2(M+1)e \alpha T\right). \nonumber \\
    && \label{PropEq:Accuracy_quasienergy_6}
\end{eqnarray}
The sum over $l \in \bbZ \backslash [L]$ in Eq. (\ref{PropEq:Accuracy_quasienergy_3}) is evaluated in a similar way, and results in the same form as Eq. (\ref{PropEq:Accuracy_quasienergy_6}).
Therefore, $\ket{\tilde{\phi}_n^L(0)}$ is an approximate eigenstate of $U(T;0)$ in a sense that
\begin{eqnarray}
    && \norm{\left( U(T;0) -  e^{-i \tilde{\epsilon}_n^L T} 
 \right) \ket{\tilde{\phi}_n^L(0)}} \nonumber \\
    && \quad \leq 32 (2M+1)^2 \exp \left( - \frac{L-|\tilde{\epsilon}_n^L|/\omega}{2M+1} + 2(M+1)e \alpha T\right) \nonumber \\
    &&
\end{eqnarray}
holds.
The right hand side scales as $e^{-\Theta(L-|\tilde{\epsilon}_n^L/\omega|-\alpha T)}$.
Since the norm of $\ket{\tilde{\phi}_n^L(0)}$ scales as $\norm{\ket{\tilde{\phi}_n^L(0)}} = 1 + e^{- \Theta (L-|\tilde{\epsilon}_n^L|/\omega - \alpha T)}$ by Proposition \ref{A_Prop:norm_approximate_eigenst}, we can achieve
\begin{equation}
    \frac{\norm{\left(U(T;0)- e^{-i \tilde{\epsilon}_n^L T}\right) \ket{\tilde{\phi}_n^L(0)}}}{\norm{\ket{\tilde{\phi}_n^L(0)}}} \leq \varepsilon
\end{equation}
under the choice of the cutoff $L \in \Theta (\alpha T + |\tilde{\epsilon}_n^L|/\omega + \log (1/\varepsilon))$.
This completes the proof of Proposition \ref{A_Prop:Est_Eigenst_from_truncated} for $t=0$.

Proposition \ref{A_Prop:Est_Eigenst_from_truncated} for generic time $t \in [0,T)$ is easily proved by the result for $t=0$.
The state $\ket{\tilde{\phi}_n^L(t)}$ organized from the truncated Sambe space by Eq. (\ref{A_Eq:Est_Eigenst_from_truncated}) is related to $\ket{\tilde{\phi}_n^L(0)}$ by $\ket{\tilde{\phi}_n^L(t)}=e^{i\tilde{\epsilon}_n^L t} U(t;0) \ket{\tilde{\phi}_n^L(0)}$.
Using the relation $U(t+T;t)U(t;0)=U(t;0)U(T;0)$, which is valid for a time-periodic Hamiltonian $H(t)$, the equality
\begin{eqnarray}
    && \norm{\left(U(t+T;t)- e^{-i \tilde{\epsilon}_n^L T}\right) \ket{\tilde{\phi}_n^L(t)}} \nonumber \\
    && \qquad \qquad \quad = \norm{\left(U(T;0)- e^{-i \tilde{\epsilon}_n^L T}\right) \ket{\tilde{\phi}_n^L(0)}}
\end{eqnarray}
is derived.
The approximate normalization of $\ket{\tilde{\phi}_n^L(t)}$ follows immediately from $\norm{\ket{\tilde{\phi}_n^L(t)}}=\norm{\ket{\tilde{\phi}_n^L(0)}}$.
Thus, the bound on Eq. (\ref{A_Eq:Error_approx_eigenst}) for generic time $t$ is exactly the same as the case of $t=0$. $\quad \square$

%=================================
% Section : Irrelvant boundary effects
%=================================
\section{Block-encoding of Floquet Hamiltonian}\label{A_Sec:boundary_condition}

\renewcommand{\thetheoremC}{\ref*{A_Sec:boundary_condition}\arabic{theoremC}}

In the Sambe space formalism, we often use the truncated Floquet Hamiltonian $H_\mr{F}^L$ defined by
\begin{equation}
    H_\mr{F}^L = \sum_{|m| \leq M} \sum_{l \in [L]; l+m \in [L]} \ket{l+m}\bra{l}_f - \sum_{l\in [L]} l\omega \ket{l}\bra{l}_f \otimes I,
\end{equation}
and the main text follows this definition.
However, for block-encoding toward QSVT, another truncated Floquet Hamiltonian $H_\mr{F,pbc}^L$ by
\begin{eqnarray}
    H_\mr{F,pbc}^L &=& \sum_{|m| \leq M} \mr{Add}_m^{[L]} \otimes H_m - \sum_{l \in [L]} l \omega \ket{l}\bra{l}_f \otimes I, \nonumber \\
    && \label{A_Eq:modified_H_F} \\
    \mr{Add}_m^{[L]} &=& \sum_{l \in [L]} \ket{(l\oplus m)_{[L]}}\bra{l}_f,
\end{eqnarray}
is rather feasible, as shown in Section \ref{Subsec:Modified_H_F}.
Here, we provide an efficient block-encoding of $H_\mr{F,pbc}^L$, and prove that using either $H_\mr{F}^L$ or $H_\mr{F,pbc}^L$ results in the same eigenvalues and eigenstates with negligible errors.
Namely, the discussion based on $H_\mr{F}^L$ in the main text is valid, while the actual quantum algorithms run with queries to block-encoding of $H_\mr{F,pbc}^L$ by $C[O_{H_m}]$.

\subsection{Block-encoding under periodic boundary conditions}\label{A_Subsec:modified_H_F}

We briefly review the block-encoding of $H_\mr{F,pbc}^L$ in Ref. \cite{Mizuta_Quantum_2023}, and how Proposition \ref{Prop:block_encode_H_F_pbc} in the main text is confirmed.
The block-encoding unitaries for the first and the second terms in Eq. (\ref{A_Eq:modified_H_F}) are constructed separately and their combination forms the one for $H_\mr{F,pbc}^L$.
Preparing a $\order{1}$-ancilla qubits expressed by $\{ \ket{m}_M \}_{|m| \leq M}$, the block-encoding of the first term $O_1$ is provided by
\begin{eqnarray}
    && O_1 = G_M^\dagger \left( \sum_{|m| \leq M} \ket{m}\bra{m} \otimes \mr{Add}_m^{[L]} \otimes O_{H_m} \right) G_M, \nonumber \\
    &&
\end{eqnarray}
with a unitary gate $G_M$ on this anclla system such that
\begin{equation}
    G_M \ket{0}_M = \sum_{|m| \leq M} \sqrt{\frac{\alpha_m}{\sum_{|m^\prime| \leq M} \alpha_{m^\prime}}} \ket{m}_M. 
\end{equation}
The block-encoding $O_1$ embeds the first term of Eq. (\ref{A_Eq:modified_H_F}),
\begin{equation}
    (\bra{0}_M \bra{0}_a) O_1 (\ket{0}_M \ket{0}_a) = \frac{\sum_{|m| \leq M} \mr{Add}_l^{[L]} \otimes H_m}{\sum_{|m| \leq M} \alpha_m},
\end{equation}
requiring one query respectively for $C[O_{H_m}]$ and at most $\order{\log L}$ primitive gates.
The second term of Eq. (\ref{A_Eq:modified_H_F}) has a block-encoding $O_2$ such that
\begin{equation}
    \braket{0| O_2 | 0}_{f^\prime} = \frac{- \sum_{l \in [L]} l\omega \ket{l}\bra{l}_f \otimes I}{L\omega},
\end{equation}
with a $\Theta (\log L)$ ancilla qubit system $f^\prime$, where we use only $\Theta (\log L)$ primitive gates \cite{Mizuta_Quantum_2023}.

The block-encoding of $H_\mr{F,pbc}^L$ is organized one query respectively to $O_1$ and $O_2$.
Introducing additional two qubits $c$, it is defined by
\begin{eqnarray}
    O_{H_\mr{F,pbc}^L} &=& G_c^\dagger \biggl( \ket{00}\bra{00}_c \otimes O_1 + \ket{01}\bra{01} \otimes O_2 \biggr. \nonumber \\
    && \qquad + (\ket{10}\bra{11}+\ket{11}\bra{10})_c \otimes I \biggl. \biggr) G_c, \nonumber \\
    && \\
    G_c \ket{00}_c &=& \sqrt{\frac{\sum_{|m| \leq M} \alpha_{m}}{\tilde{\alpha}}} \ket{00}_c + \sqrt{\frac{L\omega}{\tilde{\alpha}}} \ket{00}_c \nonumber \\
    && \quad + \sqrt{\frac{\tilde{\alpha}-\sum_{|m| \leq M} \alpha_{m}-L\omega}{\tilde{\alpha}}} \ket{10}_c,
\end{eqnarray}
where some identity matrices are omitted.
The above block-encoding is well-defined for arbitrary $\tilde{\alpha} \geq (2M+1)\alpha + L\omega$, which is ensured to be larger than $\sum_{|m| \leq M} \alpha_{m}+L\omega$.
By summarizing the ancilla system $a$ and the additional $\order{\log L}$ qubits as $a^\prime$, this embeds the truncated Floquet Hamiltonian $H_\mr{F,pbc}^L$ with the denominator $\tilde{\alpha}$ as Proposition \ref{Prop:block_encode_H_F_pbc}.
Therefore, QSVT algorithms working with the query complexity $q$ in $O_{H_\mr{F,pbc}^L}$ (e.g., the standard QPE based on the truncated Floquet Hamiltonians in the main text) can be executed by $\order{qM}=\order{q}$ queries to $C[O_{H_m}]$.

\subsection{Equivalence between different boundary conditions}

Here, we show the equivalence of the quantum algorithms working with $H_\mr{F}^L$ and $H_\mr{F,pbc}^L$.
In Section \ref{Sec:Floquet_QPE_Sambe}, we apply the QPE under the Hamiltonian $H_\mr{F}^{pL}$ to the input state $\ket{\Psi_0^L}$, Eq. (\ref{Eq_QPE:initial_st_truncated}).
The input state is approximately a superposition of Floquet eigenstates $\ket{\Phi_n^l}$ for $l \in [6L]$ according to Eqs. (\ref{Eq_QPE:Initial_decomp_finite_generic}) and (\ref{Eq_QPE:high_term_decomp}) based on Proposition \ref{Prop:FloquetQPE_Initial_st}.                                              
The calculation relies solely on the facts that each $\ket{\Phi_n^l}$ is an approximate eigenstate of $H_\mr{F}^L$ satisfying $\norm{(H_\mr{F}^L-(\epsilon_n - l \omega))\ket{\Phi_n^l}} \leq e^{- \Theta (L-\alpha T)}$ by Theorem \ref{Thm:Accuracy_quasienergy} and that it causes only a small error in the QPE result $\delta_\mr{approx}$ like Eq. (\ref{Eq_QPE:delta_approx}) as shown by Proposition \ref{A_Prop:approximate_QSVT}.
To prove the equivalence, it is sufficient to show the former fact for the alternative Floquet Hamiltonian $H_\mr{F,pbc}^{pL}$ as follows.

\begin{propositionC}\label{A_Prop:Equivalence_pbc}
\textbf{}

Let $\ket{\Phi_n^l} \in \mcl{H}^\infty$ be a Floquet eigenstate such that $H_\mr{F} \ket{\Phi_n^l}=(\epsilon_n-l\omega)\ket{\Phi_n^l}$.
It is an approximate eigenstate of the truncated Floquet Hamiltonian $H_\mr{F,pbc}^L$ by
\begin{eqnarray}
    && \norm{(H_\mr{F,pbc}^L-(\epsilon_n - l \omega)) \ket{\Phi_n^l}} \nonumber \\
    && \quad \leq 54 (2M+1)^2 \exp \left( - \frac{L-|l|}{2M+1} + \frac{\sinh 1}{2 \pi} \alpha T\right) \nonumber \\
    && \quad \leq e^{-\Theta (L-\alpha T)}, \label{A_Eq:Equivalence_pbc}
\end{eqnarray}
where $H_\mr{F,pbc}^L$ is defined by Eq. (\ref{A_Eq:modified_H_F}),
\end{propositionC}

\textit{Proof of Proposition \ref{A_Prop:Equivalence_pbc}.---} 
The proof is completely parallel to the proof for $H_\mr{F}^L$, which is shown as Proposition \ref{Prop:Accuracy_Floquet_Hamiltonian_eigenvalue} in the main text.
Setting $l=0$ for simplicity, we obtain the relation,
\begin{eqnarray}
    && (H_\mr{F,pbc}^L-\epsilon_n ) \ket{\Phi_n} \nonumber \\
    && \quad = \sum_{l^\prime \in [L]} \ket{l^\prime}_f \left( \sum_{|m| \leq M} H_m \ket{\phi_n^{(l^\prime \ominus m)_{[L]}}} - (\epsilon_n + l^\prime \omega) \ket{\phi_n^{l^\prime}}\right) \nonumber \\
    && \quad = \sum_{l^\prime \in [L]} \ket{l^\prime}_f \left( \sum_{|m| \leq M} H_m \ket{\phi_n^{l^\prime-m}} - (\epsilon_n + l^\prime \omega) \ket{\phi_n^{l^\prime}}\right) \nonumber \\
    && \qquad \quad + \sum_{l^\prime \in [L]} \ket{l^\prime}_f \sum_{|m| \leq M} H_m \left( \ket{\phi_n^{(l^\prime \ominus m)_{[L]}}} - \ket{\phi_n^{l^\prime-m}} \right). \nonumber \\
    &&
\end{eqnarray}
The symbol $(l \ominus m)_{[L]}=(l \oplus (-m))_{[L]}$ means $l-m$ modulo $2L$ defined on $[L]$.
In the last line, the first term disappears by the eigenvalue equation $H_\mr{F}\ket{\Phi_n}=\epsilon_n\ket{\Phi_n}$.
In the second term, only the contributions from $|l^\prime| \geq L-M$ can survive since $(l \ominus m)_{[L]}=l-m$ is otherwise satisfied.
As a result, the norm is bounded by
\begin{eqnarray}
    && \norm{(H_\mr{F,pbc}^L-\epsilon_n ) \ket{\Phi_n}} \nonumber \\
    && \quad \leq \sum_{|l^\prime| \geq L-M} \sum_{|m|\leq M} \left( \norm{\ket{\phi_n^{(l^\prime \ominus m)_{[L]}}}} +  \norm{\ket{\phi_n^{l^\prime-m}}} \right) \nonumber \\
    && \quad \leq 2 (2M+1) \alpha \sum_{l^\prime \bbZ \backslash [L-2M]} \norm{\ket{\phi_n^{l^\prime}}} \nonumber \\
    && \quad \leq 18 (2M+1)^2 \exp \left( - \frac{L-2M}{2M+1} + \frac{\sinh 1}{2 \pi} \alpha T\right), \nonumber \\
    &&
\end{eqnarray}
where we use Eq. (\ref{Eq:phi_n_l_contribute_out}) in the main text for deriving the last inequality.
Using $e^{2M/(2M+1)} < 3$, we arrive at the inequality, Eq. (\ref{A_Eq:Equivalence_pbc}), for $l=0$.
The result for a generic integer $l \in \bbZ$ is obtained similarly. $\quad \square$

In the main text, the algorithm in Section \ref{Sec:Floquet_QPE_Sambe} deals with the truncated Hamiltonian $H_\mr{F,pbc}^{pL}$ ($p \geq 7$), and each Floquet eigenstate $\ket{\Phi_n^l}$ for $l \in [6L]$ is involved in the initial state $\ket{\Psi_0^L}$ as Eqs. (\ref{Eq_QPE:Initial_decomp_finite_generic}) and (\ref{Eq_QPE:high_term_decomp}).
The above proposition guarantees the relation,
\begin{equation}
    \norm{(H_\mr{F,pbc}^{pL}-(\epsilon_n-l\omega)) \ket{\Phi_n^l}} \leq e^{-\Theta (L-\alpha T)}.
\end{equation}
The standard QPE based on $H_\mr{F,pbc}^{pL}$ exactly as described in Section \ref{Sec:Floquet_QPE_physical}.
Combined with the efficient block-encoding of $H_\mr{F,pbc}^{pL}$ in Section \ref{A_Subsec:modified_H_F}, the Floquet QPE algorithms for $(\epsilon_n,\ket{\Phi_n})$ are efficiently executed by the controlled block-encoding $C[O_{H_m}]$ and its inverse.
%==========================================
% Appendix Section: Approximate QSVT
%==========================================
\section{QSVT for approximate eigenstates}\label{A_Sec:Approximate_QSVT}

\renewcommand{\thetheoremD}{\ref*{A_Sec:Approximate_QSVT}\arabic{theoremD}}

In the QPE algorithm of Sec. \ref{Sec:Floquet_QPE_Sambe}, we use QSVT based on the truncated Floquet Hamiltonian $H_\mr{F}^L$, but we apply it to a state expanded by the set of approximate eigenstates $\{ \ket{\Phi_n} \}$ as Eq. (\ref{Eq_QPE:QPE_H_F}).
Here, we show that the influence of this deviation $\delta_\mr{approx}$ is bounded by Eq. (\ref{Eq_QPE:delta_approx}).

Let us consider a generic time-independent Hamiltonian $H$ with a spectral decomposition $H= \sum_n E_n \ket{\phi_n}\bra{\phi_n}$.
The Hamiltonian $H$ is assumed to be renormalized as $\norm{H} \leq 1$, and thus $E_n \in [-1,1]$.
The QSVT based on $H$ applies a degree-$q$ polynomial $f_q (H)$ to a state expanded by approximate eigenstates $\{ \ket{\tilde{\phi}_n}\}_n$.
Denoting their approximate eigenvalues by $\{ \tilde{E}_n \in [-1,1]\}_n$, we examine how the QSVT based on $H$ reproduces the one based on
\begin{equation}
    \tilde{H} = \sum_{n=1}^{n_\imax} \tilde{E}_n \ket{\tilde{\phi}_n}\bra{\tilde{\phi}_n}.
\end{equation}
We summarize the result be the following proposition.

\begin{propositionD}\label{A_Prop:approximate_QSVT}
\textbf{(Approximate QSVT)}

Suppose that a given state $\ket{\psi}$ is expanded by $\ket{\psi} = \sum_{n=1}^{n_\imax} c_n \ket{\tilde{\phi}_n}$ ($\sum_n |c_n|^2 = 1$), where the approximate eigenstates $\{ \ket{\tilde{\phi}_n} \}_n$ are characterized by
\begin{equation}
    \norm{(H - \tilde{E}_n) \ket{\tilde{\phi}_n}} \leq \eta, \quad ^\exists \tilde{E}_n \in [-1,1],
\end{equation}
and $\braket{\tilde{\phi}_n | \tilde{\phi}_{n^\prime}} = \delta_{nn^\prime}$.
Then, the difference between $f_q (H)$ and $f_q (\tilde{H})$ when applied to $\ket{\psi}$ is bounded by
\begin{equation}\label{A_Eq:approximate_QSVT}
    \norm{f_q (H) \ket{\psi} - f_q (\tilde{H}) \ket{\psi}} \leq q^2 \eta \sqrt{n_\imax},
\end{equation}
for any degree-$q$ polynomial $f_q$ realized by QSVT.
\end{propositionD}

\textit{Proof of Proposition \ref{A_Prop:approximate_QSVT}.---}
We first focus on the difference of Eq. (\ref{A_Eq:approximate_QSVT}) for each approximate eigenstate $\ket{\psi}=\ket{\tilde{\phi}_n}$.
Using a degree-$(q-1)$ polynomial $g_{q-1}(x,y)$ defined by factorization $f_q (x)-f_q (y) = (x-y) g_{q-1} (x,y)$, it is evaluated as follows;
\begin{eqnarray}
    && \norm{(f_q (H)- f_q (\tilde{H}) )\ket{\tilde{\phi}_n}} \nonumber \\
    && \quad = \norm{\sum_{n^\prime} g_{q-1}(E_{n^\prime},\tilde{E}_n) (E_{n^\prime} - \tilde{E}_n) \ket{\phi_{n^\prime}}\braket{\phi_{n^\prime} | \tilde{\phi}_n}} \nonumber \\
    && \quad \leq \max_{x \in [-1,1]} (|g_{q-1}(x,\tilde{E}_n)|) \norm{(H-\tilde{E}_n) \ket{\tilde{\phi}_n}}.
\end{eqnarray}
The mean value theorem gives an upper bound on $g_{q-1}$  by 
\begin{eqnarray}
    |g_{q-1}(x,\tilde{E}_n)| &\leq& \sup_{x \in (-1,1)} (|f^{\prime}_q (x)|) \leq q^2.
\end{eqnarray}
The second inequality follows from the fact that achievable degree-$q$ polynomials in QSVT should be renormalized as $|f_q (x)| \leq 1$ ($^\forall x \in [-1,1]$), and then their derivatives cannot exceed $q^2$ by the Markov theorem \cite{Sachdeva2013-approx}.
Therefore, for generic state $\ket{\psi} = \sum_{n=1}^{n_\imax} c_n \ket{\tilde{\phi}_n}$, the difference, $\norm{f_q (H) \ket{\psi} - f_q (\tilde{H}) \ket{\psi}}$, has an upper bound $\sum_{n=1}^{n_\imax} |c_n| q^2 \eta$.
Using the inequality $\sum_{n=1}^{n_\imax} |c_n| \leq \sqrt{n_\imax}$, which is derived from the Cauchy-Schwartz inequality, we obtain Eq. (\ref{A_Eq:approximate_QSVT}). $\quad \square$

While the above results are for QSVT for hermitian matrices, their extension to generic matrices can be immediately obtained.
In the case of QPE under the truncated Floquet Hamiltonian in Section \ref{Sec:Floquet_QPE_Sambe}, we apply QSVT based on $H_\mr{F}^{pL}$ or $H_\mr{F,pbc}^{pL}$ to the input state $\ket{\Psi_0^L}$, which is a superposition of $\ket{\Phi_n^l}$ by Eqs. (\ref{Eq_QPE:Initial_decomp_finite_generic}) and (\ref{Eq_QPE:high_term_decomp}).
Then, Proposition \ref{Prop:FloquetQPE_Initial_st} suggests that the number $n_\imax$ for the terms other than the negligible state $\ket{\Psi_\mr{neg}^L}$ is bounded by
\begin{equation}
    n_\imax \leq \mr{dim}(\mcl{H}) \times 12 L,
\end{equation}
consisting of $\ket{\Phi_n^l}$ for $n=1,2,\hdots,\mr{dim}(\mcl{H})$ and $l \in [6L]$.
The error $\eta$, which provides the upper bound on $\norm{(H_\mr{F}^{pL} - (\epsilon_n-l\omega)) \ket{\Phi_n^l}}$ scales as $e^{- \Theta (L-\alpha T)}$ by Eq. (\ref{PropEq:Accuracy_Floquet_Hamiltonian_eigenvalue_2}).
This is also true for the modified Hamiltonian $H_\mr{F,pbc}^{pL}$ as discussed in Section \ref{A_Sec:boundary_condition}.
Therefore, when the QPE based on QSVT is executed with the query complexity $q_\mr{QPE}$, the error of regarding $\ket{\Phi_n^l}$ as the exact eigenstate of the truncated Floquet Hamiltonians amounts to at most
\begin{equation}
    \delta_\mr{approx} \leq q_\mr{QPE}^2 e^{- \Theta (L-\alpha T - N)},
\end{equation}
as shown in Eq. (\ref{Eq_QPE:delta_approx}).
%=============================
% Appendix: Extension
%=============================
\section{Extension to Hamiltonian with exponentially-decaying Fourier components}\label{A_Sec:Exp_decay}

\renewcommand{\thetheoremE}{\ref*{A_Sec:Exp_decay}\arabic{theoremE}}

In the main text, we focus on the cases where the Fourier indices in the Hamiltonian $H(t)$ are bounded by $|m| \leq M$ as in Eq. (\ref{Eq:Periodic_Hamiltonian}).
Here, we consider the case whose Hamiltonian is given by
\begin{equation}\label{A_Eq:Exp_decay_Hamiltonian}
    H(t) = \sum_{m=-\infty}^\infty H_m e^{-im \omega t}, \quad \norm{H_m} \leq \alpha e^{-|m|/\zeta},
\end{equation}
with positive constants $\alpha, \zeta > 0$.
For instance, these Hamiltonians can describe the Gaussian wave packet of laser light \cite{Mizuta_Quantum_2023}.
Our results in the main text, i.e. the guaranteed quasienergy from Sambe space and the quantum algorithms for quasienergy and Floquet eigenstates, can be easily extended to these cases.

To extend our results in the main text to this class of time-periodic Hamiltonians, we note the points to be confirmed: 
\begin{enumerate}[(\ref*{A_Sec:Exp_decay}-1)]
\item Exponential decay of  $\braket{l|e^{-iH_Ft}|l^\prime}_f$ in the distance $|l-l^\prime|$ (Lieb-Robinson bound)

\item Exponential decay of $\ket{\phi_n^l}$ in the Fourier index $l$

\item Efficient block-encoding of the truncated Floquet Hamiltonian $H_F^L$
\end{enumerate}

(\ref*{A_Sec:Exp_decay}-1) is required for the Floquet QPE to compute pairs of $(\epsilon_n, \ket{\phi_n(t)})$ as Section \ref{Sec:Floquet_QPE_physical}.
According to Ref. \cite{Mizuta_Quantum_2023}, a time-periodic Hamiltonian by Eq. (\ref{A_Eq:Exp_decay_Hamiltonian}) also possesses a bound, $\norm{\braket{l|e^{-iH_Ft}|l^\prime}_f} \leq e^{-\Theta (|l-l^\prime|-\alpha T)}$.
This results in the proper cutoff $L_\mr{LR} \in \Theta (\alpha T + \log (1/\varepsilon))$ for the Sambe space formalism of the Floquet operator $U(T;0)$.
(\ref*{A_Sec:Exp_decay}-2) is required for the Floquet QPE to compute pairs of $(\epsilon_n, \ket{\Phi_n})$ as Section \ref{Sec:Floquet_QPE_physical}.
The accuracy of quasienergy and Floquet eigenstates by the truncated Floquet Hamiltonian $H_\mr{F}^L$, i.e., Theorem \ref{Thm:Accuracy_quasienergy}, relies solely on the exponential decay of $\ket{\phi_n^l}$, which tells us a proper cutoff $L \in \Theta (\alpha T + \log (1/\varepsilon))$ to achieve the allowable error $\varepsilon$.
Namely, the proof of (\ref*{A_Sec:Exp_decay}-2) validates the QPE under $H_\mr{F}^L$ in the Floquet QPE algorithm.
Finally, (\ref*{A_Sec:Exp_decay}-3) is required for the Floquet QPE algorithm to run efficiently with designated oracles.
Our algorithms use the block-encoding of the truncated Floquet Hamiltonian for realizing the Floquet operator $U(T;0)$ (Section \ref{Sec:Floquet_QPE_physical}) or for executing the QPE (Section \ref{Sec:Floquet_QPE_Sambe}).
It should be constructed efficiently by queries to some block-encoding of the Hamiltonian to preserve the efficiency of time-independent cases.

(\ref*{A_Sec:Exp_decay}-1) and (\ref*{A_Sec:Exp_decay}-3) are resolved in Ref. \cite{Mizuta_Quantum_2023}: 
A time-periodic Hamiltonian $H(t)$ by Eq. (\ref{A_Eq:Exp_decay_Hamiltonian}) has the Lieb-Robinson bound on the propagation, given by
\begin{equation}
    \norm{\braket{l|e^{-iH_\mr{F}t}|l^\prime}_f} \leq \exp \left( - \frac{|l-l^\prime|}{\zeta^\prime} + 2 \zeta^{\prime \prime} \alpha t + \frac{2}{\zeta^{\prime \prime}} \right),
\end{equation}
with the two constants $\zeta^\prime = (1/\zeta - 1 + e^{-1/\zeta})^{-1}$ and $\zeta^{\prime\prime} = (1-e^{-1/\zeta})^{-1}$.
For the efficient block-encoding, we assume that the Hamiltonian $H(t)$ by Eq. (\ref{A_Eq:Exp_decay_Hamiltonian}) is written by
\begin{equation}
    H(t) = \sum_{j=1}^J \alpha_j (t) H_j, \quad \alpha_j (t+T) = \alpha_j (t), 
\end{equation}
and that we can organize block-encoding $O_{H_j}$ respectively for each time-independent operator $H_j$.
Then, in a similar manner to Section \ref{Subsec:Modified_H_F}, we can organize the modified Floquet Hamiltonian suitable for block-encoding by
\begin{equation}
    H_\mr{F,pbc}^L = \sum_{l \in [L]} \left( \sum_{j=1}^J \mr{Add}_l^{[L]} \otimes (\alpha_j^l H_j) - l\omega \ket{l}\bra{l}_f \otimes I \right),
\end{equation}
where $\alpha_j^l = T^{-1} \int_0^T \dd t \, \alpha_j(t) e^{il\omega t}$ denotes the Fourier component of $\alpha_j(t)$.
The block-encoding of $H_\mr{F,pbc}^L$ can be constructed by one query to $C[O_{H_j}]$ respectively for $j=1,2,\hdots,J$ and some other cheap primitive gates like Proposition \ref{Prop:block_encode_H_F_pbc}.
At the same time, running the algorithms with $H_\mr{F,pbc}^L$ is essentially the same as running with the original Floquet Hamiltonian $H_\mr{F}^L$ under $L \in \Theta (\alpha T + \log (1/\varepsilon))$.
The remaining task for the extension is to prove condition (\ref*{A_Sec:Exp_decay}-2).
We end up with this section by showing the counterpart of Theorem \ref{Thm:Tail_Eigenstates} as follows.

\begin{theoremE}\label{A_Thm:Tail_Eigenstate_exp_decay}
\textbf{}

Let $\ket{\phi_n(t)}$ be a Floquet eigenstate of a time-periodic Hamiltonian with exponentially-decaying Fourier components by Eq. (\ref{A_Eq:Exp_decay_Hamiltonian})
Then, its Fourier component $\ket{\phi_n^l}$ exponentially decays in the Fourier index $l$ by
\begin{equation}\label{A_Eq:Tail_Eigenstate_exp_decay}
    \norm{\ket{\phi_n^l}} \leq \exp \left( - \frac{|l|-1/2}{4 \zeta}+ \frac{\coth (1/4\zeta)}{8\pi \zeta} \alpha T \right),
\end{equation}
when the quasienergy $\epsilon_n$ belongs to $\mr{BZ}=[-\omega/2,\omega/2)$.
\end{theoremE}

\textit{Proof of Theorem \ref{A_Thm:Tail_Eigenstate_exp_decay}.---}
We follow the proof of Theorem \ref{Thm:Tail_Eigenstates} in Section \ref{A_Sec:Tails_Eigenstate}.
We can prove Proposition \ref{A_Prop:Relation_eigenspaces} in the same way, and this part does not affect the bound itself.
The difference appears in the evaluation of $\norm{H_{\mr{Add},I}^L(\tau)}$ in Eq. (\ref{A_Eq:H_add_int_bound}) as follows.
\begin{eqnarray}
    \norm{H_{\mr{Add},I}^L(\tau)} &\leq& \sum_{m \in \bbZ} e^{m\omega \tau} \norm{H_m} \nonumber \\
    &\leq& \alpha \left( 2 \sum_{m=0}^\infty e^{-m (\zeta^{-1} - \omega \tau)} -1 \right),
\end{eqnarray}
in Proposition \ref{A_Prop:decay_truncated_eigenstates}.
For $\tau \in [0, 1/(2 \zeta \omega)]$, this gives a bound, $\norm{H_{\mr{Add},I}^L(\tau)} \leq \alpha \coth (1/4\zeta)$. 
By choosing the parameter $\lambda = 1/(4\zeta \omega)$ in Eq. (\ref{PropEq:decay_truncated_1}), we arrive at Eq. (\ref{A_Eq:Tail_Eigenstate_exp_decay}). $\quad \square$

Compared to Theorem \ref{Thm:Tail_Eigenstates}, the characteristic scale $\zeta \in \order{1}$ plays a role of the maximum Fourier index $M$ in a time-periodic Hamiltonian by Eq. (\ref{Eq:Periodic_Hamiltonian}).
This exponential decay implies the approximation of the quasienergy $\epsilon_n$ by the eigenvalue of the truncated Floquet Hamiltonian $H_\mr{F}^L$ with an error up to $e^{-\Theta (L-\alpha T)}$ as in Theorem \ref{Thm:Accuracy_quasienergy}.
As a consequence, all the results in the main text are valid also for the class of Hamiltonians by Eq. (\ref{A_Eq:Exp_decay_Hamiltonian}).
The cost of the Floquet QPE for $(\epsilon_n, \ket{\phi_n(t)})$ [Section \ref{Sec:Floquet_QPE_physical}] or $(\epsilon_n,\ket{\Phi_n})$ [Section \ref{Sec:Floquet_QPE_Sambe}] remains given by Theorem \ref{Thm:algorithm_phys} and Theorem \ref{Thm:algorithm_Sambe} respectively.
The cost of the Floquet eigenstate preparation discussed in Section \ref{Sec:Floquet_preparation} is summarized by Table \ref{Table:eigenstate_preparation}.
We conclude that the quantum algorithm for computing quasienergy and Floquet eigenstates can achieve near optimal query complexity even for time-periodic Hamiltonians having exponentially-decaying Fourier components.
This result is reminiscent of Hamiltonian simulation \cite{Mizuta_Quantum_2023}, in which real-time dynamics can be simulated with nearly optimal query complexity both for time-periodic Hamiltonians with a finite number of Fourier components and for those with exponentially-decaying Fourier components.

\end{document}